\documentclass[11pt,a4paper]{article}
\pdfoutput=1
\usepackage{amssymb,amsmath,amsfonts,mathrsfs,enumerate,wrapfig}
\usepackage[utf8]{inputenc}

\usepackage{multirow}
\usepackage{array}
\usepackage{makecell}
\usepackage{mathtools, nccmath, textcomp}

\newcolumntype{C}[1]{>{\centering\arraybackslash}m{#1}}
\usepackage{amsmath}
\usepackage{slashed}
\usepackage{jheppub}
\usepackage{latexsym}
\usepackage{multirow}
\usepackage[usenames,dvipsnames,table]{xcolor}
\usepackage{graphicx,subfigure}
\usepackage{epsfig}  
\usepackage{epsf}    
\usepackage{dcolumn}
\usepackage{bm}
\usepackage{dcolumn}
\usepackage{textcomp}
\usepackage{float}
\usepackage{lipsum}
\usepackage{hypcap}
\usepackage[]{hyperref}
\usepackage{makecell}
\usepackage{color}
\usepackage{pifont}
\usepackage{appendix}
\usepackage{url}
\usepackage{cancel}
\allowdisplaybreaks[1]
\usepackage{import}
\usepackage{subfiles}
\usepackage{filecontents}
\usepackage{threeparttable, tablefootnote}
\usepackage{slashed}
\usepackage{placeins}
\usepackage{tikz}
\usepackage{caption}
\usetikzlibrary{snakes}\usepackage{subcaption}
\usepackage{subcaption}
\let\Oldsection\section
\renewcommand{\section}{\FloatBarrier\Oldsection}

\let\Oldsubsection\subsection
\renewcommand{\subsection}{\FloatBarrier\Oldsubsection}

\let\Oldsubsubsection\subsubsection
\renewcommand{\subsubsection}{\FloatBarrier\Oldsubsubsection}

\newcommand{\mzp}{M_{Z^\prime}}

\newcommand{\mdm}{M_\mathrm{DM}}

\usepackage{multicol}
\usepackage{color}
\usepackage{comment}
\usepackage{multicol}
\usepackage{mathtools}
\newcounter{example}[section]

   \newcounter{question}[section]

\DeclareMathAlphabet      {\mathbfit}{OML}{cmm}{b}{it}

\usepackage{wrapfig}
\definecolor{light-gray}{gray}{0.95}
\usepackage{tcolorbox}
\usepackage[normalem]{ulem}
\definecolor{darkgreen}{cmyk}{1,0,1,0.4}

\definecolor{pink}{cmyk}{0.4,1,0.3,0}

\def\com2#1{\textcolor{red}{\it{#1}}}
\hypersetup{
  bookmarks=true,         
  unicode=false,          
  pdftoolbar=true,        
 pdfmenubar=true,        
 pdffitwindow=true,     
 pdfstartview={FitH},    
 pdfsubject={GUT and Topological defects},   
 pdfnewwindow=true,      
 pdfcreator={RevTeX},
 colorlinks=true,       
 linkcolor=red,          
 citecolor=blue,        
 filecolor=black,      
 urlcolor=blue,           
  }

\makeatletter
\renewcommand{\fnum@table}{\textbf{\tablename~\thetable}}
\renewcommand{\fnum@figure}{\textbf{\figurename~\thefigure}}
\makeatother
\title
{Multipartite dark matter in a gauge theory of leptons}
 \author[a]{{Utkarsh Patel}}
 \author[b]{{, Avnish}}
 \author[a]{{, Sudhanwa Patra}}
 \author[c,d]{{, Kirtiman Ghosh}}
 \affiliation[a]{Department of Physics, Indian Institute of Technology Bhilai, Durg 491002, India}
 \affiliation[b]{Department of Physics, Indian Institute of Technology Guwahati, North Guwahati, Assam 781039, India}
 \affiliation[c]{Institute of Physics, Sachivalaya Marg, Bhubaneswar}
 \affiliation[d]{Homi Bhabha National Institute, Training School Complex, Anushakti Nagar, Mumbai 400085, India}
 \emailAdd{utkarshp@iitbhilai.ac.in}
 \emailAdd{ avnish.ephy@gmail.com}
 \emailAdd{ sudhanwa@iitbhilai.ac.in}
 \emailAdd{ kirti.gh@gmail.com, kirtiman.ghosh@iopb.res.in}

\abstract
{The classical conservation of the lepton number is an accidental symmetry present in the Standard Model (SM). Thus, we consider here a scenario where the SM is extended with a U(1) gauge group, promoting the lepton number to a local symmetry. The gauge anomaly cancellations necessitate the extension of the particle spectrum with several beyond the SM (BSM) particle fields. The extended lepton gauge group breaks around the TeV scale via spontaneous symmetry breaking, and a $Z_2$ symmetry remains, which ensures the stability of the light $Z_2$ odd BSM particles. Interestingly, the particle spectrum of the model has two distinct dark sectors with one having a Dirac type DM and the other one containing a Majorana type DM, thus resulting in a multipartite dark matter scenario. We have explored the available parameter space consistent with the observed dark matter relic density and direct detection measurements for both of the DM particles. Having a Majorana dark matter, we have also studied for the gamma line signatures to constraint the parameter space from the indirect dark matter detection experiments like FermiLAT and CTA.}

\begin{document}
\maketitle
\flushbottom
\newpage

\section{Introduction}
\label{sec:intro}
After the recent experimental detection of 125 GeV massive Higgs boson
at the Large Hadron Collider~(LHC) experiment \cite{CMS:2012qbp, ATLAS:2012yve}, yet
undetermined nature of dark matter (DM) is among the most pressing
fundamental questions that are motivating the searches for physics beyond the
Standard Model (SM). In the literature, there is a plethora of
theoretical frameworks to address the issue of DM \cite{Tao:1996vb,
Ma:2006km, Moroi:1991mg, Moroi:1992zk, Kang:2007ib, Babu:2008ge,
Heinemeyer:2022anz, Antoniadis:2001cv, Hosotani:2004wv, Agashe:2004rs,
Couture:2017mbd, Duffy:2009ig, Arkani-Hamed:2002ikv, Han:2003wu,
Perelstein:2003wd, Schmaltz:2005ky, Dobrescu:1997nm, Chivukula:1998wd,
Collins:1999rz, He:2001fz, Villanueva-Domingo:2021spv}. On the
experimental frontier, in the last decades, massive efforts have been
put into detecting these non-luminous and elusive particles in various
experiments: direct detection \cite{Liu:2017drf, Schumann:2019eaa,
Misiaszek:2023sxe}, indirect detection \cite{Gaskins:2016cha,
PerezdelosHeros:2020qyt}, and collider experiments \cite{Boveia:2018yeb,
Argyropoulos:2021sav, PerezAdan:2023phe}. See Ref. \cite{Arbey:2021gdg}
for a detailed discussion on DM.

Among the various extensions of the SM, simple gauge extensions of the
SM gauge group ${\rm SU}(3)_C \times {\rm SU}(2)_L \times {\rm U}(1)_Y$
provide the most straightforward avenue to investigate beyond the
Standard Model (BSM) physics. In the SM, the classical conservation of
the baryon and lepton numbers are accidental symmetries. Recently, Ref.
\cite{FileviezPerez:2010gw} has promoted these accidental symmetries as local symmetries to obtain cosmologically viable candidate for dark matter. A significant
interest in such extensions is observed in recent years~\cite{Chao:2010mp, Dulaney:2010dj, Dong:2010fw, Ko:2010at, FileviezPerez:2011pt, Duerr:2013dza, Schwaller:2013hqa, Duerr:2013lka, FileviezPerez:2013zsr, FileviezPerez:2014lnj, Lee:2014tba, Duerr:2014wra, Chang:2018vdd, Caron:2018yzp, Madge:2018gfl, Carena:2019xrr, Altmannshofer:2019xda, Aguilar-Saavedra:2019iil, Fornal:2020esl, Ma:2020quj, Restrepo:2022cpq, Bosch:2023spa}. Usually, in these theories with typical gauge group
structure ${\rm SU}(3)_C \times {\rm SU}(2)_L \times {\rm U}(1)_Y \times
{\rm U}^\prime(1)$, the extended gauge group $U^\prime(1)$ is broken around the TeV scale to have collider detectable signatures and a remnant $Z_2$ symmetry ensures the stability of cold dark matter candidate. Arguably, these simplified extensions of the SM must be self-consistent by being anomaly-free. The particle spectrum of these BSM scenarios are usually motivated from the requirements of anomaly cancellation.

In this article, we have explored the dark matter phenomenology in a simplified model with gauged lepton number. The gauge anomaly cancellation is ensured by introducing a number of fermions instead of a three-generation right-handed neutrino within the particle spectrum given in the appendix of Ref.~\cite{FileviezPerez:2019cyn}. These fermions are charged under $U(1)_{\ell}$ and get mass through the spontaneous breaking of $U(1)_{\ell}$. The remnant $Z_2$ symmetry ensures the
stability of the lightest $Z_2$--odd fermion. The
most striking feature of this framework is that both Dirac and
Majorana-type dark matter candidates are available. In this
multi-particle dark matter framework, there is a significant
contribution in dark matter relic density (RD) from both Dirac and
Majorana-type dark matter candidates. Interestingly, in the case of Majorana-type dark matter, we get gamma lines as smoking gun signatures in the dark matter indirect detection experiments. The light neutrino masses in our model can be realized by following the procedure outlined in Ref.~\cite{Patra:2016ofq}. In this approach, the addition of a heavy scalar field, $\Delta$, can induce a Type-II like seesaw mechanism to generate neutrino masses. Remarkably, the already added $\phi_2$ scalar for breaking $U(1)_\ell$ in our framework will then facilitate this mechanism by replacing the trilinear term $\mu_\Delta \Delta H H$ with the coupling $H \Delta H \phi_2$ to induce a VEV for $\Delta$, ensuring the process remains consistent with the gauged $\ell$ symmetry of the model. The vacuum expectation value (VEV) of $\phi_2$ will thus connect the neutrino mass to the $\ell$ breaking scale, providing a natural way to account for light neutrino masses in our work. Also, as pointed in Ref.~\cite{Patra:2016ofq}, the light neutrino masses via this procedure can be generated with $M_\Delta\sim 10^8\text{ Gev}$, thus ensuring that it does not affect the dark matter phenomenology to be discussed in this work. To keep our discussion streamlined with DM phenomenology, we refrain from performing a collider analysis in this work and plan to address it in a future extension.

The structure of this article is as follows: Sec. \ref{sec:II-model} introduces the gauge theory of leptons with various gauge anomalies and presents the particle spectrum for this work. Sec. \ref{sec:IntLag} describes the interaction Lagrangian and discusses mass generation and mixings among the fields. Phenomenological aspects associated with the DM candidates from relic analysis and direct-indirect searches are addressed in Sec.~\ref{sec:DMPheno}. Finally, we conclude this article in conclusion. An appendix containing the relevant Feynman diagrams is included at the end.
\section{Model Framework}
\label{sec:II-model}
The model framework in this work is based on the gauge theory of leptophilic forces. The gauge group of such a construct is given as
\begin{equation}
 G_{LF} =SU(3)_C \otimes SU(2)_L \otimes U(1)_Y \otimes U(1)_{\ell}\,,
 \label{eq:mf1}
\end{equation}
where, the usual SM has been extended by a $U(1)_{\ell}$ gauge symmetry. The SM fermionic fields and their transformations under this gauge group are presented in Table~\ref{Tab:1}.

The addition of extra gauge symmetry other than the usual SM gauge group allows room for non-zero triangle anomalies in a quantum theory. Thus, the cancellation of such anomalies needs to be ensured before proceeding with the framework. In a $U(1)_{\ell}$ model as described above, the newly introduced gauge anomalies are listed as:
\begin{gather*}
\mathcal{A}[SU(3)^2_C \otimes U(1)_\ell],  \hspace{.5in} \mathcal{A}[SU(2)^2_L \otimes U(1)_\ell], \\
\mathcal{A}[U(1)^2_Y \otimes U(1)_\ell],  \hspace{.5in} \mathcal{A}[U(1)_Y \otimes U(1)^2_\ell],\\
\mathcal{A}[U(1)^2_Y \otimes U(1)_\ell],  \hspace{.5in} \mathcal{A}[U(1)_Y \otimes U(1)^2_\ell].
\label{eq:mf2}
\end{gather*}
\noindent
Out of these, the vanishing gauge anomalies are:
\begin{gather}
 \mathcal{A}[SU(3)^2_C \otimes U(1)_{\ell}],\hspace{0.5in}\mathcal{A}[U(1)_Y \otimes U(1)^2_{\ell}].
 \label{eq:mf3}
\end{gather}
and the non-vanishing ones are given as:
\begin{gather}
\hspace{-0.65in}\mathcal{A}[SU(2)^2_L \otimes U(1)_\ell] =\frac{3}{2}, \hspace{.5in} \mathcal{A}[U(1)^3_\ell] =3, \nonumber \\
\mathcal{A}[U(1)^2_Y \otimes U(1)_\ell] =-\frac{3}{2}, \hspace{.5in} \mathcal{A}[\text{gravity}^2 \otimes U(1)_\ell] =3.
\label{eq:mf4}
\end{gather}
There have been attempts in the literature previously, to extend the SM with locally gauged Baryon and Lepton numbers~\cite{FileviezPerez:2014lnj,Duerr:2013dza}. Also, the simplest possible self-consistent gauge theory that could be constructed using such local symmetries is the extension of SM with a $U(1)$ symmetry of the difference between Baryon and Lepton number, i.e., $U(1)_{B-{\ell}}$. Such extension is readily incorporated in higher-scale theories such as left-right symmetric theories and unification models~\cite{Aulakh:1982sw,Mohapatra:1980qe,Clark:1982ai,Aulakh:1997ba,Bajc:2004xe,Aulakh:1998nn,Pandita:1998dq,Kuchimanchi:1993jg}. In this context, for the case where only Lepton number is promoted as a gauge symmetry, the generated vanishing and non-vanishing anomalies are provided in Eqs.~(\ref{eq:mf3}) and (\ref{eq:mf4}). Refs.~\cite{Chao:2010mp,FileviezPerez:2019cyn,Jeong:2015bbi,Lee:2014tba,Schwaller:2013hqa} are some of the notable works that have considered a leptophilic extension of SM.

\par In Ref.~\cite{FileviezPerez:2019cyn}, anomaly cancellation for leptophilic gauge group as shown in Eq.~(\ref{eq:mf1}) is tackled by introducing three generations of SM singlet right-handed neutrinos, $\nu_{Ri}$, a pair of $SU(2)_L$ doublet fermions $(\Psi_{L}$ \text{and} $\Psi_{R})$, a left-handed isospin triplet $(\Sigma_{L})$ and a left-handed singlet under SM $(\rho_{L})$ along with an extended scalar sector consisting of Higgs and a scalar $(S_L)$ required to break the high scale Lepton symmetry. These fermions along with their transformations under $SU(3)_{C}\otimes SU(2)_{L}\otimes U(1)_{Y}\otimes U(1)_{\ell}$ gauge group are listed as:
\begin{equation*}
 \Psi_L=\begin{pmatrix} \Psi^+_L \\ \Psi^0_L \end{pmatrix} \equiv [1,2,1/2,3/2],\hspace{1.0in} \Psi_R=\begin{pmatrix} \Psi^+_R \\ \Psi^0_R \end{pmatrix} \equiv [1,2,1/2,-3/2],
\end{equation*}
\begin{equation*}
\Sigma_L=\frac{1}{\sqrt{2}} \begin{pmatrix} \Sigma^0_L & \sqrt{2} \Sigma^+_L \\
           \sqrt{2} \Sigma^-_L & - \Sigma^0_L \end{pmatrix} \equiv [1,3,0,-3/2],
\end{equation*}
\vspace{.01in}
\begin{equation*}
\hspace{-0.8in} \nu_{Ri}~(i=1,2,3) \equiv [1,1,0,1],\hspace{1.1in}~~~ \rho_L \equiv [1,1,0,-3/2].
\end{equation*}
Taking motivation from the works~\cite{Chao:2010mp,FileviezPerez:2019cyn}, we in the following propose an alternative way of canceling the gauge anomalies without invoking right-handed neutrinos. In doing so, this novel scenario provides us with a dark sector that has two stable DM candidates with one being Dirac and the other Majorana in nature.

\begin{table}
\centering
\begin{tabular}{ccccc}
\hline\hline
 & $\text{SU(3)}_C$ & $\text{SU(2)}_L$ & $\text{U(1)}_Y$  & $\text{U(1)}_{\ell}$\\
 Gauge fields & $\vec{G}_\mu$ & $\vec{W}_\mu$ & $B_\mu$ & $Z'_{\mu}$\\[1mm]
\hline \underline{SM Fermions:} \\
$Q_{L} =  \begin{pmatrix}

             u_L \\ d_L

            \end{pmatrix}$	& $\textbf{3}$ & $\textbf{2}$ & $1/6$ & $0$	\\

 $u_R$	& $\textbf{3}$ & $\textbf{1}$ & $2/3$ & $0$\\

 $d_R$	& $\textbf{3}$ & $\textbf{1}$ & $-1/3$ & $0$\\

$\ell_L =  \begin{pmatrix}

             \nu_L \\ e_L

            \end{pmatrix}$	& $\textbf{1}$ & $\textbf{2}$ & $-1/2$ & $1$	\\

$e_{R}$ & $\mathbf{1}$ & $\mathbf{1}$ & $-1$ & $1$ \\
\hline \hline
\end{tabular}
\caption{The SM quarks and leptons with their transformations under the leptophilic gauge group extension given as $SU(3)_C \otimes SU(2)_L\otimes U(1)_Y \otimes U(1)_{\ell}$. The generation indices for all fermions have been omitted here and the electric charge~$(Q)$ is expressed as, $Q=T^3+Y$ with $T^3$ associated with the isospin and $Y$ being the hypercharge.}
\label{Tab:1}
\end{table}

\subsection{A new $U(1)_{\ell}$ gauged model without right-handed neutrinos}
The right-handed neutrinos~$\nu_{Ri}~(i=1,2,3)$ are revoked by introducing instead a set of four exotic chiral fermions~$(\xi_{L},\eta_{L},\zeta_{1R},\zeta_{2R})$ as listed in Table~\ref{Tab:4}. Table~\ref{Tab:4} also contains the complete list of exotic fermions present in our framework. In addition to this, the model also modifies the scalar sector by introducing two $SU(2)_L$ singlet scalars $\phi_1$ and $\phi_2$ to provide masses to the exotic fermions via their VEVs, along with the usual $SU(2)_L$ doublet Higgs~($H$) scalar as presented in Table~\ref{Tab:5}. In this framework, the ${\rm U}(1)_\ell$ symmetry is completely broken when both $\phi_1$ and $\phi_2$ acquire nonzero vacuum expectation values (VEVs). However, an accidental $Z_2$ symmetry emerges due to the specific choice of gauge quantum numbers for the fields, similar to the scenario described in Ref.~\cite{Patra:2016ofq}. Under this $Z_2$ symmetry, the fields $(\xi_L, ~\eta_L, ~\zeta_{1,R}, ~\zeta_{2,R})$ and $(\Psi_L, ~\Psi_R, ~\Sigma_L, ~\rho_L)$ are odd, while all other fields are even.\footnote{It is possible to assign the fields $(\xi_L, ~\eta_L, ~\zeta_{1,R}, ~\zeta_{2,R})$ and $(\Psi_L, ~\Psi_R, ~\Sigma_L, ~\rho_L)$ to two separate $Z_2$ symmetries without any loss of generality. However, in the present work, they are considered odd under the same $Z_2$ symmetry.} Since these two sets of $Z_2$-odd fermions do not interact with each other at tree level, they are treated as two distinct parts of the same dark sector, each providing a plausible dark matter candidate through its lightest member. The accidental $Z_2$ symmetry, arising from the fractional ${\rm U}(1)_\ell$ charges of the beyond Standard Model (BSM) fermions, remains stable and is not violated by higher-dimensional operators, at least up to dimension six. This ensures the stability of the dark matter candidates. Thus, the substitution of right-handed neutrinos with exotic chiral fermions fulfills a three-fold motivation summarized below.
\begin{itemize}
 \item The newly substituted fermions exactly cancel the non-zero gauge anomalies remaining in the framework and thus provide an alternative constitution for an anomaly-free leptophilic extension of SM.
 \item These four chiral fermions~$(\xi_{L},\eta_{L},\zeta_{1R},\zeta_{2R})$ after mass matrix diagonalization, give a set of 2 Dirac fermions, where the lightest among them is a plausible DM candidate. The leptophilic quantum numbers of these exotic fermions are chosen such that they can decay only into a lighter exotic fermion in association with one or more SM particles. In the absence of any kinematically allowed decay mode for the lightest exotic fermion, it remains stable and hence, can be a candidate for DM.
 \item The DM candidate (the lighter among the $2$ Dirac fermions) is a WIMP type and thus its correct relic density value is obtained around the TeV scale DM mass range. This ensures that the $U(1)_{\ell}$ breaking scale should also lie close to TeV and, therefore, not far from the LHC reach allowing the testability of the framework from the current and near-future collider experiments.
\end{itemize}
\section{The Lagrangian and particle interactions}
\label{sec:IntLag}
In this section, we present the relevant interaction Lagrangian for the fields in our model. We also derive here the expressions for the physical eigenstates of the various scalars and fermions present in the theory. The masses and mixings of the DM states are also discussed here.
\begin{table}
\centering
\begin{tabular}{cccc}
\hline\hline
 & $\text{SU(2)}_L$ & $\text{U(1)}_Y$  & $\text{U(1)}_{\ell}$\\
 $\Psi_L=\begin{pmatrix} \Psi^+_L \\ \Psi^0_L \end{pmatrix}$ & $\mathbf{2}$ & $1/2$ & $3/2$\\[.15in]
 $\Psi_R=\begin{pmatrix} \Psi^+_R \\ \Psi^0_R \end{pmatrix}$ & $\mathbf{2}$ & $1/2$ & $-3/2$\\[.15in]
 $\Sigma_L=\frac{1}{\sqrt{2}}\begin{pmatrix} \Sigma^0_L & \sqrt{2}\Sigma^+_L\\ \sqrt{2}\Psi^-_L & -\Sigma^0_L \end{pmatrix}$ & $\mathbf{3}$ & $0$ & $-3/2$\\
 $\rho_L$ & $\mathbf{1}$ & 0 & $-3/2$\\
\hline\hline \underline{Newly Introduced fermions:} \\
		 $\xi_{L}$	    & $\mathbf{1}$	& 0   &  $-4/3$	\\
		 $\eta_{L}$	    & $\mathbf{1}$	& 0	  &   $-1/3$	\\
		 $\zeta_{1R}$	& $\mathbf{1}$	& 0	  &   $2/3$	\\
		 $\zeta_{2R}$	& $\mathbf{1}$	& 0	  &   $2/3$
\\[1mm] \hline
\hline
\end{tabular}
\caption{Complete list of exotic fermions in our framework for a anomaly free leptophilic gauge group given in Eq.~(\ref{eq:mf1}). The newly introduced four neutral fermions with fractional leptonic charges replace the 3-generation right-handed neutrino~$(\nu_{R})$ in the model from Ref.~\cite{Chao:2010mp}. All the particles in this table are color singlets.}
\label{Tab:4}
\end{table}
\begin{table}
\centering
\begin{tabular}{cccc}
\hline\hline
 & $\text{SU(2)}_L$ & $\text{U(1)}_Y$  & $\text{U(1)}_{\ell}$\\
\hline
\underline{Scalar fields}\\
$H=\begin{pmatrix}

             H^0 \\ ~H^-

            \end{pmatrix}$ & $\mathbf{2}$ & $\phantom{+}1/2$  & $0$ \\
$\phi_1$ & $\mathbf{1}$ & $0$ & $1$ \\
$\phi_2$ & $\mathbf{1}$ & $0$ & $2$ \\
\hline\hline
\end{tabular}
\caption{Scalar sector for our framework. Here, $H$ is the electroweak~(EW) symmetry breaking Higgs particle and $\phi_1$, $\phi_2$ are the newly introduced scalars that take non-zero VEVs and break $U(1)_{\ell}$ symmetry. All these particles are color singlets.}
\label{Tab:5}
\end{table}
\subsection{The scalar sector and symmetry breaking}
\label{subsc:sc}
The relevant part of the most general Lagrangian associated with exotic bosons can be given by:
\begin{align}
\label{eq:TheModel}
	\mathscr{L}=
	& - \frac{1}{4} F_{Z^\ell}^{\mu \nu}F^{Z^\ell}_{\mu \nu}
	  - \frac{1}{2} \kappa F_{Z^\ell}^{\mu \nu} F^{B}_{\mu \nu}
	\nonumber \\
	&+ \left( \mathcal{D}^\mu H \right)^\dagger \left( \mathcal{D}_\mu H \right) +
	|\left( \partial_\mu + \,i\,g_{\ell}\,Z^\ell_\mu \right) \phi_1|^2
		 + |\left( \partial_\mu +2 \,i\,g_{\ell}\,Z^\ell_\mu \right) \phi_2|^2
	     - V( H,  \phi_1,  \phi_2 )
	  \, ,
\end{align}
where $\kappa$ denotes the kinetic mixing term between $U(1)_{Y}$ and $U(1)_{\ell}$ gauge bosons $Z$ and $Z^{\ell}$, respectively. Also, the relevant scalar potential involving $H$, $\phi_1$ and $\phi_2$ invariant under the gauge symmetry reads as,
 \begin{align}
V(H,\phi_1,\phi_2)= & \mu^2_H  H^\dagger H + \lambda_H (H^\dagger H)^2
     + \mu^2_1 \phi^\dagger_1 \phi_1 + \lambda_1 (\phi^\dagger_1 \phi_1)^2
      +\mu^2_2 \phi^\dagger_2 \phi_2  \nonumber \\
      &+ \lambda_2 (\phi^\dagger_2 \phi_2)^2  +\lambda_{4} (H^\dagger H) (\phi^\dagger_1 \phi_1)
      +\lambda_{5} (H^\dagger H) (\phi^\dagger_2 \phi_2)\nonumber \\
      &+\lambda_{3} (\phi^\dagger_1 \phi_1) (\phi^\dagger_2 \phi_2)
      +\mu_{12} \left( \phi_2 \phi^{\dagger^2}_1 + \phi_2^\dagger \phi^{^2}_1 \right).
      \label{eq:V}
 \end{align}
The necessary conditions to bound this potential from below are given as,
\begin{equation}
 \lambda_H, \lambda_1, \lambda_2 \geq 0,\;\;
 \lambda_4 + \sqrt{\lambda_H \lambda_1} \geq 0\, , \;\; \lambda_5 + \sqrt{\lambda_H \lambda_2} \geq 0\, , \;\;
  \lambda_3 + \sqrt{\lambda_1 \lambda_2} \geq 0 \,.
\end{equation}
The spontaneous symmetry breaking~(SSB) of the gauge group $SU(2)_L \times U(1)_Y \times U(1)_{\ell}$ to the electromagnetism~(EM) gauge group $U(1)_{\rm EM}$ proceeds in two stages. At first, the gauge group $SU(2)_L \times U(1)_Y \times U(1)_{\ell}$ break down to the EW symmetry $SU(2)_L \times U(1)_Y$ through non-zero vacuum expectation values (VEVs) to the scalars $\phi_1$ and $\phi_2$ at a scale much higher than the EWSB. Subsequently, $SU(2)_L \times U(1)_Y$ breaks down to $U(1)_{\rm EM}$ via a non-zero VEV of the neutral component~$(H^0)$ of the Higgs scalar doublet.

One can parameterize the scalar fields $H^0,~\phi_1,~{\rm and}~ \phi_2$ in terms of the real and pseudo scalars as:
\begin{align}
&H^0 =\frac{1}{\sqrt{2} }(v+h)+  \frac{i}{\sqrt{2} } G^0\, , \nonumber \\
& \phi_1 = \frac{1}{\sqrt{2} }(v_1+h_1)+  \frac{i}{\sqrt{2} } A_1\,, \nonumber \\
& \phi_2 = \frac{1}{\sqrt{2} }(v_2+h_2)+  \frac{i}{\sqrt{2} } A_2\, .
\end{align}
where $\langle H^0\rangle=v/\sqrt2$, $\langle \phi_1\rangle=v_1/\sqrt2$, $\langle \phi_2\rangle=v_2/\sqrt2$ are denoting the VEVs of the scalar fields $H^0,~\phi_1,~{\rm and}~ \phi_2$, respectively. One important parameter relevant for dark matter phenomenology is $\tan\beta =\frac{v_1}{v_2}$, which is the ratio between the VEVs of the exotic scalars $\phi_1$ and $\phi_2$.

The minimization of scalar potential results in the following
\begin{align}
\mu^2_H& = -\left(\lambda_H v^2 + \frac{\lambda_4}{2} v^2_1 +   \frac{\lambda_5}{2} v^2_2 \right),  \nonumber \\
\mu^2_1& = -\left(\lambda_1 v^2_1  +   \frac{ \lambda_3}{2} v^2_2 + \frac{\lambda_4 }{2} v^2 +\sqrt{2} v_2 \mu_{12} \right) , \nonumber \\
\mu^2_2& = -\left(\lambda_2 v^2_2 + \frac{\lambda_3}{2} v^2_1 +   \frac{ \lambda_5}{2} v^2 +\frac{1}{\sqrt{2}} \frac{v^2_1 \mu_{12}}{v_2} \right).
\end{align}
Being charged under the $U(1)_\ell$, non-zero VEVs of $\phi_1$ and $\phi_2$ breaks the $U(1)_\ell$ symmetry and lead to a massive gauge boson, $Z^{\ell}$. Under a negligible kinetic mixing approximation\footnote{The bounds on the kinetic mixing parameter~$(\epsilon)$ between the SM hypercharge gauge boson~$Z$ and any new abelian vector boson~$Z'$ from precison measurements can be referred from Ref.~\cite{Hook:2010tw}. The strong constraints on $\epsilon$ parameter specifically in the context of a $U(1)_{\ell}$ extension of the SM can be referred from Ref.~\cite{Schwaller:2013hqa}.} for the groups $U(1)_Y$ and $U(1)_\ell$, the gauge and mass eigen states of $U(1)_\ell$ gauge boson aligns to each other i.e. $Z^{\ell} \simeq Z^{\prime}$. Hence, throughout the remainder of this article, we will use $Z^{\prime}$ to denote the exotic gauge state unless explicitly stated otherwise. Under this zero kinetic mixing approximation, $Z^{\prime}$ mass simply reads as,
\begin{eqnarray}
	M^2_{Z^\prime}= g^2_{\ell} \left( v_1^2+ 4 v_2^2 \right).
\end{eqnarray}
For our future convenience, we recast it in terms of $\tan\beta$, the ratio between the VEVs of the scalars $\phi_1$ and $\phi_2$. Hence, $Z'$ mass is given by,
\begin{equation}
M^2_{Z^\prime}= g^2_{\ell} v_2^2 \left(4+\tan^2\beta\right).
\end{equation}
Since, $Z^{\prime}$ directly couples to the SM leptons, given they are charged under $U(1)_\ell$, the constraints on its mass can be determined at the collider experiments. The bounds on $M_{Z'}$ from the LEP II experiment~\cite{Carena:2004xs} reads as,
\begin{equation}
  \frac{M_{Z^\prime}}{g_{\ell}} \gtrsim 7 ~{\rm TeV }.
\end{equation}
where, $g_{\ell}$ is the gauge coupling strength associated with $U(1)_{\ell}$. The scalar potential terms associated with $\lambda_4$ and $\lambda_5$ in Eq.~(\ref{eq:V}) invoke the mixing of the SM Higgs boson with the other two scalar fields. Since the properties~(decay modes, branching ratios, production rates etc.) of the CMS and the ATLAS observed scalar boson with mass 125 GeV is in fair agreement with the properties of the SM Higgs boson, these mixings must be minuscule. Hence, for simplicity, we have neglected such mixing in our work by effectively fixing the values of $\lambda_4$ and $\lambda_5$ to zero.

In addition to the SM Higgs boson~$(H)$, the CP-even scalar spectrum consists of two other states that mix with each other and their mass matrix before diagonalization is given as,
\begin{align}
\label{eq:scalarMass}
\mathcal{M}^2_{\rm Higgs} =
\begin{pmatrix}
2 \lambda_1 v^2_1     &   v_1( \lambda_3  v_2 + \sqrt{2}  \mu_{12} )   \\
v_1( \lambda_3  v_2 + \sqrt{2} \mu_{12})   &   2 \lambda_2 v^2_2 -\frac{\mu v^2_1}{\sqrt{2}v_2}
\end{pmatrix}
\end{align}
in the basis $(h_1,h_2)$. The resultant mass eigenstates of CP-even scalars are denoted by $H_1$ and $H_2$ with $M_{H_1}$ and $M_{H_2}$, respectively. The eigenstates $h_{1,2}$ and $H_{1,2}$ are associated to each other as
\begin{equation}
\begin{pmatrix} h_1\\h_2 \end{pmatrix}=\begin{pmatrix}\cos\theta &\sin\theta\\
-\sin\theta &\cos\theta \end{pmatrix}\begin{pmatrix}H_1\\H_2 \end{pmatrix},
\end{equation}
where $\theta$ is the mixing angle. In this work, we have taken the physical masses and mixing angle as the free parameters in the scalar sector and hence, the couplings can be re-expressed in their terms as following
\begin{align}
\lambda_1=& \frac{1}{2v_1^2}\left[\cos^2\theta\,M^2_{H_1}+\sin^2\theta\,M^2_{H_2}\right],\\
\lambda_2=& \frac{1}{2v_2^2}\left[\sin^2\theta\,M^2_{H_1}+\cos^2\theta\,M^2_{H_2}+\frac{\mu_{12} v_1^2}{\sqrt 2v_2}\right],\\
\text{and }\lambda_3  =&\frac{1}{v_1v_2}\left[\sin\theta\cos\theta(M_{H_2}^2-M_{H_1}^2)-\sqrt 2\mu_{12} v_1\right].
\end{align}
There are also two CP-odd scalars denoted by $A_1$ and $A_2$ for which the mass matrix can be given by:
\begin{align}
\mathcal{M}^2_{\mbox{\small CP-odd}} =
\begin{pmatrix}
-2 \sqrt{2} v_2 \mu_{12}  &  \sqrt{2} v_1  \mu_{12} \\
\sqrt{2} v_1 \mu_{12}    &   -\frac{v^2_1}{\sqrt{2}} \frac{\mu_{12}}{v_2}
\end{pmatrix},
\label{eq:ss1}
\end{align}
As expected, one of the resulting mass eigenstates has zero mass eigenvalue and is recognized as the Goldstone boson of $Z^{\prime}$. Here, the mixing angle~($\alpha$) between the CP-odd scalars~$A_1$ and $A_2$ is completely determinable in terms of the VEVs: $\sin\alpha= \sqrt{4v_2^2/(v_1^2+4v_2^2)}$. The only physical CP-odd eigenstate is represented by $A$ and has mass $M_A$. The parameter $\mu_{12}$ in Eq.~(\ref{eq:scalarMass}) is expressed as
\begin{equation}
\mu_{12} = -\frac{M_A^2\,\sin^2\alpha}{2\sqrt 2 v_2}\,.
\end{equation}
Hence, the scalar sector of this model contains three BSM states: $H_1$, $H_2$, and $A$, which might be probed at the collider experiments for having gauge interactions with the $Z^\prime$, and the Yukawa interactions with exotic fermions.
\subsection{Exotic fermion sector}
\label{subsec:FS}
Apart from the SM Lagrangian, the kinetic terms associated with the new fermion fields required for a consistent gauge theory are given by,
\begin{align}
\label{eq:TheModel}
\mathscr{L}_{\ell}
=   &  i \, \overline{\Psi_{L}} \slashed{D} \Psi_{L} +i \, \overline{\Psi_{R}} \slashed{D} \Psi_{R} +i \, \overline{\rho_{L}} \slashed{D} \rho_{L}  +i \, \overline{\Sigma_{L}} \slashed{D} \Sigma_{L} \nonumber\\
& +i \, \overline{\xi_{L}} \slashed{D} \xi_{L} + i \, \overline{\eta_{L}} \slashed{D} \eta_{L} + i \, \overline{\zeta}_{iR} \slashed{D} \zeta_{iR}.
\end{align}

\noindent Here, $\slashed{D}$ denotes the covariant derivative and its general form is given as: \begin{equation}
\slashed{D}=\slashed{\partial} + ig_L\frac{\sigma^i}{2}W^i_{\mu}+ig_YYB_{\mu}+i\,g_\ell n_{\ell} \,Z_\mu^\ell
\label{eq:covar}
\end{equation}
with $g_{L}$, $g_Y$, and $g_{\ell}$ being weak, hypercharge, and leptophilic couplings, respectively. In addition to this, $Y$ represents usual hypercharge, and $n_{\ell}$ represents corresponding $U(1)_{\ell}$ charge. In our framework, we replace right-handed neutrinos~(RHNs) $\nu_R$ with four exotic chiral fermions. This substitution provides a unique scenario with two independently stable dark sectors. The four new fermions~$(\xi_{L},\eta_{L},\zeta_{1R},\zeta_{2R})$ interact among each other but doesn't interact with the other exotic fermions~$(\Psi_L,\Psi_R,\Sigma_L,\rho_L)$ in the model. Thus, both the sectors evolve independently of each other. Consequently, the lightest stable state from both these sectors serves as viable dark matter (DM) candidates. Thus, in this work, we analyze the DM interactions and their relevant phenomenological aspects for a two candidate DM scenario. The exact nature of these DM states is determined by the relative strengths of various Yukawa and gauge couplings involved. In \ref{subsec:DDM} and \ref{subsec:MDM}, we discuss the two independent fermionic DM sectors separately in detail.
\subsubsection{Dirac type dark matter}
\label{subsec:DDM}
The relevant Yukawa interaction terms concerning the exotic fermions~$(\eta_L, \xi_L, \zeta_{1\,R}, \zeta_{2\,R})$ are given by
\begin{eqnarray}
\mathscr{L}^{\rm DM_1}_{\rm Yuk} &=& - \left(\alpha_{i}\, \overline{\xi_L} \zeta_{i\,R}\,\phi^*_2
     + \beta_{i}\, \overline{\eta_L} \zeta_{i\,R}\, \phi^*_1+ h.c.\, \right).
     \label{eq:DMM1}
\end{eqnarray}
Here, we are keeping the notations the same as the ones used in the Ref.~\cite{Patra:2016ofq} for the clarity of the reader. Thus, $\alpha_i$, $\beta_i$ are new fermion Yukawa couplings, and the rest of the notations have their usual meaning as defined in the preceding sections. The exotic fermions in Eq.~(\ref{eq:DMM1}) being charged under $U(1)_{\ell}$ can not have bare mass term. The spontaneous breaking of $U(1)_{\ell}$ by the VEVs of $\phi_1$ and $\phi_2$ leads to the masses for these exotic fermions. Therefore, the masses of these fermions are determined by $U(1)_{\ell}$ symmetry breaking scale. The mass matrix for these fermions is presented in Eq.~(\ref{eq:diracnu}), where the lightest mass eigenstate remains stable and thus can serve as a DM candidate.
\begin{align}
\label{eq:diracnu}
M_{\rm NF}^{\rm DM_1} &=\left(\begin{array}{cc}\overline{\xi_L}& \overline{\eta_{L}}\end{array}\right) 	\left( \begin{array}{cc}  \alpha_1\,\langle \phi_2\rangle & \alpha_2\, \langle \phi_2\rangle       \\
\beta_1\,\langle\phi_1\rangle & \beta_2\langle\phi_1\rangle
		\end{array}
	\right) \left(\begin{array}{c}\zeta_{1\,R}\\ \zeta_{2\,R} \end{array}\right) + h.c.
\end{align}
The chiral gauge eigenstates~$(\eta_L, \xi_L)$ and $(\zeta_{1\,R}, \zeta_{2\,R})$ given in Eq.~(\ref{eq:diracnu}) form chiral mass eigenstates via mixing matrices $U_L$ and $U_R$ that diagonalize the fermion mass matrix as:
\begin{equation}
\begin{pmatrix}\xi_L\\ \eta_L \end{pmatrix}=U_L\begin{pmatrix}\chi_{1\,L}\\ \chi_{2\,L}\end{pmatrix},\qquad
\begin{pmatrix}\zeta_{1\,R}\\ \zeta_{2\,R}\end{pmatrix}=U_R\begin{pmatrix}\chi_{1\,R}\\ \chi_{2\,R}\end{pmatrix},
\label{eq:DMM2}
\end{equation}
These chiral mass eigenstates can be expressed as two Dirac eigenstates given by $\chi_1$ and $\chi_2$, with $\chi_1=\chi_{1\,L}+\chi_{1\,R}$ and $\chi_2=\chi_{2\,L}+\chi_{2\,R}$, and $U_{L,R}$ can each be parameterized by a mixing angle $\theta_{L,R}$ as
\begin{equation}
U_{L,R}=\begin{pmatrix}\cos\theta_{L,R} &\sin\theta_{L,R}\\
-\sin\theta_{L,R} &\cos\theta_{L,R}\end{pmatrix}.
\label{eq:DMM3}
\end{equation}
Taking the couplings $\alpha_i$ and $\beta_i$ (with $i\in \{1,2\}$) as free parameters to decide the final fermion masses is a convenient measure. Thus, we write couplings $\alpha_{1,2}$ and $\beta_{1,2}$ in terms of $M_{\chi_1,\chi_2}$ and the two mixing angles $\theta_{L,R}$; hence given as:
\begin{align}
\alpha_1  =& \frac{\sqrt 2}{v_2}\left(\cos\theta_L\cos\theta_R \,M_{\chi_1}+\sin\theta_L\sin\theta_R \,M_{\chi_2}\right),\\
\alpha_2  =& \frac{\sqrt 2}{v_2}\left(-\cos\theta_L\sin\theta_R \,M_{\chi_1}+\sin\theta_L\cos\theta_R \,M_{\chi_2}\right),\\
\beta_1  =& \frac{\sqrt 2}{v_1}\left(-\sin\theta_L\cos\theta_R \,M_{\chi_1}+\cos\theta_L\sin\theta_R \,M_{\chi_2}\right),\\
\text{and }\beta_2  =& \frac{\sqrt 2}{v_1}\left(\sin\theta_L\sin\theta_R \,M_{\chi_1}+\cos\theta_L\cos\theta_R \,M_{\chi_2}\right).
\label{eq:DMM4}
\end{align}
The interaction terms between the Dirac fermions~$(\chi_{1,2})$ and the $Z^\prime$ are given, in the mass eigenstate basis, by
\begin{align}
\mathscr{L}_{\chi Z^\prime}=&\frac{g_{\ell}}{6}\bigg[
\overline{\chi_{1}} \gamma^\mu
\bigg\{\left(5+ 3 \cos 2\theta_L \right) P_L- 4 P_R \bigg\} \chi_{1} \nonumber \\
&+\overline{\chi_{2}} \gamma^\mu
\bigg\{\left(5- 3 \cos 2\theta_L \right) P_L- 4 P_R \bigg\} \chi_{2} \nonumber \\
&+\overline{\chi_{1}} \gamma^\mu
\left(3 \sin 2\theta_L \right)P_L \chi_{2}
+\overline{\chi_{2}} \gamma^\mu
\left(3 \sin 2\theta_L  \right)P_L \chi_{1}
\bigg]
Z^\prime_\mu\,,
\label{eq:DMM5}
\end{align}
where we see no dependence on $\theta_R$ in the final expression. To keep things simple for the reader, we have considered $\chi_1$ to be lighter than $\chi_2$ throughout this work and, therefore, is a viable DM candidate. Also, as $\chi_1$ is Dirac type, so for the phenomenological analysis, we refer to the mass of $\chi_1$ as $M^{D}_{\text{DM}}$ interchangeably with $M_{\chi_1}$ in the later sections of this article. Thus, $M^{D}_{\text{DM}}$ and $M_{\chi_1}$ convey the same meaning. Now, the obtained vector ($g_{\chi V}$) and axial ($g_{\chi A}$) couplings of the Dirac DM particle~$\chi_1$ to the $Z^\prime$ are similar in their form with the ones given in Ref.~\cite{Patra:2016ofq}, except an overall negative sign in the axial-vector coupling. These play a crucial role in the dark matter phenomenology, and we express them here as:
\begin{align}
g_{\chi V} = \frac{g_{\ell}}{12} \left(6 \cos^2 \theta_{L} -2 \right) ,~~g_{\chi A} = \frac{g_{\ell}}{2} \left(\cos^2\theta_{L} +1 \right).
\label{eq:DMM6}
\end{align}

\subsubsection{Majorana type dark matter}
\label{subsec:MDM}
\noindent The relevant Yukawa interactions associated to the other exotic fermions $(\Psi_L, \Psi_R, \Sigma_L, \rho_L)$ are given as,
\begin{align}
\label{eq:TheModel}
\mathscr{L}^{\rm DM_2}_{\rm Yuk} &=  Y_{\Psi \rho} \overline{\Psi^c}_L \widetilde{H} \rho_L +Y^\prime_{\Psi \rho} \overline{\Psi}_R H \rho_L
+ Y_{\Psi \Sigma} \overline{\Sigma^c}_L \widetilde{H} \Psi_L
+ Y^\prime_{\Psi \Sigma} \overline{\Psi}_R H \Sigma_L \nonumber \\
&+ \frac{\lambda_\Psi}{\Lambda}\overline{\Psi}_R \Psi_L \phi^*_1 \phi^*_2
+ \frac{\lambda_\rho}{\Lambda}\overline{\rho^c}_L \rho_L \phi_1 \phi_2
+ \frac{\lambda_\Sigma}{\Lambda} \mbox{Tr}\big[\overline{\Sigma^c}_L \Sigma_L\big] \phi_1 \phi_2.
\end{align}
These exotic fermions not only solves anomaly conditions but also provides a novel possibility of an additional DM candidate. As the possibility of a Dirac DM has already been discussed in \ref{subsec:DDM}, so here we focus on the possibility of a Majorana-type dark matter arising from the combinations of $\Psi_{L},\Psi_{R},\Sigma_{L}\text{ and }\rho_{L}$. From Eq.~(\ref{eq:TheModel}), it can be seen that dimension-5 operators (suppressed by some new physics scale $\Lambda$) are required to generate bare masses to $\Psi_{L,R}, \Sigma_L,\rho_L$ after $U(1)_{\ell}$ breaking~\footnote{Alternatively, one can introduce a scalar $\Phi_3$ with a ${\rm U}(1)_\ell$ charge of 3 to generate Dirac mass terms at the renormalizable level. In this scenario, ${\rm U}(1)_\ell$ will again be completely broken once all the scalars acquire non-zero vacuum expectation values (VEVs). The scalar potential in this case will include a few additional terms, leading to a distinct phenomenology. While exploring this possibility is beyond the scope of the present work, it could be an intriguing direction for future studies.
}. The resulting mass matrix of neutral fermions reads as,
\begin{eqnarray}
M_{\rm NF}^{\rm DM_2}= \left(\begin{array}{cccc|c} \overline{\Psi_L^0}   &   \overline{{\Psi^0_R}^c} & \overline{\Sigma^0_L} & \overline{\rho_L}  \\  \hline
0 & \frac{\lambda_\Psi v_1v_2}{2\Lambda} & \frac{Y_{\Psi \Sigma} v}{2} & \frac{Y_{\Psi \rho} v}{\sqrt{2}} & {\Psi^0_L}^c \\
 \frac{\lambda_\Psi v_1v_2}{2\Lambda} & 0 & \frac{Y^\prime_{\Psi \Sigma} v}{2} & \frac{Y^\prime_{\Psi \rho} v}{\sqrt{2}} & {\Psi^0_R}\\
\frac{Y_{\Psi \Sigma} v}{2} & \frac{Y^\prime_{\Psi \Sigma} v}{2} &  \frac{\lambda_\Sigma v_1v_2}{\Lambda} & 0 & {\Sigma^0_L}^c \\
\frac{Y_{\Psi \rho} v}{\sqrt{2}} &\frac{Y^\prime_{\Psi \rho} v}{\sqrt{2}} & 0 &\frac{\lambda_\rho v_1v_2}{\Lambda} & {\rho_L}^c  \\
\end{array}\right)\,
\label{eq:ss1}
\end{eqnarray}
For the limit, where all the Yukawa couplings~$(Y\text{ and }Y')$ $\rightarrow 0$ in Eq.~(\ref{eq:ss1}), the diagonalization of the mass matrix~($M_{\rm NF}^{\rm DM_2}$) leads to a Dirac fermion~($\Psi=\Psi_L+\Psi_R$) with mass $M_{\Psi}=\frac{\lambda_\Psi v_1v_2}{\Lambda}$, and two Majorana fermions with masses $M_{\Sigma}=\frac{\lambda_\Sigma v_1v_2}{\Lambda}$ and $M_{\rho}=\frac{\lambda_\rho v_1v_2}{\Lambda}$. Now, if we allow non-zero Yukawa couplings for these exotic fermions, then it leads to a contribution from off-diagonal terms to the final mass eigenstates for these Dirac and Majorana fermions. Although, as the value of SM Higgs VEV~$(v)$ is much smaller than the VEVs $v_1,v_2$ of scalars $\phi_1$ and $\phi_2$, respectively, so the contributions from the off-diagonal terms~(which are proportional to the value of $v$) to the final mass eigenstates are fairly sub-dominant. We represent the final mass eigenstates as~$N^0_1,~N^0_2,~N^0_3\text{ and }N^0_4$ with an ascending mass ordering, and the diagonalized matrix in these basis is labelled as $M_{\rm NF}^{\text{diag}}$. Thus, by choosing a suitable combination of values for the parameters~$\lambda_{\Psi}, \lambda_{\Sigma}, \lambda_{\rho}$ and by setting all the Yukawa couplings at zero in Eq.~(\ref{eq:ss1}), we keep a mass hierarchy where the mass of Majorana particle $\rho_L$ is lowest i.e. $M_{N^0_1}\simeq M_{\rho_L}$ and thus it becomes a viable Majorana type DM candidate. Similarly, for the case of non-zero Yukawa couplings in Eq.~$(\ref{eq:ss1})$, we may again carefully choose the various parameter values in a way that for the lowest mass eigenstate $M_{N_1^0}$, the major dominance comes from the Majorana mass term of $\rho_L$ given by $\frac{\lambda_{\rho}v_1v_2}{\Lambda}$. Also, as the nature of $N^0_1$ is majorly Majorana type, so for the phenomenological analysis, we refer to the mass of $N_1^0$ as $M^{M}_{\text{DM}}$ interchangeably with $M_{N_1^0}$ in the later sections of this article. Thus, $M^{M}_{\text{DM}}$ and $M_{N_1^0}$ convey the same meaning. Similar to the neutral fermions, the mass matrix structure for the charged exotic fermions in Eq.~(\ref{eq:TheModel}) can be expressed as,
\begin{eqnarray}
M_{\rm CF}= \left(\begin{array}{cc|c} \overline{\Psi_L^+}   &   \overline{{\Sigma^+_L}} \\  \hline
\frac{\lambda_\Psi v_1v_2}{2\Lambda} & \frac{Y^\prime_{\Psi \Sigma} v}{\sqrt{2}}  & \Psi_R^+ \\
 \frac{Y_{\Psi \Sigma} v}{\sqrt{2}} & \frac{\lambda_\Sigma v_1v_2}{\Lambda} & {\Sigma^+_L}^c\\
\end{array}\right)\,.
\label{eq:ss2}
\end{eqnarray}
After the diagonalization of matrix $M_{\rm CF}$, we have a diagonalized matrix~$M_{\rm CF}^{\text{diag}}$ with two mass eigenstates namely $\Psi_1^+$ and $\Psi_2^+$. The diagonalizations for both the neutral sector and the charged sector mass matrices~(expressed in Eqs.~(\ref{eq:ss1}) and~(\ref{eq:ss2}), respectively) can be generally viewed in a structure given as,
\begin{eqnarray}
\nonumber M_{\rm NF}^{\text{diag}}&=&R^{T}M_{\rm NF}^{\rm DM_2}R,\\
\text{and }M_{\rm CF}^{\text{diag}}&=&V_{L}^{\dagger}M_{\rm CF}V_R,
\label{eq:ss3}
\end{eqnarray}
where $R$ is an orthogonal matrix, and $V_{L}$ and $V_{R}$ are the unitary matrices. For having zero $U(1)_{\ell}$ charge, the SM quarks do not interact with new gauge boson $Z'$ directly at tree level, while the interaction of $Z'$ with the charged SM leptons having unit leptonic charge is given by,
\begin{equation}
 g_{\ell} Z'_{\ell_{\mu}}\sum_l \Bigg(\overline{{\ell}_L}\gamma^{\mu}{\ell}_L+\overline{{\ell}_R}\gamma^{\mu}{\ell}_R\Bigg).
\end{equation}
As we have now discussed the possible interactions among exotic fermions and scalars, and have also laid down the mass matrices containing Yukawa couplings and other relevant parameters for both the DM sectors, thus in the next section we present our results for the phenomenological analysis of both DM candidates from relic density and direct-indirect searches.
\section{Dark Matter Phenomenology}
\label{sec:DMPheno}
In this section, we first focus on constraining the model's parameter space from the observed DM relic abundance requirements. Additionally, we perform direct and indirect detection searches for DM particles. Direct detection involves observing DM particles interacting with ordinary matter in controlled laboratory settings, offering valuable experimental insights, and the indirect techniques entail observing the consequences of dark matter annihilation into the SM particles or decay in cosmic rays or astrophysical phenomena.
\subsection{Relic Density}
\label{subsec:Relic}
In this section, we perform the relic density analysis for both the DM candidates separately while fixing the mass of other one at a particular value. For proper treatment, we have relied on the micrOMEGAs~\cite{Belanger:2013oya} for all the numerical analysis associated to DM phenomenology by first implementing our model on SARAH~\cite{Staub:2008uz} package and then generating the compatible calcHEP~\cite{Belyaev:2012qa} and SPheno~\cite{Porod:2011nf} input files.
\subsubsection{Relic Contributing channels}
\label{subsubsec:CC}
Before discussing the relic abundances for DM candidates explicitly, here we first focus our discussion on the various parameters that may directly or indirectly affect this relic abundance in our framework. As mentioned in~\ref{subsec:FS}, the model contains two separately stable dark matter states, namely $\chi_1$ as a Dirac type candidate and $N_1^0$ as a Majorana type. The exotic fermions added instead of right-handed neutrinos, mix together to form $\chi_1$ and $\chi_2$ mass eigenstates, among which the lighter component $\chi_1$ being neutral and stable is a viable cold DM candidate. Similarly, among the remaining neutral exotic fermions, mass eigenstate~$N_1^0$ can be made lightest, stable, and dominantly Majorana by choosing the right set of exotic Yukawa couplings. The state $\chi_1$ and $\chi_2$ don't have any direct interaction with the SM particles and thus they annihilate into either the Majorana DM particle,~$N^0_1$ or to the exotic gauge Boson,~$Z'$. This is a known scenario in multi component dark matter models, where one DM state annihilates partially or completely into the other DM state. Few of the recent and notable works in literature showcasing such multi-component DM scenarios can be referred here~\cite{Belanger:2020hyh,Yaguna:2021rds,Ho:2022erb,Khan:2023uii,Khan:2024biq}. The Feynman diagrams at tree level for all possible $\chi_1$ annihilations are depicted in Fig.~\ref{fig:Feyn1} and for all possible $\chi_1$ co-annihilations with $\chi_2$ are depicted in Fig.~\ref{fig:Feyn2} in Appendix~\ref{app:DDM}. Majorana DM $N_1^0$ may annihilate to SM states via both $Z$ and $Z'$ vector bosons and can also annihilate via scalar interactions mediated by the SM Higgs~$(h)$ or the new scalars $H_1$ and $H_2$. The strength of these interactions~(DM annihilations and co-annihilations) relative to the Hubble's expansion rate~$(H_m)$ in the early universe plays a role in deciding the DM freeze-out temperature and its relic abundance. We depict these interactions at tree level via various possible Feynman diagrams in Figs.~\ref{fig:Feyn3}-~\ref{fig:Feyn5} for $N_1^0$ in Appendix~\ref{app:MDM}. For DM phenomenology associated to $\chi_1$, the relevant model parameters are listed below:
\begin{equation}
 M_{\chi_1},~~~ M_{Z'},~~~ \theta_{L},~~~ g_{\ell},~~~ \tan\beta.
\label{eq:dm0}
\end{equation}
Here, $M_{\chi_1}$ is the mass of Dirac DM candidate~($\chi_1$), $M_{Z'}$ denotes the mass of exotic gauge boson~($Z'$), $\theta_L$ refers to the DM mixing angle, $g_{\ell}$ is the exotic gauge coupling and $\tan\beta$ is the VEVs ratio for the SM singlet scalars $\phi_1$ and $\phi_2$. These parameters affect the various annihilation channels for Dirac DM and thus can be tweaked around to study their effect on the resultant DM relic density obtained via freeze-out. The major annihilation channels available for $\chi_1$ in our framework are:
\begin{equation}
 \overline{\chi_1}\chi_1\rightarrow~\overline{N_1^0}N_1^0,~Z'Z',
\label{eq:dm1}
\end{equation}
where all the symbols have their usual meaning. For DM phenomenology associated to $N_1^0$, the relevant parameters are:
\begin{equation}
 M_{N_1^0},~~~ M_{Z'},~~~ g_{\ell},~~~ \lambda_\Psi,~~~ \lambda_\rho,~~~ Y_\Psi,~~~ Y_\Psi',~~~ Y_{\Psi_\rho},~~~ Y_{\Psi_\rho}',
\label{eq:dmMaj1}
\end{equation}
where $M_{N_1^0}$ is the mass of Majorana type DM candidate~$(N_1^0)$, $\lambda_\Psi$ is the dimension $5$ coupling involving $\Psi_L$ and $\Psi_R$, $\lambda_\rho$ is the bare mass coupling for exotic fermion $\rho_L$, $Y_\Psi$ and $Y_\Psi'$ are the Yukawa couplings for the interaction of $\Sigma_L$ with $\Psi_L$ and $\Psi_R$, respectively, and $Y_{\Psi_\rho}$ and $Y_{\Psi_\rho}'$ are the Yukawa couplings for the interaction of $\rho_L$ with $\Psi_L$ and $\Psi_R$, respectively. These parameters can affect the freeze-out number density of $N_1^0$ and are thus important for studying its relic phenomenology. The major annihilation channels available for $N_1^0$ in our framework are:
\begin{equation}
 \overline{N_1^0}N_1^0\rightarrow \overline{l}l,~\overline{\nu}\nu,~Z'Z',~Z'H_{i},~W^{+}W^{-},~ZZ.
\label{eq:dmMaj2}
\end{equation}
The channels listed in Eq.~($\ref{eq:dmMaj2}$) can be either scalar or vector boson mediated depending on the mass of DM candidate~$N_1^0$ with respect to the masses of various scalars and vector bosons available in the framework. In the case of vector-mediated interactions, there can arise two important cases:
\begin{itemize}
 \item DM mass is smaller than the mass of $Z'$~($M_{N_1^0}<M_{Z'}$): In this case, the available annihilation channels are:
 \begin{equation}
 \overline{N_1^0}N_1^0\xrightarrow{Z/W^+/W^-} \text{ allowed final states}.
\label{eq:dm2}
\end{equation}
The Feynman diagrams corresponding to these decay modes can be referred from Figs.~\ref{fig:Feyn3i} and~\ref{fig:Feyn3j} for $N_1^0$ in Appendix~\ref{app:MDM}. Among these, the $W^{+}W^{-}$ and $ZZ$ channels are heavily suppressed as the couplings $\lambda_{4,5}$ are zero~(refer \ref{subsc:sc}) and are thus negligible for DM relic density calculations.

\item DM mass is larger than the mass of $Z'$~($M_{N_1^0}>M_{Z'}$): Here, an additional annihilation channel~($\overline{N_1^0}~N_1^0\xrightarrow{Z'} \text{ final states})$ opens along with the other channels mentioned in the previous case. The importance of $Z'Z'$ channel lies in the fact that it is not velocity-suppressed and thus may contribute significantly to DM relic density~($\Omega h^2$) at higher masses of DM, if allowed through gauge interactions.
\end{itemize}
The vector boson~$(Z')$ mediated annihilation channel~($\overline{N_1^0}N_1^0\rightarrow Z'\rightarrow\overline{f}f $) has an interaction cross-section, which in the non-relativistic and low lepton mass limit can be given from Ref.~\cite{Patra:2016ofq} as:
\begin{align}
\sigma v \left(\overline{N_1^0}N_1^0 \xrightarrow{Z'^{*}} \bar{f}f \right)
&=
\frac{N_c^f \mdm^2 g_{\chi V}^2g_{fV}^2}{\pi
\big[ \left(4 M^2_{\rm DM}- M^2_{Z^\prime} \right)^2 + M^2_{Z^\prime} \Gamma^2_{Z^\prime}\big]
 } \,,
 \label{eq:dm2}
\end{align}
where $M_{\rm DM}$ denotes the mass of DM candidate~($N_1^0$) in GeV, $\Gamma_{Z^\prime}$ denotes the total decay width of $Z^\prime$ boson, and ($N_c^f$, $g_{fV}$) is equal to ($1$, -$g_\ell$) for leptons. Thus, DM annihilation mediated via $Z^\prime$ boson depends on four independent parameters: $\mdm$, $\mzp$, $g_\ell$ and $\theta_L$ (via $g_{\chi V}$). Also, there is no direct channel for DM annihilating into quarks as quarks are singlets under leptophilic gauge group as shown in Table~\ref{Tab:1}. Similar conditions on annihilation channels can also be deduced for Dirac DM~$(\chi_1)$.

Apart from the $Z'$ mediated annihilations, scalar mediated channels are also depicted in Figs.~\ref{fig:Feyn1a}-\ref{fig:Feyn1f} for $\chi_1$ in Appendix~\ref{app:DDM} and in Figs.~\ref{fig:Feyn3a}-\ref{fig:Feyn3h} for $N_1^0$ in Appendix~\ref{app:MDM}. Here, the scalar masses along with other scalar sector parameters like $\tan\beta$ and $\theta_L$ dictate the contribution of these channels to the DM relic density. DM can also annihilate to final states containing vector bosons via t-channel processes as depicted in Fig.~\ref{fig:Feyn1g} for $\chi_1$ in Appendix~\ref{app:DDM} and in Figs.~\ref{fig:Feyn3b}-\ref{fig:Feyn3k} for $N_1^0$ in Appendix~\ref{app:MDM}.
\subsubsection{Numerical Analysis}
\label{subsubsec:RelicRnD}
Here, we present our results for the DM relic density analysis performed for both the DM candidates separately, keeping the other relevant model parameters fixed at some particular values. These values have been carefully chosen to highlight striking features and dependencies of relic abundance on them, which are discussed in the subsequent paragraphs and illustrated in various plots to follow. To keep the current discussion simpler, we have intentionally avoided incorporating the effects of the next-to-lightest DM state on DM relic density. To achieve this, we have maintained a mass splitting of 200~GeV between the Dirac DM and its next-to-lightest partner~$(\Delta M(\chi_1,\chi_2)$, ensuring negligible effects on DM phenomenology. Similarly, in the Majorana DM sector, the mass splitting between the DM and the next-to-lightest particle~$(\Delta M(N_1^0,N_2^0)$ is also set to 200~GeV to suppress co-annihilation effects. We plan to address the effects of the next-to-lightest DM particle on DM relic density in an extended study, where we will also investigate collider bounds on these other exotic states. This future work will provide a comprehensive analysis of the properties of these exotic states, considering both relic density requirements and collider constraints.

\begin{figure*}[tbp]
  \centering
  \includegraphics[width=.8\textwidth,scale=1.8]{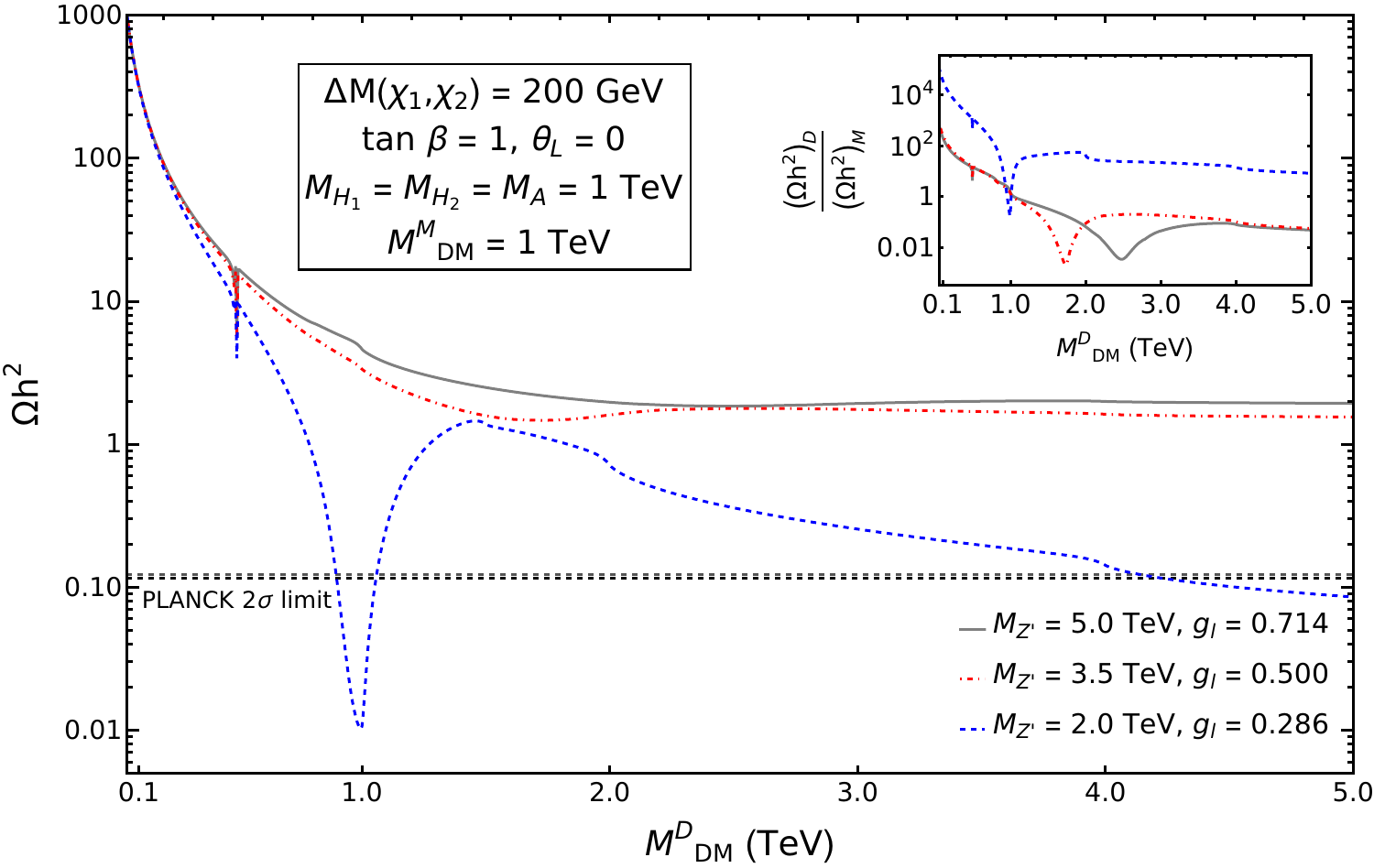}
  \caption{Total DM relic density~$(\Omega h^2)$ as a function of Dirac DM mass~$(M^D_{\rm DM})$ for three different combinations of input parameters~$(M_{Z'},g_l)$: (5.0 TeV, 0.714) in solid line, (3.5 TeV, 0.500) in dot-dashed line, and (2.0 TeV, 0.286) in dashed line. The mass of the Majorana DM candidate~$(M^D_{\rm DM})$ is set at 1 TeV, $\tan\beta=1$, and no mixing is considered in the Dirac DM mass matrix, i.e., $\theta_L=0$.  Values of all the other relevant input parameters are mentioned within the plot. Two horizontal black dashed lines denote the region consistent with the observed DM relic density within $2\sigma$ bounds. The inset-plot depicts the ratio of relic density contributions arising from Dirac DM vs the Majorana DM as a function of Dirac DM mass.}
  \label{fig:ex1}
\end{figure*}

\begin{figure*}[tbp]
  \centering
  \includegraphics[width=.8\textwidth,scale=1.8]{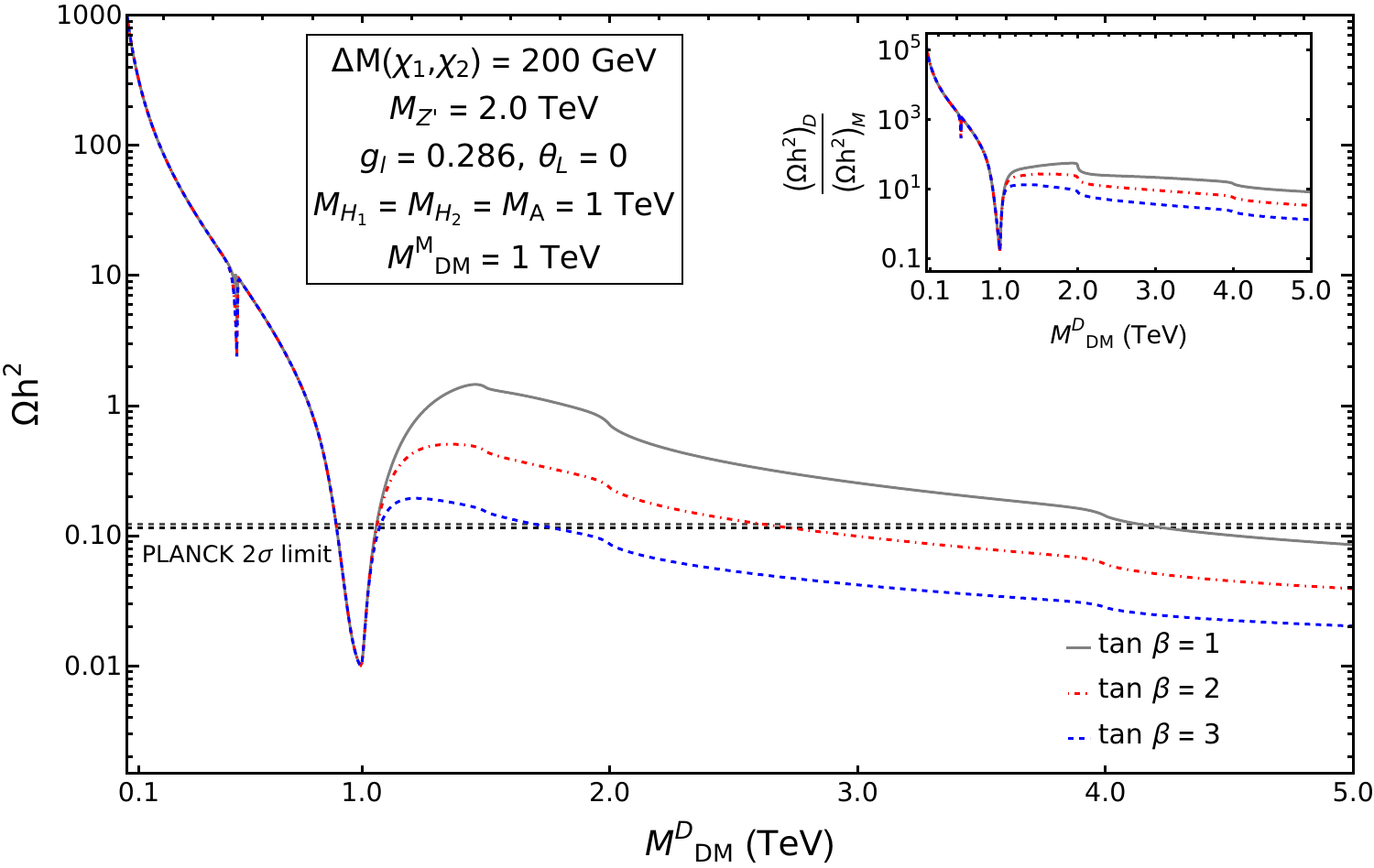}
  \caption{Total DM relic density~$(\Omega h^2)$ as a function of Dirac DM mass~$(M^D_{\rm DM})$ for three different values of input parameter $\tan\beta$: $(\tan\beta=1)$ in solid line, $(\tan\beta=2)$ in dot-dashed line, and $(\tan\beta=3)$ in dashed line. The mass of the Majorana DM candidate is set at 1 TeV. To generate this plot, we have chosen $g_l=0.286$, $M_{Z'}=2\text{ TeV}$ and no mixing is considered in the Dirac DM mass matrix, i.e., $\theta_L=0$.  Values of all the other relevant parameters are mentioned within the plot. Two horizontal gray dashed lines correspond to the region consistent with the observed DM relic density within $2\sigma$ bounds. The inset-plot depicts the ratio of relic density contributions arising from Dirac DM vs the Majorana DM as a function of Dirac DM mass.}
  \label{fig:ex2}
\end{figure*}

\begin{figure*}[tbp]
  \centering
  \includegraphics[width=.8\textwidth,scale=1.8]{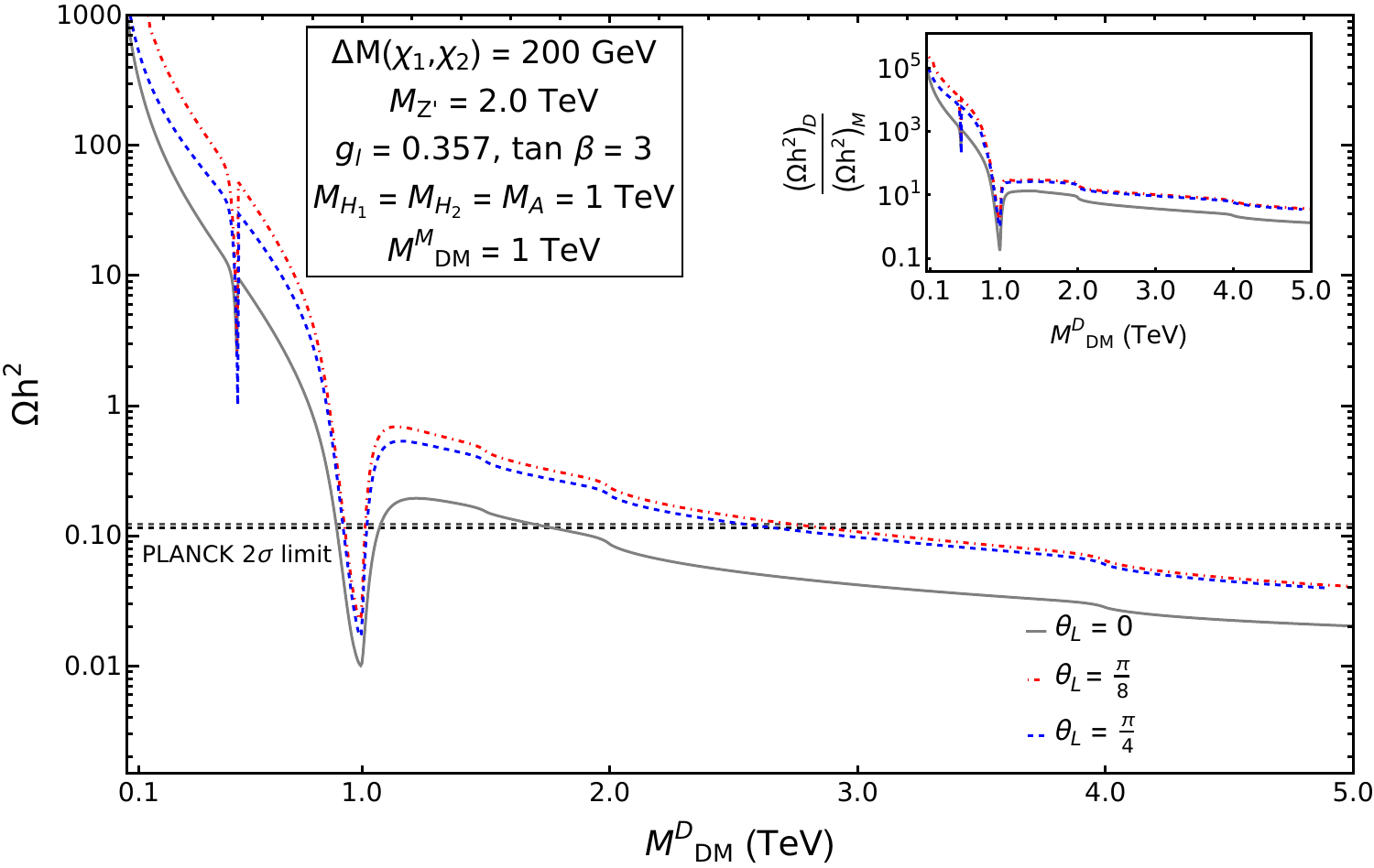}
  \caption{Total DM relic density~$(\Omega h^2)$ as a function of the Dirac dark matter mass~$(M^D_{\rm DM})$ for three different values of input parameter $\theta_L$: $(\theta_L=0)$ in solid gray line, $(\theta_L=\pi/8)$ in dot-dashed red line, and $(\theta_L=\pi/4)$ in dashed blue line. The mass of Majorana DM candidate is set at 1 TeV. To generate this plot, we have chosen $g_l=0.357$, $\tan\beta=3$, and the mass of exotic gauge boson, $M_{Z'}$ is fixed at $2 \text{ TeV}$. Values of all the other relevant input parameters are mentioned within the plot. Two horizontal gray dashed lines correspond to the region consistent with the observed DM relic density within $2\sigma$ bounds. The inset-plot depicts the ratio of relic density contributions arising from Dirac DM vs the Majorana DM as a function of Dirac DM mass.}
  \label{fig:ex3}
\end{figure*}

Figure~\ref{fig:ex1} shows the dependence of total DM relic density~$(\Omega h^2)$ on the mass of Dirac DM~($M_{\rm DM}^D$). The mass of Majorana DM~$M_{\rm DM}^M$ is fixed at~$1~\rm TeV$. The figure shows three different lines corresponding to a set of values for independent parameters $M_{Z'}$ and $g_{L}$: $M_{Z'}=5.0~\text{TeV}, g_{L}=0.714$ in the solid gray line, $M_{Z'}=3.5~\text{TeV}, g_{L}=0.500$ in the red dashed-dot line and $M_{Z'}=2.0~\text{TeV}, g_{L}=0.286$ in the blue dotted line. The other relevant parameters like $\theta$, $\theta_L$, and $\tan\beta$ have been fixed to $1$ along with scalar masses $M_{H_1}$, $M_{H_2}$ and $M_A$ set to $1$ TeV and the mass splitting between the Dirac DM~$(\chi_1)$ and the next-to-lightest~$(\chi_2)$ given by~$\Delta M(\chi_1,\chi_2)$ is set at $200~\text{GeV}$. The inset-plot shows the total relic abundance ratio for the two DM candidates as a function of $M_{\rm DM}^D$. From the plot, we see that the total relic abundance is sensitive to the value of $M_{Z'}$ for a fixed value of $M_{\rm DM}^M$ at $1$ TeV. For the blue line, where $M_{Z'}=2~\text{TeV}$, the resonance condition for the Majorana candidate is satisfied, and thus the relic abundance is significantly reduced. For other values of $M_{Z'}$ as shown in dot-dashed red and solid gray lines, the plot features are subdued due to an ever-present higher contribution to the overall relic density from the Majorana DM~$(N_1^0)$. All the lines show sharp resonance funnel at $M_{H_1, H_2, A}/2$. One may see the resonance feature of the relic density from the inset-plot as well, where there is a significant drop in the relative contribution of the Dirac DM to the total relic at $M_{\rm DM}^D=M_{Z'}/2$ and at $M_{\rm DM}^D=M_{H_1, H_2}/2$ for all the three plots. Notably, the resonance dip due to the SM-like Higgs~(i.e. at around $M_H/2$) is absent in the plots. This behavior can be understood from the fact that the Dirac DM candidate predominantly annihilates into the Majorana DM candidate. In the absence of scalar sector mixing or kinetic mixing, such annihilation processes are mediated exclusively by exotic particles, such as the $Z'$ boson or exotic scalars. Thus, no annihilation interaction between the two DM candidates mediated by SM Higgs is possible. This explains the absence of the SM-like Higgs resonance, while dips at $M_{H_1, H_2}/2$ and $M_{Z'}/2$ are clearly seen for all plot curves in \textcolor{red}{Figures~1--3}. Two horizontal gray dashed lines show the value of DM relic density within $2\sigma$ bounds obtained via the Planck data~\cite{Planck:2018vyg}.

In Figure~\ref{fig:ex2}, one can see the dependence of $\Omega h^2$ on $M_{\rm DM}^D$ for three different values of scalar VEV ratio~($\tan\beta$):~$\tan \beta=1$ in the solid gray line, $\tan \beta=2$ in the red dashed-dot line and $\tan \beta=3$ in the blue dotted line. Here, $M_{\rm DM}^M$ is set at~$1~\rm TeV$ and the other relevant parameters have been fixed at a particular value with $M_{Z'}=2.0~\rm TeV$, $\Delta M(\chi_1,\chi_2)=200~\text{GeV}$, $g_l=0.286$, $(\theta_L)=0$, and $M_{H_1}$, $M_{H_2}$ and $M_A$ equal to $1$ TeV. The inset-plot shows the relic abundance ratio for the two DM candidates as a function of $M_{\rm TeV}^D$. From the figure, one may see that the change in $\tan\beta$ does not play a significant role when the Dirac DM mass is smaller than the scalar masses~$(M_{H_1}=M_{H_2}=M_{A}=1\text{ TeV})$, as the parameter~$\tan\beta$ primarily modifies the interaction strengths of the channels that are associated with DM annihilation into scalars, and such channels are kinematically absent for DM mass smaller than scalar masses. Also, with the increase of the $\tan\beta$ parameter from $1$ to $3$ as shown in the figure, the interaction strength of these scalar mediated DM annihilation channels increase, thus decreasing the value of obtained relic abundance accordingly. From the figure, it can also be inferred that DM relic density can be satisfied for a wider range of DM mass by properly tuning the value of $\tan\beta$ parameter. Again, the resonance dip due to the SM-like Higgs~(i.e. at around $M_H/2$) is absent in the plots, due to the reasons as explained in Figure~\ref{fig:ex1}.

Figure~\ref{fig:ex3} explores the dependence of total DM relic density on the Dirac DM mass~$(M_{\rm DM}^D)$ for various values of Dirac DM mixing angle~($\theta_L$): $\theta_L=0$ in the solid gray line, $\theta_L=\pi/8$ in the red dot-dashed line and $\theta_L=\pi/4$ in the blue dashed line. The mass of Majorana DM $(M_{\rm DM}^M)$ is set at~$1~\rm TeV$. All the other relevant parameters have been fixed at a particular value with $M_{Z'}=2.0~\rm TeV$, $\Delta M(\chi_1,\chi_2)=200~\text{GeV}$, $g_l=0.357$, $\tan\beta=3$, and $M_{H_1}$, $M_{H_2}$ and $M_A$ set to $1$ TeV. The inset-plot shows the relic abundance ratio for the two DM candidates as a function of $M_{\rm DM}^D$. From the figure, one can see that as we increase $\theta_L$, the value of relic density obtained for a particular $M^D_{\text{DM}}$ increases alongside, thus suggesting a decrease in the strength of DM annihilation cross-section. For $\theta_L=0$ or no mixing case, Dirac DM mass state~$\chi_1$ is solely obtained from the $\xi$ fermion which has a higher $U(1)_{\ell}$ charge than $\eta$ fermion, so the strength of DM annihilation into leptons is higher in comparison with the case of non-zero $\theta_L$, where $\chi_1$ is a mixture of both $\xi$ and $\zeta$ fermions, and thus relic abundance falls with decreasing $\theta_L$ for a fixed $M^D_{\text{DM}}$. Additionally, a sharp resonance feature can be seen in the plots at $M_{H_1, H_2, A}/2$. Once again, the resonance dip due to the SM-like Higgs~(i.e. at around $M_H/2$) is absent in the plots, due to the reasons as explained in Figure~\ref{fig:ex1}.

\begin{figure*}[tbp]
  \centering
  \includegraphics[width=.8\textwidth,scale=1.8]{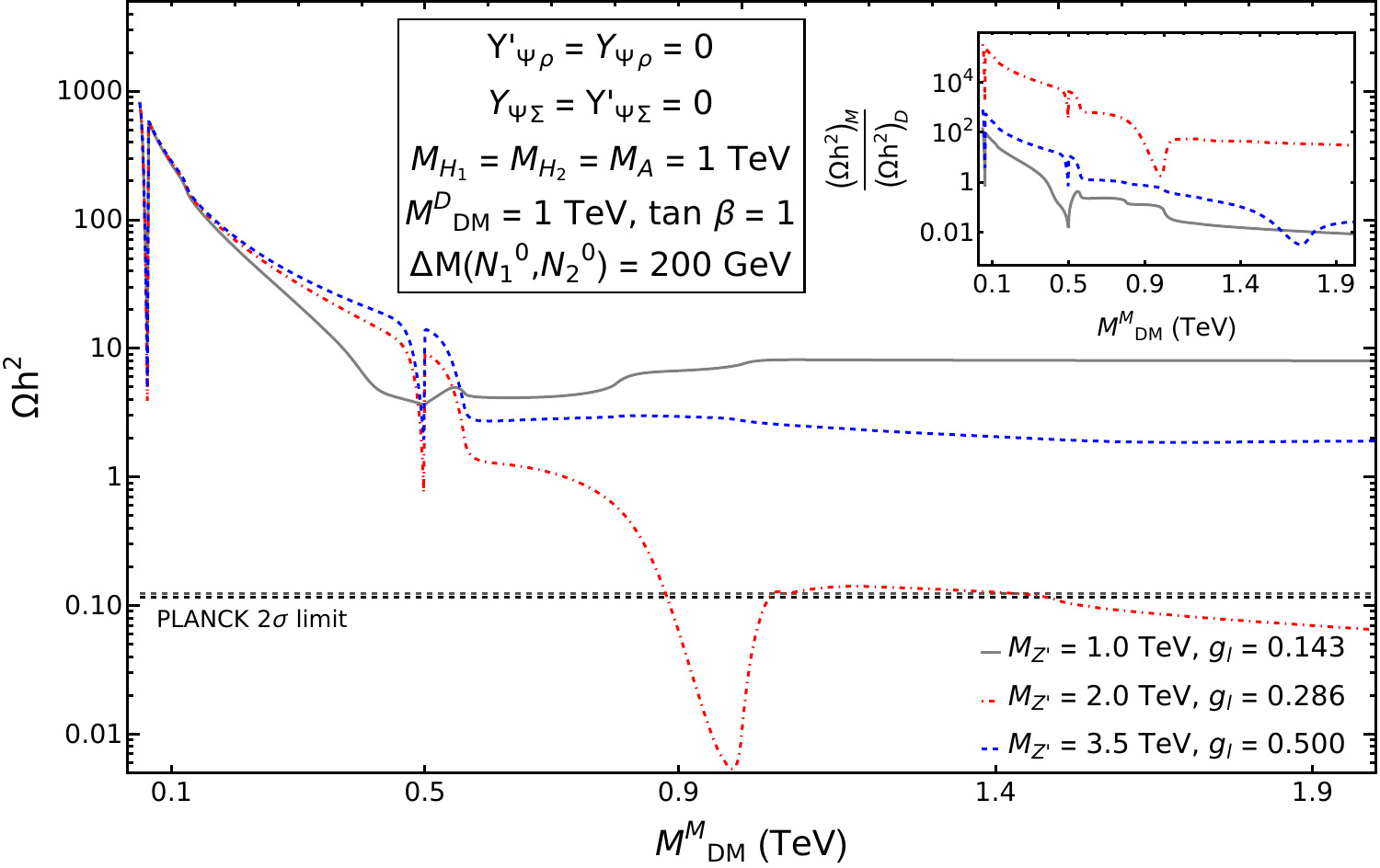}
  \caption{Dark matter total relic density~$(\Omega h^2)$ as a function of Majorana dark matter mass~$(M^M_{\rm DM})$ for three different combinations of input parameters~$(M_{Z'},g_l)$: (1.0 TeV, 0.143) in solid gray line, (2.0 TeV, 0.286) in dot-dashed red line, and (3.5 TeV, 0.500) in dashed blue line. The mass of the Dirac DM candidate~$(M^D_{\rm DM})$ is set at 1 TeV, and the relevant Yukawa couplings are considered to be zero, i.e. $Y'_{\Psi\rho}=Y_{\Psi\rho}=0$ and $Y_{\Psi\Sigma}=Y'_{\Psi\Sigma}=0$. Values of all the other relevant input parameters are mentioned within the plot. Two horizontal gray dashed lines correspond to the region consistent with the observed DM relic density within $2\sigma$ bounds. The inset-plot depicts the ratio of relic density contributions arising from Dirac DM vs the Majorana DM as a function of Majorana DM mass.}
  \label{fig:ex4}
\end{figure*}

\begin{figure*}[tbp]
  \centering
  \includegraphics[width=.8\textwidth,scale=1.8]{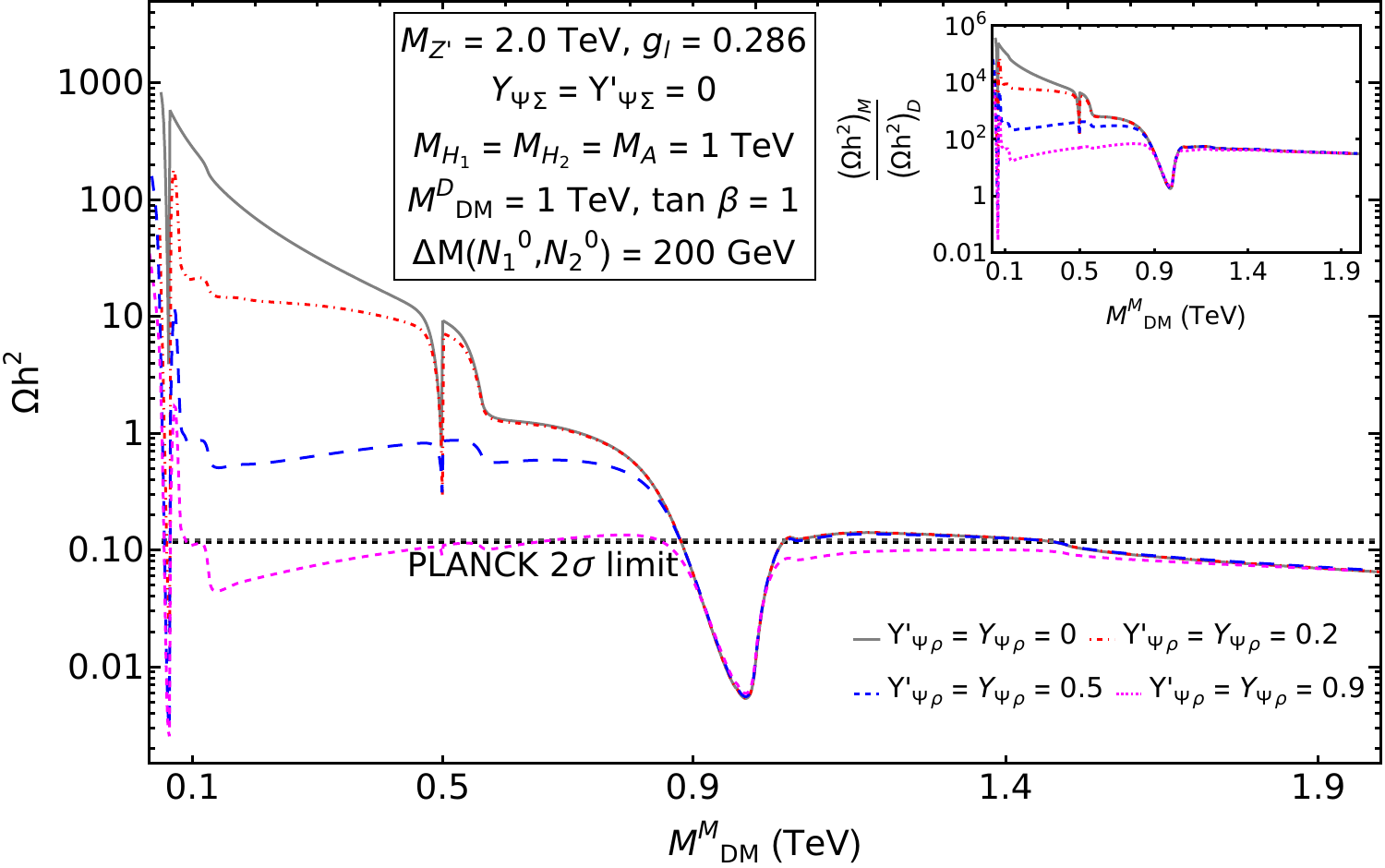}
  \caption{Dark matter total relic density~$(\Omega h^2)$ as a function of Majorana dark matter mass~$(M^M_{\rm DM})$ for four different combinations of input parameters~$Y'_{\Psi\rho}\text{ and }Y_{\Psi\rho}$: $(Y'_{\Psi\rho}=Y_{\Psi\rho}=0)$ in solid gray line, $(Y'_{\Psi\rho}=Y_{\Psi\rho}=0.1)$ in dot-dashed red line, $(Y'_{\Psi\rho}=Y_{\Psi\rho}=0.5)$ in dashed blue line, and $(Y'_{\Psi\rho}=Y_{\Psi\rho}=0.9)$ in dotted magenta line. The mass of the Dirac DM candidate~$(M^D_{\rm DM})$ is set at 1 TeV, $g_l=0.286$, $M_{Z'}=2.0\text{ TeV}$, and the other Yukawa couplings in the exotic sector mass matrix are set to be zero, i.e. $Y_{\Psi\Sigma}=Y'_{\Psi\Sigma}=0$. Values of all the other relevant input parameters are mentioned within the plot. Two horizontal gray dashed lines correspond to the region consistent with the observed DM relic density within $2\sigma$ bounds. The inset-plot depicts the ratio of relic density contributions arising from Dirac DM vs the Majorana DM as a function of Majorana DM mass.}
  \label{fig:ex5}
\end{figure*}

\begin{figure*}[tbp]
  \centering
  \includegraphics[width=.8\textwidth,scale=1.8]{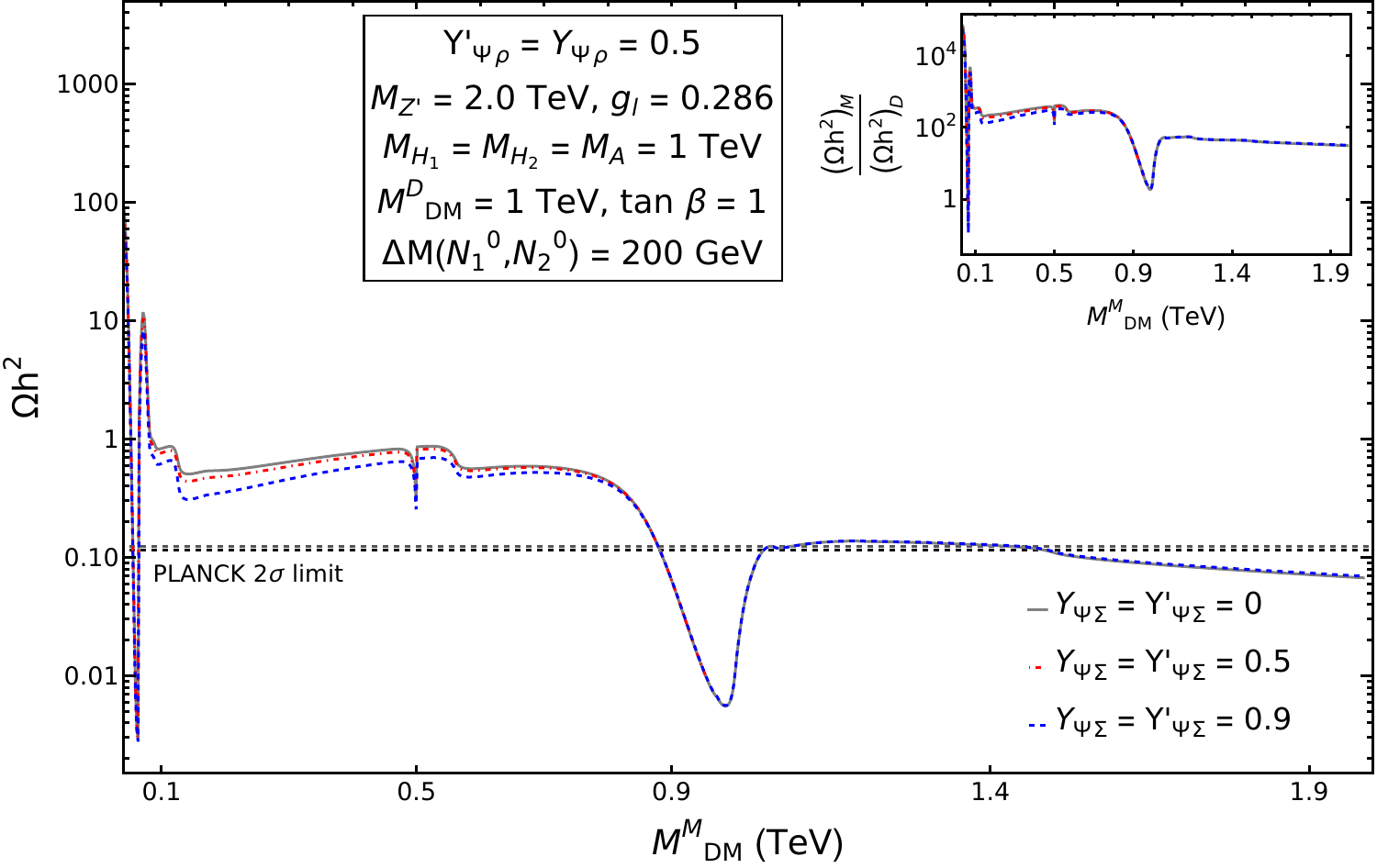}
  \caption{Dark matter total relic density~$(\Omega h^2)$ as a function of Majorana dark matter mass~$(M^M_{\rm DM})$ for three different combinations of input parameters~$Y_{\Psi\Sigma}\text{ and }Y'_{\Psi\Sigma}$: $(Y_{\Psi\Sigma}=Y'_{\Psi\Sigma}=0)$ in solid gray line, $(Y_{\Psi\Sigma}=Y'_{\Psi\Sigma}=0.5)$ in dot-dashed red line, and $(Y_{\Psi\Sigma}=Y'_{\Psi\Sigma}=0.9)$ in dashed blue line. The mass of the Dirac DM candidate~$(M^D_{\rm DM})$ is set at 1 TeV, $g_l=0.286$, $M_{Z'}=2.0\text{ TeV}$, and $Y'_{\Psi\rho}=Y_{\Psi\rho}=0.5$. Values of all the other relevant input parameters are mentioned within the plot. Two horizontal gray dashed lines correspond to the region consistent with the observed DM relic density within $2\sigma$ bounds. The inset-plot depicts the ratio of relic density contributions arising from Dirac DM vs the Majorana DM as a function of Majorana DM mass.}
  \label{fig:ex6}
\end{figure*}

In the preceding few paragraphs, we explored the dependence of total DM relic density as a function of Dirac DM mass~$(M_{\rm DM}^D)$ for different combinations of relevant input parameters, while keeping the mass of Majorana DM~$(M_{\rm DM}^M)$ fixed at a particuar value. Thus, it is sequitur to extend our discussion to analyze the dependence of the total relic abundance on $M_{\rm DM}^M$ while keeping the value of $M_{\rm DM}^D$ fixed. For simplicity, we keep $M^D_{\rm DM}$ fixed at $1$~TeV for this analysis performed in subsequent paragraphs.

In Figure~\ref{fig:ex4}, we depict the total DM relic density as a function of the Majorana DM mass~($M_{\rm DM}^M$). The figure shows three different lines associated to a set of values for parameters $M_{Z'}$ and $g_{L}$: $M_{Z'}=1.0~\text{TeV}, g_{L}=0.143$ in the solid gray line, $M_{Z'}=2.0~\text{TeV}, g_{L}=0.286$ in the dashed-dot red line and $M_{Z'}=3.5~\text{TeV}, g_{L}=0.5$ in the dashed blue line. The other relevant parameters like exotic Yukawa couplings $Y'_{\Psi\rho},~ Y_{\Psi\rho},~Y_{\Psi\Sigma},~Y'_{\Psi\Sigma}$ are completely switched off for simplicity, scalar masses $M_{H_1}$, $M_{H_2}$ and $M_A$ are fixed at 1 TeV and the mass splitting between the Majorana DM~$(N_1^0)$ and the next-to-lightest~$(N_2^0)$ given by~$\Delta M(N_1^0,N_2^0)$ is set at $200~\text{GeV}$. The inset-plot shows the relic abundance ratio for the two DM candidates as a function of $M_{\rm DM}^M$. From the Figure~\ref{fig:ex4}, one can see that the total relic abundance is highly sensitive to the mass of $Z'$. For the dashed-dot red line, where $M_{Z'}=2~\text{TeV}$, the resonance condition for the Dirac candidate is satisfied, and thus the relic abundance is significantly reduced. For other values of $M_{Z'}$ as shown in dashed blue and solid gray lines, the plot features are subdued due to an ever-present higher contribution to the overall relic density from the Dirac DM~$(\chi_1)$. 
We here observe sharp resonance dips at half the SM-like Higgs' mass i.e. at around~$(62 \text{ GeV})$, unlike the case for Dirac DM candidate as shown in Figures~\ref{fig:ex1}-~\ref{fig:ex3}. This is expected because the Majorana DM candidate annihilates into SM particles, and such interactions can be mediated by both scalar and vector-type mediators. All the plot lines also reflect sharp resonance funnels at $M_{H_1, H_2, A}/2$. One may see the resonance feature of the relic density from the inset-plot as well, where there is a significant drop in the relative contribution of the Majorana DM to the total relic at $M_{\rm DM}^M=M_{Z'}/2$ and at $M_{\rm DM}^M=M_{H_1, H_2}/2$ for all the three plots. Two horizontal dashed gray lines show the value of DM relic density within $2\sigma$ bounds obtained via the Planck data~\cite{Planck:2018vyg}.

In Figure~\ref{fig:ex5}, we show the dependence of total DM relic density on the Majorana DM mass~$(M_{\rm DM}^M)$ for different values of exotic Yukawa couplings $Y'_{\Psi\rho}$ and $Y_{\Psi\rho}$.  The figure shows four different lines based on a set of values for Yukawa couplings $Y'_{\Psi\rho}$ and $Y_{\Psi\rho}$: $Y'_{\Psi\rho}=Y_{\Psi\rho}=0$ in the solid gray line, $Y'_{\Psi\rho}=Y_{\Psi\rho}=0.1$ in the red dashed-dot line, $Y'_{\Psi\rho}=Y_{\Psi\rho}=0.5$ in the blue dashed line, and $Y'_{\Psi\rho}=Y_{\Psi\rho}=0.9$ in the violet dotted line. All the other relevant parameters have been fixed at a particular value with $M_{Z'}=2.0~\rm TeV$, $g_l=0.286$, couplings $Y_{\Psi\sigma}$ and $Y'_{\Psi\sigma}$ equal to $0$, and $M_{H_1}$, $M_{H_2}$ and $M_A$ set to $1$ TeV. The inset-plot shows the relic abundance ratio for the two DM candidates as a function of $M_{\rm TeV}^M$. From the Figure~\ref{fig:ex5}, it can be observed that the value of obtained DM relic abundance is independent of the Yukawa couplings $Y'_{\Psi\rho}$ and $Y_{\Psi\rho}$ after the $Z'$ resonance point: $M_{\rm DM}^M=M_{Z'}/2$. For DM masses less than $M_{Z'}/2$, as we increase the coupling strengths $Y'_{\Psi\rho}$ and $Y_{\Psi\rho}$ from 0 to 0.9, new channels via the annihilation of $\Psi_L$ and $\Psi_R$ doublets become more significant and thus the final relic abundance decrease with the increase of new accessible annihilation channels. After the resonance, the channel pertaining to the annihilation of DM into $Z'$ states become the dominant channel, and thus, any other effects on the final abundance due to the change in Yukawa couplings become insignificant and are feebly noticeable only for the $Y'_{\Psi\rho}=Y_{\Psi\rho}=0.9$ curve for $M_{\rm DM}^M$ in range $1.0$ to $1.5$ TeV. All the plot lines also possess sharp resonance funnels at $M_{H, H_1, H_2, A}/2$, as expected.

In Figure~\ref{fig:ex6}, we explore the dependence of total relic density on the Majorana DM mass~$(M_{\rm DM}^M)$ for different values of Yukawa couplings $Y_{\Psi\Sigma}$ and $Y'_{\Psi\Sigma}$. The figure shows three different plot-lines associated to a set of values for Yukawa couplings $Y_{\Psi\Sigma}$ and $Y'_{\Psi\Sigma}$:  $Y_{\Psi\Sigma}=Y'_{\Psi\Sigma}=0$ in the solid gray line, $Y_{\Psi\Sigma}=Y'_{\Psi\Sigma}=0.5$ in the red dashed-dot line, and $Y_{\Psi\Sigma}=Y'_{\Psi\Sigma}=0.9$ in the blue dashed line. All the other relevant parameters have been fixed at a particular value with $M_{Z'}=2.0~\rm TeV$, $g_l=0.286$, $Y'_{\Psi\rho}=Y_{\Psi\rho}=0.5$, and $M_{H_1}$, $M_{H_2}$ and $M_A$ set to $1$ TeV. The inset-plot shows the relic abundance ratio for the two DM candidates as a function of $M_{\rm TeV}^M$. From the plot, it can be observed that the value of obtained relic abundance is independent of the Yukawa couplings $Y_{\Psi\Sigma}$ and $Y'_{\Psi\Sigma}$ for most of the considered parameter space, although a small dependence on these couplings is observed for $M_{\rm DM}^M$ values less than $500$ GeVs. This could be understood by the fact that at $M_{\rm DM}^M<M_{H_{1/2},A,Z'}$, there are no kinematically allowed DM annihilation channels into scalars $H_1$, $H_2$ and gauge boson $Z'$ containing final states. Thus, the contribution to $N_{1}^0$ mass eigenstates from $\Sigma$ and $\Psi$ particles become slightly significant, which allows a small dependence of total DM relic density on the values of couplings $Y_{\Psi\Sigma}$ and $Y'_{\Psi\Sigma}$. But once the scalar channels are available~(i.e. after the scalar resonance at $0.5$ TeV), the effect of $Y_{\Psi\Sigma}$ and $Y'_{\Psi\Sigma}$ on the obtained relic abundance become insignificant. From the figure, we also see a resonance behaviour around $M_{\rm DM}^M=M_{Z'}/2$, and sharp resonance funnels are obtained at $M_{H, H_1, H_2, A}/2$, as expected.
\subsection{Direct Detection prospects}
\label{subsec:DD}
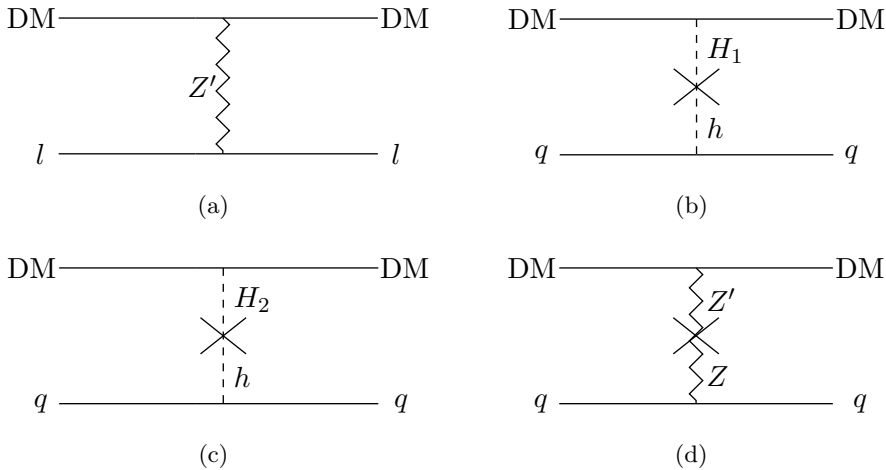
\begin{figure}[H]
		\centering
		\subfigure[]{
		\label{fig:DD1a}
		\begin{tikzpicture}[line width=0.5 pt, scale=1.2]
        \draw[solid] (-3.5,1.0)--(-2.0,1.0);
        \draw[solid] (-3.5,-0.5)--(-2.0,-0.5);
        \draw[snake](-1.7,1.0)--(-1.7,-0.5);
        \draw[solid] (-2.0,1.0)--(0.0,1.0);
        \draw[solid] (-2.0,-0.5)--(0.0,-0.5);
        \node at (-3.8,1.0) {$\text{DM}$};
        \node at (-3.7,-0.5) {$l$};
        \node [right] at (-2.2,0.25) {$Z'$};
        \node at (0.3,1.0) {$\text{DM}$};
        \node at (0.2,-0.5) {$l$};
        \end{tikzpicture}
        }~~~
        \subfigure[]{
        \label{fig:DD1b}
        \begin{tikzpicture}[line width=0.5 pt, scale=1.2]
        \draw[solid] (1.7,1.0)--(3.2,1.0);
        \draw[solid] (1.7,-0.5)--(3.2,-0.5);
        \draw[dashed](3.2,1.0)--(3.2,-0.5);
        \draw[solid] (2.95,0.45)--(3.45,0.05);
        \draw[solid] (2.95,0.05)--(3.45,0.45);
        \draw[solid] (3.2,1.0)--(4.7,1.0);
        \draw[solid] (3.2,-0.5)--(4.7,-0.5);
        \node at (1.4,1.0) {$\text{DM}$};
        \node at (1.5,-0.5) {$q$};
        \node [right] at (3.2,0.65) {$H_1$};
        \node [right] at (3.2,-0.2) {$h$};
        \node at (5.0,1.0) {$\text{DM}$};
        \node at (4.9,-0.5) {$q$};
        \end{tikzpicture}
        }\\
        \subfigure[]{
        \label{fig:DD1c}
        \begin{tikzpicture}[line width=0.5 pt, scale=1.2]
        \draw[solid] (-3.5,1.0)--(-2.0,1.0);
        \draw[solid] (-3.5,-0.5)--(-2.0,-0.5);
        \draw[dashed](-1.7,1.0)--(-1.7,-0.5);
        \draw[solid] (-1.95,0.45)--(-1.45,0.05);
        \draw[solid] (-1.95,0.05)--(-1.45,0.45);
        \draw[solid] (-2.0,1.0)--(0.0,1.0);
        \draw[solid] (-2.0,-0.5)--(0.0,-0.5);
        \node at (-3.8,1.0) {$\text{DM}$};
        \node at (-3.7,-0.5) {$q$};
        \node [right] at (-1.7,0.65) {$H_2$};
        \node [right] at (-1.7,-0.2) {$h$};
        \node at (0.29,1.0) {$\text{DM}$};
        \node at (0.25,-0.5) {$q$};
        \end{tikzpicture}
        }~~~
        \subfigure[]{
        \label{fig:DD1d}
        \begin{tikzpicture}[line width=0.5 pt, scale=1.2]
        \draw[solid] (1.7,1.0)--(3.2,1.0);
        \draw[solid] (1.7,-0.5)--(3.2,-0.5);
        \draw[snake](3.2,1.0)--(3.2,-0.5);
        \draw[solid] (2.95,0.45)--(3.45,0.05);
        \draw[solid] (2.95,0.05)--(3.45,0.45);
        \draw[solid] (3.2,1.0)--(4.7,1.0);
        \draw[solid] (3.2,-0.5)--(4.7,-0.5);
        \node at (1.4,1.0) {$\text{DM}$};
        \node at (1.5,-0.5) {$q$};
        \node [right] at (3.2,0.65) {$Z'$};
        \node [right] at (3.2,-0.2) {$Z$};
        \node at (5.0,1.0) {$\text{DM}$};
        \node at (5.0,-0.5) {$q$};
        \end{tikzpicture}
        }
     \caption{Processes that could potentially lead to scattering of DM particles with leptons and nuclie in our framework. Here, \ref{fig:DD1a} shows the interaction of DM with leptons at the tree level, and \ref{fig:DD1b}-\ref{fig:DD1d} depicts the Higgs mixing and kinetic mixing portal interaction of DM with nuclie.}
\label{fig:DD1}
\end{figure}

In a laboratory setup, the presence of a dark matter particle can, in principle, scatter coherently off the nuclei of a target atom through a tree-level or loop-mediated process, and this can provide a recoil momentum to the target atom. Such recoils, if detected, can directly infer the existence of dark matter. For our framework, DM candidates are leptophilic and have zero hypercharges, thus they can directly interact with leptons at tree level as shown in Figure~\ref{fig:DD1a} but don't directly interact with quarks/nucleons at tree level via either $Z$ boson or via $U(1)_{\ell}$ mediator $Z'$. They can only interact with nucleons through either a kinetic mixing or a Higgs portal, as shown in Figs.~\ref{fig:DD1b}-\ref{fig:DD1d}. Since for dark matter particles considered in this framework, the particle spin is not a crucial factor in investigation, the pertinent interactions with nucleus for our study are those that are spin-independent. Thus, in general, the dark matter-nucleon spin-independent direct detection~(SIDD) total cross section for our framework contains the contributions from the individual boson mediated channels and also via their interference terms. This could be expressed as~\cite{Patra:2016ofq}
\begin{eqnarray}\label{DM-nucleon-h-Z-Z'}
\sigma_{\rm SI}^{\text{TOT}} &=& \sigma_{\rm SI}^{h_i}\oplus\sigma_{\rm SI}^{Z} v^2\oplus\sigma_{\rm SI}^{Z'}v^2,
\end{eqnarray}
where $i=1,2,3$ for the three scalar mass eigenstates in the framework, and $v$ is the DM velocity. The explicit expressions for various scalar and vector-mediated cross-sections in Eq.~(\ref{DM-nucleon-h-Z-Z'}) can be referred from Refs.~\cite{Goodman:1984dc,Essig:2007az,FileviezPerez:2019jju,Bhattacharya:2018fus,MurguiGalvez:2020mcc}. For our framework containing two different DM candidates, the SIDD cross-section for both needs to be analyzed separately. From section~\ref{subsc:sc}, one may see that for the purpose of this work, we have assumed zero kinetic and Higgs portal mixings by setting couplings $\lambda_4$, $\lambda_5$ in the scalar sector and coupling $\kappa$ in the kinetic Lagrangian equal to zero. This leaves no direct way for our DM candidates to interact with quarks, and hence, at first glance, our framework seems to be insensitive to the direct detection probes. Interestingly, on closer analysis, we find that through some mixings within the exotic fermion states, a sizable interaction with quarks is possible.
With the assumption of zero kinetic and scalar portal mixings, for the case of Dirac DM candidate~$(\chi_1)$, the mixing between the various left-handed gauge eigenstates~$(\xi_{L},\eta_{L})$ or between the right-handed states~$(\zeta_{1R},\zeta_{2R})$ as shown in Eq.~(\ref{eq:DMM2}) cannot lead to a final DM particle that can couple to quarks directly because neither of these states couple directly to either the SM Higgs or to the SM $Z$ boson. Thus, we are always working in direct detection allowed parameter space for Dirac DM~$(\chi_1)$. With the same assumptions, if we inspect the Majorana DM candidate~$(N_1^0)$, we find that from the structure of mass matrix given in Eq.~(\ref{eq:ss1}), the framework allows mixing of $\rho_L$ with the neutral component of chiral exotic fermions $\Psi_{L,R}$ via the terms $\frac{Y_{\Psi\rho}v}{\sqrt{2}}$ and $\frac{Y'_{\Psi\rho}v}{\sqrt{2}}$, respectively. From the structures of $\Psi_L$ and $\Psi_R$ given in section~\ref{sec:II-model}, we see that they have non-zero hypercharge and can interact to nucleus via neutral currents. Thus, $\Psi_L$ and $\Psi_R$ can interact with quarks via the exchange of SM $Z$ bosons. With this, a substantial mixing of $\rho_L$ with $\Psi_L$ and $\Psi_R$ by keeping non zero $Y_{\Psi\rho}, Y'_{\Psi\rho}$ can induce non-zero SIDD cross-section values for the Majorana DM~$(N_1^0)$. Ref.~\cite{Essig:2007az} shows the explicit expression for the SIDD cross-section per nucleon for a $Z-$ boson mediated interaction. We see that for a fixed value of $M_{Z'}$, $\tan{\beta}$ and $g_L$, which decides the values for $v_1$ and $v_2$, the direct detection cross-section of the dark matter~$(N_1^0)$ mainly depends on the parameters $Y_{\Psi\rho}$, $Y'_{\Psi\rho}$ and $\lambda_{\rho}$. The dependence on $\lambda_{\rho}$ can be checked by varying SIDD cross-section as a function of $M^M_{\rm DM}$ for fixed values of $Y_{\Psi\rho}$, $Y'_{\Psi\rho}$. We show this analysis in the subsequent paragraphs.
\subsubsection{SIDD cross-section analysis}
\label{subsubsec:SIDD}
\begin{figure*}[tbp]
  \centering
  \includegraphics[width=.99\textwidth,scale=2.0]{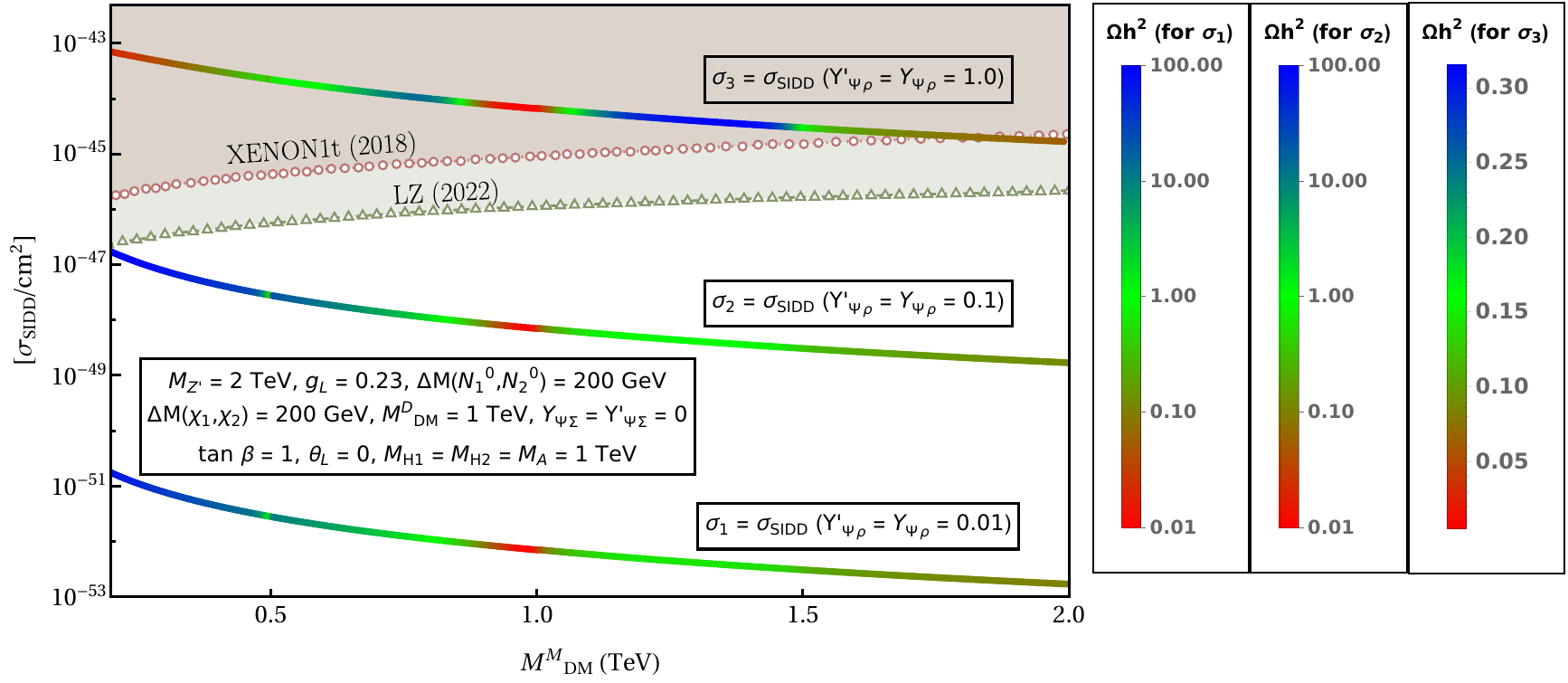}
  \caption{Spin-independent DD cross-section~$(\sigma_{\rm SIDD}/ \text{cm}^2)$ as a function of the mass of Majorana DM candidate~$(M^M_{\rm DM})$. The value of total relic density~$(\Omega h^2)$ is denoted at the color axis. The figure shows three curves corresponding to the three different combinations of input parameters, $(Y'_{\Psi\rho} \text{ and } Y_{\Psi\rho})$: $Y'_{\Psi\rho}=Y_{\Psi\rho}=0.01$ given as $\sigma_1$, $Y'_{\Psi\rho}=Y_{\Psi\rho}=0.1$ given as $\sigma_2$, and $Y'_v=Y_{\Psi\rho}=1.0$ given as $\sigma_3$. The mass of the Dirac DM candidate~$(M^D_{\rm DM})$ is fixed at 1 TeV, $\tan\beta=1$, $\theta_L=0$, and the other relevant Yukawa couplings are set at zero, i.e.~$Y_{\Psi\sigma}=Y'_{\Psi\sigma}=0$. All the other relevant parameters are mentioned within the plot. The experimental bounds on the SIDD cross-section from the XENON1t~\cite{XENON:2018voc} and LUX-ZEPLIN~\cite{LZ:2022lsv} measurements have also been included.}
  \label{fig:ex7}
\end{figure*}
We have implemented the direct detection analysis of our framework in micrOMEGAs~\cite{Belanger:2013oya} for both DM candidates. With Higgs portal and kinetic mixings being switched off~(i.e. set to zero), we are always in the SIDD allowed region for the case of Dirac DM candidate~$(\chi_1)$, for the entire parameter space, as has been discussed in the previous subsection~\ref{subsec:DD}.

For the Majorana type DM~$(N_1^0)$ in our framework, a non-zero SIDD cross-section value is obtained in  micrOMEGAs when the couplings $Y_{\Psi\rho}$ and $Y'_{\Psi\rho}$ are switched ON. To properly illustrate the dependence of the SIDD cross-section on the various input parameters relevant to Majorana DM phenomenology, we present a heat map in Figure~\ref{fig:ex7}. In the plot, SIDD cross-section $(\sigma_{\rm SIDD}/\text{cm}^2)$ is plotted as a function of Majorana DM mass $(M^M_{\rm DM})$. To generate the results in the figure, the relevant input parameters are kept at fixed values: $M_{Z'}=2\text{ TeV}$, $g_L=0.23$, $\Delta M(N_1^0, N_2^0)=200\text{ GeV}$, $\Delta M(\chi_1^0, \chi_2^0)=200\text{ GeV}$, $\tan{\beta}=1$, $\theta_L=0$. The scalar sector masses are chosen to be $M_{H_1}=M_{H_2}=M_{\rm A}=1\text{ TeV}$. The mass of Dirac DM candidate~$(M_{\rm DM}^D)$ is set at resonance point of $M_{Z'}/2$ i.e.~$1$~TeV. The other relevant Yukawa couplings~$(Y_{\Psi\Sigma}, Y'_{\Psi\Sigma})$ in the Majorana mass matrix that can affect the mass of $N^0_1$ are both taken to be zero to focus solely on the impact of $\Psi_L$ and $\Psi_R$ mixings with $\rho_L$. From the heat map, we see three colored bands associated to parameters $\sigma_1, \sigma_2, \sigma_3$ based on a specific choice of $Y'_{\Psi\rho}$ and $Y_{\Psi\rho}$. For simplicity, we keep $Y'_{\Psi\rho}=Y_{\Psi\rho}$ in our analysis. $\sigma_1$ colored band corresponds to $Y'_{\Psi\rho}=Y_{\Psi\rho}=0.01$, $\sigma_2$ colored band corresponds to $Y'_{\Psi\rho}=Y_{\Psi\rho}=0.1$ and $\sigma_3$ colored band corresponds to $Y'_{\Psi\rho}=Y_{\Psi\rho}=1$. From these bands, we observe that the cross-section $\sigma_{\rm SIDD}$ is directly proportional to the values of $Y'_{\Psi\rho}$ and $Y_{\Psi\rho}$. For the case of maximal mixing, i.e.~$Y'_{\Psi\rho}=Y_{\Psi\rho}=1.0$ in Figure~\ref{fig:ex7}, the obtained cross-section is ruled out from the XENON1t~\cite{XENON:2018voc} and LUX-ZEPLIN~\cite{LZ:2022lsv} measurements for nearly the entire considered parameter space. For a relativiely weaker mixing of $Y'_{\Psi\rho}=Y_{\Psi\rho}=0.1$ as shown for $\sigma_2$ in Figure~\ref{fig:ex7}, the obtained SIDD cross-section is below the current limits from the experimental data for the entire parameter space and thus allows the testability of the framework from the future DD experiments like XENONnT~\cite{XENON:2024wpa} that aim to set more stringent SIDD cross-section limits.

For the Figure~\ref{fig:ex7}, we also observe that the SIDD cross-section for a constant value of $Y'_{\Psi_\rho}$ and $Y_{\Psi_\rho}$ decreases with the increase of $M_{\rm DM}^M$. This can be understood from the fact that any increase in the value of mass for $N_1^0$ state can only be achieved by incrementing the value of $\lambda_{\rho}$ coupling, as the values for VEVs $v_1$ and $v_2$ are kept fixed for the analysis. This decreases the relative contribution of $\frac{Y_{\Psi\rho}v}{\sqrt{2}}$ and $\frac{Y'_{\Psi\rho}v}{\sqrt{2}}$ terms in the mass of $N_1^0$ given by~$M_{\rm DM}^M$. In the figure, we also reflect the value of obtained total DM relic density through color gradient for each of the considered plot points. The color gradient scale is present on the right side in Figure~\ref{fig:ex7}. By referring the color gradient scale, one may see that the DM relic requirements from the observational data~$(\Omega h^2\sim 0.1)$ is satisfied near the resonance point of $M_{\rm DM}^M=M_{Z'}/2$ and also towards the larger $M_{\rm DM}^M>1.5$ TeV values for all of the considered plot curves~$(\sigma_{1,2,3})$. Thus, from this analysis, we see that the framework has enough parameter space available from DD and observed relic density bounds that can testify to the model's viability from the near future DD experiments. This also paves way for the framework to be put under scanner from the indirect-detection techniques like gamma line signatures, which will be discussed in detail in a subsequent section ahead. Throughout this discussion, one should keep in mind that all the DD constraints are put on the Majorana DM candidate, and the Dirac DM candidate~$(\chi_1)$ remains insensitive to the SIDD experiments for the chosen set of mixing values associated to the kinetic mixing and Higgs portal.
\begin{figure*}[tbp]
  \centering
  \includegraphics[width=.94\textwidth,scale=2.0]{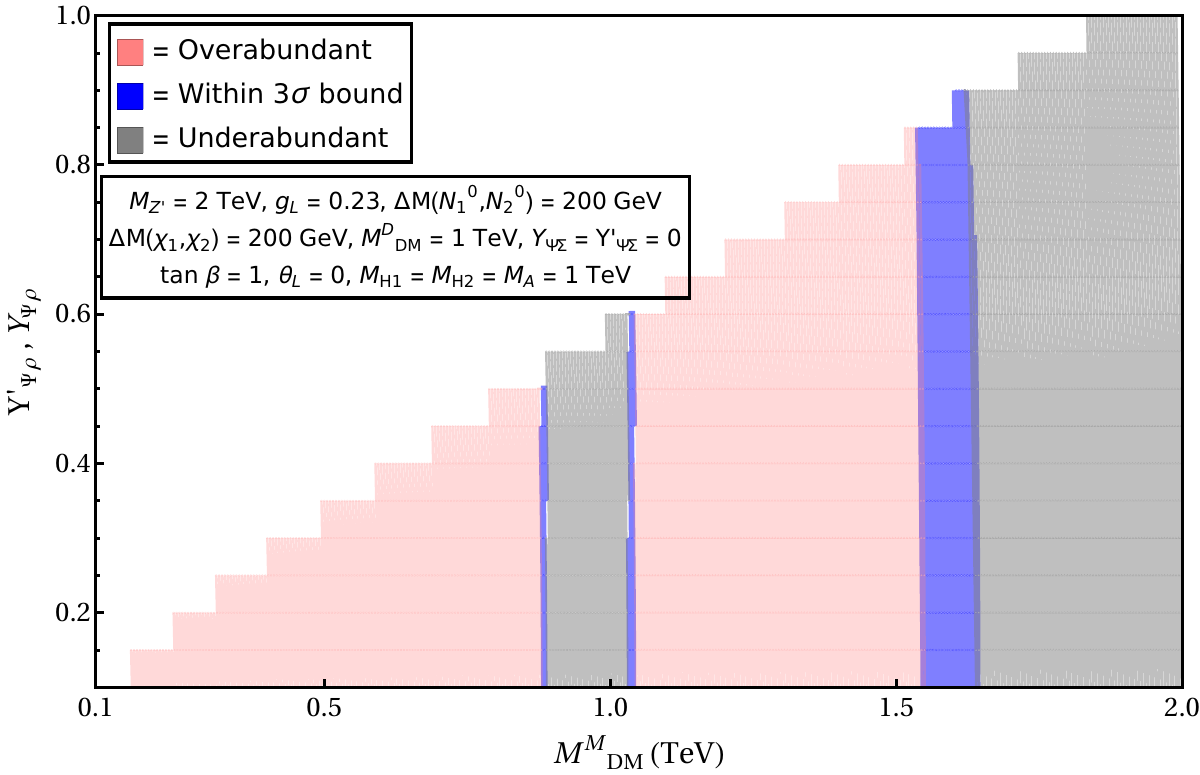}
  \caption{A scatter plot with the mass of Majorana DM in TeV at the horizontal axis and the Yukawa coupling constants, $Y'_{\Psi\rho}=Y_{\Psi\rho}$ at the vertical axis. The plot shows three regions of parameter space in the plane of $\text{M}_{\text{DM}}-Y'_{\Psi\rho},Y_{\Psi\rho}$: total DM relic density is under-abundant in the pink shaded region, over-abundant in the gray shaded region, and is within the $3\sigma$ bounds of experimental data in the blue shaded region. To generate this plot, the mass of the Dirac DM candidate is fixed at 1 TeV, parameter $\tan\beta$ is set at 1, and the mixing in the Dirac DM mass matrix is taken to be zero, i.e., $\theta_L=0$. Apart from this, we set the mass of the exotic gauge boson, $M_{Z'}= 2\text{ TeV}$, $g_L=0.23$, and $Y_3=Y_4=0$. All the other relevant input parameters are mentioned in the plot.}
  \label{fig:ex9}
\end{figure*}
\subsubsection{Allowed parameter space}
\label{subsubsec:DDpSpace}
From Figure~\ref{fig:ex7}, we see that a significant parameter space is still available for further exploration. Thus, to highlight the range of values for the input parameters~$(M_{\rm DM}^M, Y'_{\Psi\rho}, Y_{\Psi\rho})$ for which the DM relic abundance and SIDD cross-section falls within the allowed region, a scatter plot is shown in Figure~\ref{fig:ex9}. The relevant input parameters are fixed at a particular value common with the Figure~\ref{fig:ex7} to have a fair comparison with it. Figure~\ref{fig:ex9} plot contains two main regions where the colored region shows the parameter space that is allowed from the SIDD constraints, and the white region is already excluded from the DD data. The colored region is further distinguishable among three sub-regions, with the pink-colored region denoting an overabundant DM relic density. Similarly, the gray-colored region depicts the underabundant DM relic density, and the blue region shows the part of parameter space where the DM relic density is within the $3\sigma$ uncertainty of the observational value. To put things in numerical perspective, we have presented the results of Figure~\ref{fig:ex9} in a tabular form in Table~\ref{tab:bench1}. One may refer to the values of relevant input parameters from the first row of the table. For these input values, we list down few benchmark points where the model is allowed from DM relic density and SIDD requirements. It is evident from the Table~\ref{tab:bench1} that the Yukawa couplings~$(Y'_{\Psi\rho}, Y_{\Psi\rho})$ from $0.1-0.85$ for specific values of $M_{\rm DM}^M$ around $M_{Z'}/2$~(i.e. resonance) and for few other higher mass values~$(M_{\rm DM}^M>1000\rm GeV)$ provides us with a breathing space for the model from DM relic density and SIDD requirements for a fixed set of input parameters~$(M_{DM}^D=M_{Z'}/2,M_{Z'}=2\text{ TeV},g_{L}=0.23)$.
\begin{table}[H]
\centering
\begin{tabular}{|C{0.9cm}|C{2.5cm}|C{2.5cm}||C{2.5cm}|C{2.5cm}|}
\hline
\multicolumn{5}{|c|}{$M_{DM}^D=M_{Z'}/2,~~~M_{Z'}=2\text{ TeV},~~g_{L}=0.23$} \\
\hline
\multirow{2}{*}{\makecell{S. No.}}& \multicolumn{2}{c||}{$M_{DM}^M<M_{Z'}/2$}& \multicolumn{2}{c|}{$M_{DM}^M>M_{Z'}/2$ }  \\ \cline{2-5}
 & \multicolumn{1}{c|}{$M_{DM}^M~\text{(GeV)}$} & \multicolumn{1}{c||}{$Y'_{\Psi\rho}, Y_{\Psi\rho}$} & \multicolumn{1}{c|}{ $M_{DM}^M~\text{(GeV)}$} & \multicolumn{1}{c|}{$Y'_{\Psi\rho}, Y_{\Psi\rho}$}\\
\hline
\hline
1 & 889.776 & 0.10 & 1039.226 & 0.20\\
\hline
2 & 889.492 & 0.15 & 1036.898 & 0.40\\
\hline
3 & 889.095 & 0.20 & 1033.024 & 0.60\\
\hline
4 & 888.584 & 0.25 & 1564.876 & 0.10\\
\hline
5 & 887.961 & 0.30 & 1573.850 & 0.30\\
\hline
6 & 887.224 & 0.35 & 1566.790 & 0.50\\
\hline
7 & 886.375 & 0.40 & 1596.961 & 0.70\\
\hline
8 & 885.413 & 0.45 & 1601.986 & 0.80\\
\hline
9 & 884.338 & 0.50 & 1575.811 & 0.85\\
\hline
\end{tabular}
\caption{We present here the set of values for the parameters Majorana DM mass,~$M^{M}_{\text{DM}}$ and the Yukawa couplings $Y'_{\Psi\rho}, Y_{\Psi\rho}$ for which the model is within the allowed regions from DM relic density constraints and the SIDD cross-section constraints. The table has been made by selecting random data points from the dataset of Figure~\ref{fig:ex9}. The mass of the Dirac DM candidate is set at resonance point, $M^D_{\text{DM}}=M_{Z'}/2$ with the value of $M_{Z'}$ taken as $2$ TeV. The values of other relevant parameters are mentioned in the table.}
\label{tab:bench1}
\end{table}

\subsection{Indirect Searches}
\label{subsec:IDD}
After exploring the framework in the context of DD and the DM relic density constraints, we are ready to explore the indirect detection~(ID) experiments that could constrain the parameter space of our model. In the following, we present a general discussion on ID techniques and then specifically focus on the generation of distinct gamma line signatures that could provide indirect evidences for the viability of our framework.
\subsubsection{DM annihilations}
\label{subsubsec:DManni}
Apart from the possibility of a direct evidence by studying the recoil momentum of target nucleus in a controlled laboratory setup, DM can also be confirmed indirectly through its annihilation into various Standard Model (SM) states, a few of which have been illustrated in the Feynman diagrams in Figs.~\ref{fig:Feyn1} and~\ref{fig:Feyn3}. These processes must potentially be observed in places where the density of DM is still fairly high giving detectable amount of signals through DM annihilations. If such a process were to occur somewhere in the nearby galactic medium, we would possibly detect a flux of SM particles such as photons, anti-protons, and positrons coming towards us~\cite{HESS:2006vje, PAMELA:2008gwm, Blasi:2009bd}. Researchers thus search for streams of such particles unaccounted for by any other known physical phenomenon in regions assumed to have high DM density, such as galactic centers~\cite{Strong:2005zx, Maier:2008vw, Thompson:2008rw}, to indirectly confirm the presence of DM via its annihilations. These annihilations can, in general, take various forms with the possibility of different types of SM particles in the final state. In this work, we specifically focus on indirect detection techniques where the annihilation processes give off gamma photons in the final state.
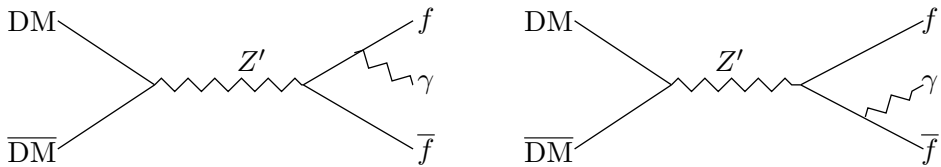
\begin{figure}[H]
\centering
    \begin{tikzpicture}[line width=0.5 pt, scale=0.85]
          \draw[solid] (-3.0,1.0)--(-1.5,0.0);
        \draw[solid] (-3.0,-1.0)--(-1.5,0.0);
         \draw[snake] (-1.5,0.0)--(0.8,0.0);
        \draw[solid] (0.8,0.0)--(2.5,1.0);
        \draw[snake] (1.59,0.5)--(2.5,0.0);
         \draw[solid] (0.8,0.0)--(2.5,-1.0);
         \node at (-3.4,1.0) {$\text{DM}$};
         \node at (-3.4,-1.0) {$\overline{\text{DM}}$};
         \node [above] at (0.0,0.05) {$Z'$};
        \node at (2.7,1.0) {$f$};
        \node at (2.7,0.0) {$\gamma$};
        \node at (2.7,-1.0) {$\overline{f}$};
         \draw[solid] (5.0,1.0)--(6.5,0.0);
        \draw[solid] (5.0,-1.0)--(6.5,0.0);
         \draw[snake] (6.5,0.0)--(8.5,0.0);
         \draw[solid] (8.5,0.0)--(10.4,1.0);
         \draw[solid] (8.5,0.0)--(10.4,-1.0);
          \draw[snake] (9.5,-0.5)--(10.4,0.0);
         \node at (4.6,1.0) {$\text{DM}$};
         \node at (4.6,-1.0) {$\overline{\text{DM}}$};
         \node [above] at (7.4,0.05) {$Z'$};
         \node at (10.5,1.0) {$f$};
        \node at (10.5,-1.0) {$\overline{f}$};
        \node at (10.5,0.0) {$\gamma$};
     \end{tikzpicture}
\caption{DM annihilations producing photon lines in the final state, also known as final state radiation~(FSR). These processes take place at the tree level. Here, $f$ can either be an SM or an exotic fermion, and $Z'$ is the mediator associated with the new gauge force present in the theory.}
\label{fig:FSR}
\end{figure}

\begin{figure}[H]
\centering
    \begin{tikzpicture}[line width=0.5 pt, scale=0.85]

          \draw[solid] (-1.5,1.2)--(-0.5,0.0);
        \draw[solid] (-1.5,-1.2)--(-0.5,0.0);
         \draw[snake] (-0.5,0.0)--(1.3,0.0);  \draw[dashed] (-0.5,0.0)--(1.3,0.0);
        \draw[solid] (1.3,0.0)--(2.6,0.7);
        \draw[solid] (1.3,0.0)--(2.6,-0.7);
        \draw[solid] (2.6,0.7)--(2.6,-0.7);
        \draw[snake] (2.6,0.7)--(3.2,1.2);
         \draw[snake] (2.6,-0.7)--(3.2,-1.2);
         \node at (-1.9,1.2) {$\text{DM}$};
         \node at (-1.9,-1.2) {$\overline{\text{DM}}$};
         \node [above] at (0.5,0.05) {$Z/Z', h_i$};

        \node at (2.1,0.0) {$f$};
        \node at (3.3,1.2) {$\gamma$};
        \node at (3.6,-1.2) {$\gamma, Z$};
         \draw[solid] (5.5,1.2)--(6.5,0.0);
        \draw[solid] (5.5,-1.2)--(6.5,0.0);
         \draw[snake] (6.5,0.0)--(8.3,0.0);  

         \draw[solid] (8.3,0.0)--(9.5,0.7);
        \draw[solid] (8.3,0.0)--(9.5,-0.7);
        \draw[solid] (9.5,0.7)--(9.5,-0.7);

         \draw[snake] (9.5,0.7)--(10.4,1.2);
         \draw[dashed] (9.5,-0.7)--(10.4,-1.2);
         \node at (5.1,1.2) {$\text{DM}$};
         \node at (5.1,-1.2) {$\overline{\text{DM}}$};
         \node [above] at (7.6,0.05) {$Z'$};
          \node at (9.0,0.0) {$f$};
         \node at (10.58,1.2) {$\gamma$};
        \node at (10.58,-1.2) {$h$};
     \end{tikzpicture}
\caption{DM annihilations into gamma rays for the processes ${\text{DM}}\hspace{0.03cm}\overline{\text{DM}} \to \gamma \gamma, \gamma Z, \gamma h$ at loop level for any BSM gauge theory involving an exotic gauge mediator~$(Z')$. Here, $f$ can either be an SM or an exotic charged fermion, $Z'$ is the mediator associated with the new gauge force, and $h_i$ denotes the scalar mass eigenstates present in the theory with $i\in\{1,2...,\text{ total number of scalars}\}$.}
\label{fig:GammaL}
\end{figure}
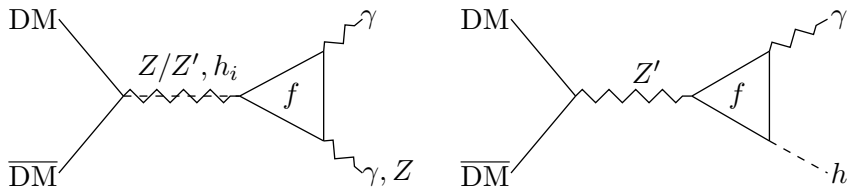

In a DM annihilation process, photons can primarily be emitted from the external legs, final-state radiation~(FSR), or via subsequent decay of the pair of particles directly produced. These photon-emitting processes can be mediated via the theory's gauge bosons or scalars at the tree level or at a loop level. The FSR fluxes, as shown in Figure~\ref{fig:FSR}, generally occur as tree-level emissions and hence are usually quite abundant across all energy ranges, making them a continuum with a statistical distribution of intensity peaks~\cite{Birkedal:2005ep, Bergstrom:2004cy}. Apart from them, there can be DM annihilation channels that produce distinct monochromatic photons via processes like $\rm DM\hspace{0.02in}\overline{\rm DM}\rightarrow Z'\rightarrow \gamma\gamma, h\gamma, Z\gamma$ as shown in Figure~\ref{fig:GammaL}. These processes are loop-level interactions and hence can, in general, be suppressed with respect to the continuum of FSR. This makes it tricky to detect their distinct monochromatic fluxes over the background of FSR. However, as pointed out in Ref.~\cite{Duerr:2015vna}, one may be able to significantly suppress FSR for the case of cold Majorana-type dark matter annihilation and also simultaneously produce distinct gamma peaks as signals to look for, at the experiments. Following Ref.~\cite{MurguiGalvez:2020mcc}, we find that a Majorana type DM couples to the exotic gauge boson solely in an axial way, and the form of squared amplitude of the FSR process in such a case can be expressed as:
\begin{equation}
|\mathcal{M}|^2_{\rm FSR}=\frac{M_f^2}{M_{Z'}^2}A+v^2B+\mathcal{O}(v^4),
 \label{eq:FSR}
\end{equation}
where, $A$ and $B$ are coefficients defined in the appendix of Ref.~\cite{MurguiGalvez:2020mcc}, and $v$ is the velocity of DM particle. From this, we can deduce that the FSR processes can be suppressed either by the mass value of $Z'$ or by the velocity of the DM candidate or both. Given the cold nature of DM today, $v$ is non-relativistic, and thus, the FSR continuum is inhibited in this case, and with sufficiently high gamma line fluxes from DM annihilations, one can hope for their distinct signatures. Now, although the rate of tree-level annihilation corresponding to FSR is small today for a Majorana DM, this does not have any consequences for the relic density annihilation rate. The processes contributing to the annihilation of DM to set its relic density occurred much earlier in the Universe, when the DM velocity was significantly higher, mitigating the velocity suppression effects. Thus, the relic density is not overabundant today, as demonstrated in Figs.~\ref{fig:ex1}-\ref{fig:ex6}. In this section, we explore the detection prospects of distinct gamma lines from the Majorana DM candidate~$(N_1^0)$ within our framework. For this search, we have considered the limits on velocity averaged cross-section~$\textlangle\sigma v\textrangle$ for the processes involving DM annihilations into $\gamma\gamma$ and into $\gamma Z$ from indirect-detection experiments such as the Fermi Gamma-ray Space Telescope~\cite{Fermi-LAT:2009ihh}, H.E.S.S.~\cite{HESS:2013rld}, and CTA~\cite{CTAConsortium:2017dvg}. We aim to constrain the parameter space of the model from indirect detection which is consistent with relic density and direct detection requirements.
\subsubsection{Gamma lines signature}
\label{subsubsec:GammaL}
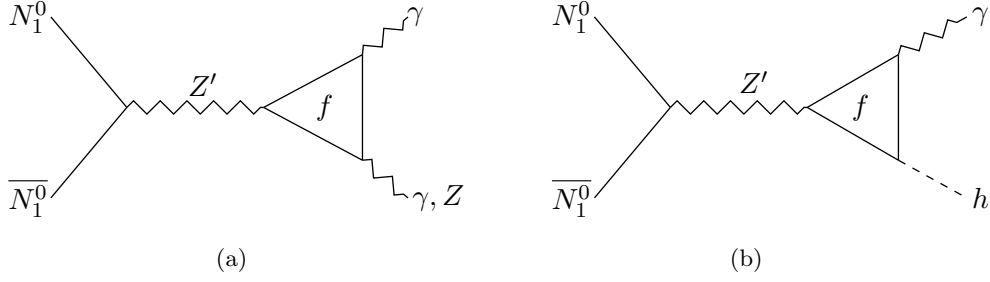
\begin{figure}[H]
        \centering
		\subfigure[]{
		\label{fig:Gamma1}
        \begin{tikzpicture}[line width=0.5 pt, scale=1]
          \draw[solid] (-1.5,1.2)--(-0.5,0.0);
        \draw[solid] (-1.5,-1.2)--(-0.5,0.0);
         \draw[snake] (-0.5,0.0)--(1.3,0.0);
         \draw[solid] (1.3,0.0)--(2.6,0.7);
        \draw[solid] (1.3,0.0)--(2.6,-0.7);
        \draw[solid] (2.6,0.7)--(2.6,-0.7);
        \draw[snake] (2.6,0.7)--(3.2,1.2);
         \draw[snake] (2.6,-0.7)--(3.2,-1.2);
         \node at (-1.8,1.2) {$N_1^0$};
         \node at (-1.8,-1.2) {$\overline{N_1^0}$};
         \node [above] at (0.5,0.02) {$Z'$};
        \node at (2.1,0.0) {$f$};
        \node at (3.3,1.2) {$\gamma$};
        \node at (3.6,-1.2) {$\gamma, Z$};
        \end{tikzpicture}
        }~~
        \subfigure[]{
        \label{fig:Gamma2}
        \begin{tikzpicture}[line width=0.5 pt, scale=1]
        \hspace{0.1in}
        \draw[solid] (5.5,1.2)--(6.5,0.0);
        \draw[solid] (5.5,-1.2)--(6.5,0.0);
         \draw[snake] (6.5,0.0)--(8.3,0.0);
         \draw[solid] (8.3,0.0)--(9.5,0.7);
        \draw[solid] (8.3,0.0)--(9.5,-0.7);
        \draw[solid] (9.5,0.7)--(9.5,-0.7);
         \draw[snake] (9.5,0.7)--(10.4,1.2);
         \draw[dashed] (9.5,-0.7)--(10.4,-1.2);
         \node at (5.2,1.2) {$N_1^0$};
         \node at (5.2,-1.2) {$\overline{N_1^0}$};
         \node [above] at (7.6,0.05) {$Z'$};
          \node at (9.0,0.0) {$f$};
         \node at (10.58,1.2) {$\gamma$};
        \node at (10.58,-1.2) {$h$};
     \end{tikzpicture}
     }
\caption{Majorana DM~$(N_1^0)$ annihilations into gamma photons at the loop level processes $N_1^0\hspace{0.03cm}\overline{N_1^0} \to \gamma \gamma, \gamma Z$ in the left panel and $N_1^0\hspace{0.03cm}\overline{N_1^0} \to \gamma h$ in the right panel. Here, fermions $f$ entering in the loop can either be a SM charged lepton~$(e,\mu,\tau)$ or a exotic charged fermion mass eigenstate~$(\Psi_L^+, \Psi_R^+, \Sigma_L^+, \Sigma_L^-)$, and $Z'$ is the $U(1)_{l}$ gauge Boson.}
\label{fig:GammaLU1L}
\end{figure}

In our analysis, we first look into the emission of loop-mediated gamma lines from the annihilation of $N_1^0$ into various $\gamma$-contributing channels. In general, the relevant terms in the Lagrangian for such an interaction can be expressed as,
\begin{equation}
\mathscr{L_{\gamma}}=g_L\overline{N_{1}^0}\gamma^{\mu}\gamma_{5}N_{1}^0Z'_{\mu}+g_L\overline{f}(n_V^f\gamma^{\mu}+n_A^f\gamma^{\mu}\gamma_{5})fZ'_\mu+\overline{f}\gamma^\mu(g_V^f+g_A^f\gamma_5)fZ_\mu,
 \label{eq:gammaLag}
\end{equation}
where, $g_L$ is the gauge coupling associated with the gauge group~$(U(1)_L)$, $f$ are the electrically-charged fermions participating in the loop from the diagram given in Figure~\ref{fig:GammaLU1L}, $n^f_{V/A}$ denotes the interaction strength between the new mediator and the charged fermions, and $g_{V/A}^f$ parameterize how these fermions interact with $Z$ boson. The subscript $V$ and $A$ refer to the vector and the axial-vector type couplings, respectively. In our model, there are four exotic electrically-charged fermionic states~$(\Psi^{+}_L,\Psi^{+}_R,\Sigma^{+},\Sigma^{-})$ that will participate in the loops within these diagrams other than the charged SM particles. These new states are all color singlets and have their electric and leptophilic charges given respectively as: $\Psi^{+}_L=[+1,3/2]$, $\Psi^{+}_R=[+1,-3/2]$, $\Sigma^{+}=[+1,-3/2]$~and $\Sigma^{-}=[-1,-3/2]$. The various SM fermions that can participate in these loops are the three charged leptons~$(e,\mu,\tau)$. Following, Refs.~\cite{Duerr:2015vna,MurguiGalvez:2020mcc}, the expressions for the processes $\overline{N_1^0}N_1^0\rightarrow Z'\rightarrow \gamma\gamma, h\gamma, Z\gamma$ for our case can be expressed as:
\begin{eqnarray}
\sigma(\overline{N_1^0}N_1^0\rightarrow \gamma\gamma) &=& \frac{81\alpha^2}{32\pi^3}\frac{g_L^4M_{N_1^0}^2(s-M_{Z'}^2)^2[M_{\Psi}^2C_0(s;M_\Psi)-M_{\Sigma}^2C_0(s;M_\Sigma)]^2}{M_{Z'}^4[(s-M_{Z'}^2)^2+\Gamma_{Z'}^2M_{Z'}^2]\sqrt{1-4M_{N_1^0}^2/s}} \nonumber, \\
\sigma(\overline{N_1^0}N_1^0\rightarrow h\gamma) &=&v^2 \frac{g_L^4n^2\alpha}{768\pi^4M_{N_1^0}^2}~\frac{4M_{N_1^0}^2-M_{Z}^2}{(4M_{N_1^0}^2-M_{Z'}^2)^2+\Gamma_{Z'}^2M_{Z'}^2}\bigg|\sum_fg_S^fn_V^fM_fQ_f\nonumber \\
&\times&\bigg(\frac{8M_{N_1^0}^2}{4M_{N_1^0}^2-M_h^2}(\Lambda^f_{N_1^0}-\Lambda^f_{h})+2+(4M_{N_1^0}^2-M_h^2+4M_f^2)C_0^h\bigg)\bigg|^2 \nonumber,\\
\sigma(\overline{N_1^0}N_1^0\rightarrow Z\gamma) &=& \frac{g_L^4g_2^2e^2n^2(2M_{N_1^0}^2+s)(s-M_{Z}^2)^3(M_Z^2+s)|\Sigma_fQ_f(C_0^f+B_3^fM_Z^2)|^2}{24576\pi^5\cos{\theta_W^2}M_{Z}^2s^{5/2}\sqrt{s-4M_{N_1^0}^2}[(s-M_{Z'}^2)^2+\Gamma^2_{Z'}M_{Z'}^2]}\nonumber.\\
\label{eq:gammaCross}
\end{eqnarray}
Here, $\alpha=e^2/4\pi$, $C_0(s;m)$ is the scalar Passarino-Veltman function\footnote{\begin{equation}
C_0(s;m)=\frac{1}{2s}\text{ln}^2\Bigg(\frac{\sqrt{1-\frac{4m^2}{s}}-1}{\sqrt{1-\frac{4m^2}{s}}+1}\Bigg)
\end{equation}
} as given in Ref.~\cite{Duerr:2015vna}, $C_0^A=C_0(0,M_A^2,s;M_f,M_f,M_f)$ is the Passarino-Veltman loop function as defined in Ref.~\cite{Patel:2015tea}, $g_2$ is the $SU(2)$ gauge coupling, $n$ is the leptophilic charge of $N_1^0$, the index $f$ runs over all the model fermions that can participate in the loop with $Q$ being their electric charge, $M$ being their masses in GeVs, $n_A$ being the associated quantum number for the axial coupling of DM with $f$ fermion, $g_V$ being the strength for the vectorial coupling of DM with $f$ fermion. Besides that, $\Gamma_{Z'}$ is the total decay width of $Z'$ and can be expressed as: \\
\begin{equation}
 \Gamma_{Z'}=g_L^2\frac{M_{Z'}}{12\pi}\Sigma_f\bigg[Q_{L}^2(f)\Big(1+2\tfrac{M^2(f)}{M_{Z'}}\Big)\sqrt{1-4\tfrac{M(f)}{M_{Z'}^2}}\bigg],
\end{equation}
where, $f$ again denotes the fermions in the loop with $Q_L$ being their leptophilic charges and $M$ being their masses in GeVs. Following Ref.~\cite{Duerr:2015wfa}, we can infer that the process $\overline{N_1^0}N_1^0\rightarrow h\gamma$~(shown in Figure~\ref{fig:Gamma2}) is velocity suppressed for a cold Majorana DM and the obtained cross section for this process is quite low. Thus, we can safely neglect it in our analysis.
\subsubsection{Results and Discussions}
\label{subsubsec:RnDIDD}
 \begin{figure*}[!h]
  \centering
  \includegraphics[width=.99\textwidth,scale=2.0]{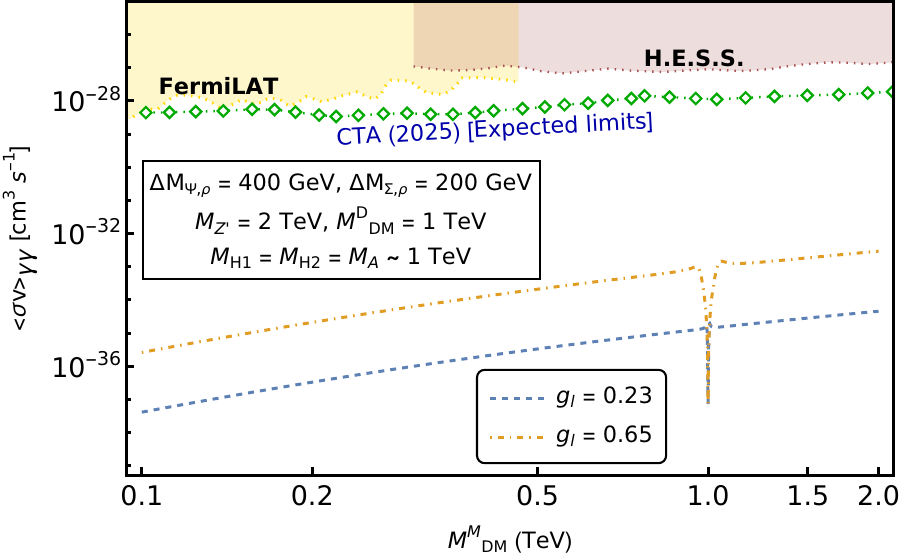}
  \caption{Velocity averaged cross section~($\text{cm}^3s^{-1}$) for the Majorana DM annihilation into two gamma photons as a function of Majorana DM mass. The cross-section is analyzed for two cases with $g_l=0.23$ (shown in the dot-dashed orange line) and with $g_l=0.65$ (shown in the dotted blue line). Other relevant parameters are kept fixed at a value with $\Delta M_{\Psi,\rho}=400\text{ GeV}$, $\Delta M_{\Sigma,\rho}=200\text{ GeV}$, $M_{Z'}=2\text{ TeV}$, $M^{D}_{\rm DM}=1\text{ TeV}$ and $M_{H1}=M_{H2}=M_{A}=1\text{ TeV}$. Experimentally probed parameter space from FermiLAT~\cite{Fermi-LAT:2009ihh} and H.E.S.S.~\cite{HESS:2013rld} are shown with yellow and violet regions, respectively and the expected new limits from CTA~\cite{CTAConsortium:2017dvg} are shown with green curve.}
  \label{fig:ex10}
\end{figure*}
\begin{figure*}[!h]
  \centering
  \includegraphics[width=.99\textwidth,scale=2.0]{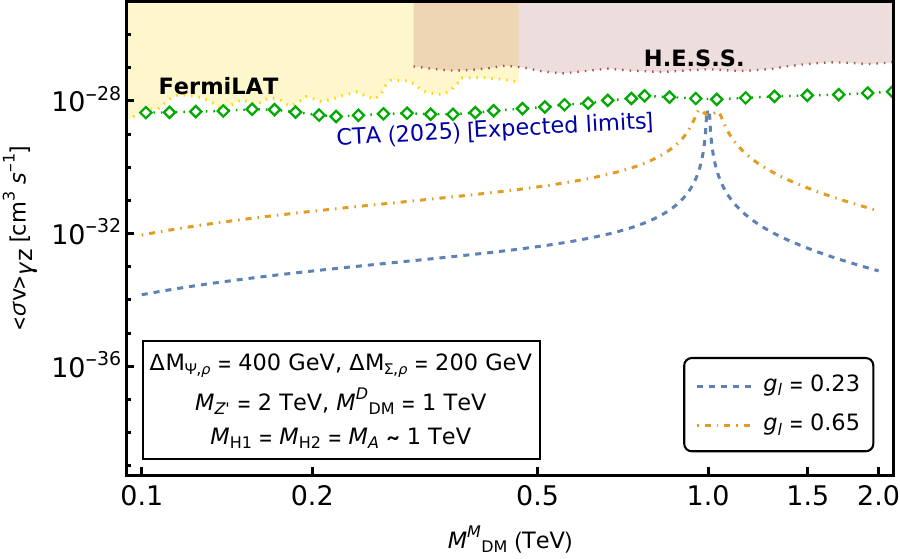}
  \caption{Velocity averaged cross section~($\text{cm}^3s^{-1}$) for the Majorana DM annihilation into a gamma photon and a $Z$ boson as a function of Majorana DM mass. The cross-section is analyzed for two cases with $g_l=0.23$ (shown in the dot-dashed orange line) and with $g_l=0.65$ (shown in the dotted blue line). To generate this plot, other relevant parameters are kept fixed at a value with $\Delta M_{\Psi,\rho}=400\text{ GeV}$, $\Delta M_{\Sigma,\rho}=200\text{ GeV}$, $M_{Z'}=2\text{ TeV}$, $M^{D}_{\rm DM}=1\text{ TeV}$ and $M_{H1}=M_{H2}=M_{A}=1\text{ TeV}$. Experimentally probed parameter space from FermiLAT~\cite{Fermi-LAT:2009ihh} and H.E.S.S.~\cite{HESS:2013rld} are shown with yellow and violet regions, respectively  and the expected new limits from CTA~\cite{CTAConsortium:2017dvg} are shown with green curve.}
  \label{fig:ex11}
\end{figure*}
Here, we present our results for the obtained velocity averaged cross-section~$(<\sigma v>)$ of Majorana DM annihilations into $\gamma\gamma$ and $\gamma Z$. For the analysis, the mass of Dirac DM is kept at $M^D_{\rm DM}=1\text{ TeV}$ and the masses of scalar sector particles $H_1, H_2$ and $A$ are set at $1$ TeV. Within the Majorana DM mass sector, we have charged fermionic states coming from the exotic particles, $\Psi$ and $\Sigma$. These charged states can enter the loops that result in gamma line productions. Thus, the masses of these particles are important parameters and for our study, we fix them at values $M_{\Psi}=M_{\rm DM}^M+400 \text{ GeV}$ and $M_{\Sigma}=M_{\rm DM}^M+200 \text{ GeV}$. For a direct comparison with the results obtained in relic or direct detection analysis, the mass of $Z'$ is fixed at the same value of $M_{Z'}=2~\text{TeV}$ as in the case of relic density and DD analysis. We show our findings in figures~\ref{fig:ex10} and \ref{fig:ex11} for the $\gamma\gamma$ and $\gamma Z$ cases, respectively. For both cases, we have considered two choices of leptophilic gauge coupling $g_l$. Following Ref.~\cite{Duerr:2015vna}, one may see that for a Majorana DM, the possible effective DM interaction, in general, could be given as
\begin{equation}
 \mathcal{L}_{\rm DM}=c_{\rm AA}\overline{N_1^0}\gamma^{\mu}\gamma^5N_1^0\overline{f}\gamma_\mu\gamma^5f+c_{\rm AV}\overline{N_1^0}\gamma^{\mu}\gamma^5N_1^0\overline{f}\gamma_\mu f
 \label{eq:IDD_cross}
\end{equation}
Here, $c_{\rm AA}$ and $c_{\rm AV}$ refer to the strength of axial and vector-axial couplings for the DM, respectively. $f$ refers to an SM fermion or a new fermion needed for anomaly cancellation. For a $c_{\rm AA}\overline{N_1^0}\gamma^{\mu}\gamma^5N_1^0\overline{f}\gamma_\mu\gamma^5f$ type interaction, it is shown in Ref.~\cite{Duerr:2015vna} that the channels $\overline{N_1^0}N_1^0\rightarrow\gamma\gamma,\gamma Z$ are not velocity suppressed, whereas channel $\overline{N_1^0}N_1^0\rightarrow\gamma h$ is velocity suppressed. Similarly, for a $c_{\rm AV}\overline{N_1^0}\gamma^{\mu}\gamma^5N_1^0\overline{f}\gamma_\mu f$ type interaction, it is shown that both $\overline{N_1^0}N_1^0\rightarrow\gamma\gamma,\gamma h$ channels are velocity suppressed and $\overline{N_1^0}N_1^0\rightarrow\gamma Z$ channel is not suppressed from DM velocity. The annihilation $\overline{N_1^0}N_1^0\rightarrow\gamma h$ is suppressed in both cases because the Yukawa couplings for the new fermions entering the loop need to be small as one cannot have heavy chiral fermions, which would change the Higgs properties~\cite{Duerr:2015wfa}. These results have been summarized in a tabular form in Table~\ref{tab:gamma}.
\begin{table}[H]
\centering
\begin{tabular}{|C{3.7cm}|C{2.5cm}||C{2.5cm}|C{2.5cm}|}
\hline
Operator & $\gamma\gamma$ & $\gamma Z$ & $\gamma h$ \\
\hline\hline
$c_{\rm AA}\overline{N_1^0}\gamma^{\mu}\gamma^5N_1^0\overline{f}\gamma_\mu\gamma^5f$ & Unsuppressed & Unsuppressed & Suppressed\\
\hline
$c_{\rm AV}\overline{N_1^0}\gamma^{\mu}\gamma^5N_1^0\overline{f}\gamma_\mu f$ & Not possible & Unsuppressed & Suppressed \\
\hline
\end{tabular}
\caption{Expected gamma lines production strengths for a Majorana DM candidate from various interaction types. Here, $N_1^0$ is the Majorana DM candidate in our framework, and $f$ could be a charged SM or exotic charged fermion that enters in the loop for these interactions.}
\label{tab:gamma}
\end{table}
Results summarized in Table~\ref{tab:gamma} have important consequences for our analysis. In Figure~\ref{fig:ex10}, we show the value of the obtained cross-section for $\gamma\gamma$ production in both $g_l=0.23$ and $g_l=0.65$ cases are many orders of magnitude smaller than the most sensitive current experimental limits coming from CTA~\cite{CTAConsortium:2017dvg}. Thus, the obtained gamma flux from such a process might not be useful to testify our model, given its highly suppressed cross-sections for the entire range of interest for the Majorana DM mass with the given input conditions. In Figure~\ref{fig:ex11}, we see that for $g_l=0.65$, the obtained cross-section for $\gamma Z$ final states may be within the reach of future experiments for the considered parameter space and input conditions. This presents interesting consequences for probing the viability of our framework from indirect gamma fluxes obtained via the annihilation of the Majorana DM candidate. Its important to note that by tuning the input parameters (masses of charged fermions and $Z'$), one may obtain higher values for these cross-sections, and thus a more in-depth analysis is required to concretely comment on the indirect detection perspective of this framework.

\section{Conclusion}
\label{sec:conc}
In this work, we have explored a leptophilic extension of SM with $U(1)_l$ being promoted to gauge symmetry. The particle spectrum of the model is the same as that in Ref.~\cite{Chao:2010mp} with the replacement of 3 right-handed neutrinos~$(\nu_R)$ by four neutral fermions having fractional leptonic charges. This paves way for two stable dark matter candidates in the framework, with one being a purely Dirac type, denoted by $\chi_1$ and the other can be made purely Majorana by a careful choice of couplings, denoted by $N_1^0$. For our phenomenological analysis, we have explored the parameter space of the model that is consistent with DM relic density requirements and is within the reach of current and/or near-future complementary direct, in-direct searches.

Firstly, we investigate the dependence of relic abundances for two dark matter (DM) states on various input parameters of the model, including particle masses and couplings governing annihilation channels. Our analysis shows that relic scans conducted separately for each DM candidate yield insights into the model's parameter space. Additionally, we explore the implications of direct DM searches. We find that for a minimal mixing of Majorana type DM with the chiral fermion $\Psi$, a sufficiently high SIDD cross-section is obtained for the considered parameter space, that can validate our framework in future DD experiments like LZ Collaboration~\cite{LZ:2022lsv} and XENONnT~\cite{XENON:2024wpa}. However, under the assumptions of zero kinetic and Higgs portal mixings, the Dirac type DM appears insensitive to DD probes due to the absence of its direct or indirect interactions with the standard model quarks. Our findings, combined with constraints from relic requirements and direct detection data, delineate a common input parameter space for the model, which is outlined in Figure~\ref{fig:ex9} and few of the random values from the dataset of this plot are presented in a tabular structure in Table~\ref{tab:bench1}. Given the presence of a Majorana DM in our framework, the gamma lines signatures from the loop level DM annihilations to $\gamma\gamma$ and $\gamma Z$ can stand out as distinct peaks above the continuum background of FSR. This provides an interesting testing ground for our model from the ongoing and future indirect detection experiments like HERD~\cite{Betti:2024wwt},~CTA\cite{CTAConsortium:2017dvg}, AMEGO~\cite{Caputo:2017sjw}. We find that the obtained velocity averaged cross-section from the DM annihilation into $\gamma Z$ channel is sufficiently high to be probed in these future IDD experiments as shown in Figure~\ref{fig:ex11}.

With the presence of two stable DM candidates, one being a Dirac type and the other being a Majorana type, the framework allows a significant parameter space where the obtained DM relic density is within the observational bounds with a 3$\sigma$ uncertainty. Within this parameter space, it also has interesting phenomenological features that allow its testability from complimentary direct and indirect searches from the current and future experiments. As an extension to our work, we would like to look at the collider signatures of the charged exotic particles present in our theory. Also, the presence of extra scalars that break $U(1)_l$ could impact early universe cosmic bubble formations. Such events produce distinct gravitational waves, the nature of which could be analyzed in a complementary study. Thus, the near-future testability prospects of this framework from the DM direct, in-direct searches and also a potential collider and gravitational analysis underscores the phenomenological relevance and significance of our work.
\section{Acknowledgement}
Utkarsh Patel (UP) would like to acknowledge the financial support from the Indian Institute of Technology, Bhilai, and the Ministry of Education, Government of India, for conducting the research work. UP would also like to mention the warm hospitality provided by Institute of Physics, Bhubaneswar during his visit, where a significant part of the analysis has been conducted. Avnish is being supported under SERB core research grant, CRG/2022/002670, hosted by IIT Guwahati.
\newpage
\large\textbf{APPENDIX}
\appendix
\section{Feynman diagrams}
\label{app:FD}
In this section, we present all the significant Feynman diagrams that are relevant to the relic density analysis of both DM candidates. These interactions will incorporate all the DM annihilations, co-annihilations via other neutral exotic fermions, and co-annihilations via charged exotic fermions, as applicable. As we have implemented the model framework on the CalcHEP package, so all the diagrams in this section have been verified from there.
\subsection{Dirac DM~$\mathbf{(\chi_1)}$}
\label{app:DDM}
 The Feynman diagrams for the phenomenologically relevant annihilation and co-annihilation channels of the Dirac DM candidate~$\chi_1$ have been shown in Figs.~$\ref{fig:Feyn1}$ and~\ref{fig:Feyn2}, respectively.
\begin{figure}[H]
		\centering
		\subfigure[]{
		\label{fig:Feyn1a}
		\begin{tikzpicture}[line width=0.5 pt, scale=0.64]
        \draw[solid] (5.0,1.0)--(6.5,0.0);
        \draw[solid] (5.0,-1.0)--(6.5,0.0);
        \draw[dashed] (6.5,0.0)--(8.5,0.0);
        \draw[solid] (8.5,0.0)--(10.4,1.0);
        \draw[solid] (8.5,0.0)--(10.4,-1.0);
        \node at (4.7,1.0) {${\chi_1}$};
        \node at (4.7,-1.0) {$\overline{\chi}_1$};
        \node [above] at (7.6,-0.05) {$H_j/A_1$};
        \node at (10.7,1.0) {$N^0_1$};
        \node at (10.7,-1.0) {$\overline{N^0_1}$};
        \end{tikzpicture}
        }
        \subfigure[]{
        \label{fig:Feyn1f}
        \begin{tikzpicture}[line width=0.5 pt, scale=0.64]
	    \draw[solid] (-3.0,1.0)--(-1.5,0.0);
        \draw[solid] (-3.0,-1.0)--(-1.5,0.0);
        \draw[dashed] (-1.5,0.0)--(0.8,0.0);
        \draw[snake] (0.8,0.0)--(2.5,1.0);
        \draw[snake] (0.8,0.0)--(2.5,-1.0);
        \node at (-3.3,1.0) {${\chi_1}$};
        \node at (-3.3,-1.0) {$\overline{\chi}_1$};
        \node [above] at (-0.2,-0.1) {$H_j/A_1$};
        \node at (2.7,1.0) {$Z'$};
        \node at (2.7,-1.0) {$Z'$};
        \end{tikzpicture}
        }
        \subfigure[]{
		\label{fig:Feyn1i}
		\begin{tikzpicture}[line width=0.5 pt, scale=0.64]
        \draw[solid] (5.0,1.0)--(6.5,0.0);
        \draw[solid] (5.0,-1.0)--(6.5,0.0);
        \draw[snake] (6.5,0.0)--(8.5,0.0);
        \draw[solid] (8.5,0.0)--(10.4,1.0);
        \draw[solid] (8.5,0.0)--(10.4,-1.0);
        \node at (4.7,1.0) {${\chi_1}$};
        \node at (4.7,-1.0) {$\overline{\chi}_1$};
        \node [above] at (7.6,-0.08) {$Z'$};
        \node at (10.7,1.0) {$N^0_1$};
        \node at (10.7,-1.0) {$\overline{N^0_1}$};
        \end{tikzpicture}
        }
        \subfigure[]{
        \label{fig:Feyn1g}
        \begin{tikzpicture}[line width=0.5 pt, scale=0.9]
	    \draw[solid] (-3.5,1.0)--(-2.0,1.0);
        \draw[solid] (-3.5,-0.5)--(-2.0,-0.5);
        \draw[solid](-2.0,1.0)--(-2.0,-0.5);
        \draw[snake] (-2.0,1.0)--(0.0,1.0);
        \draw[snake] (-2.0,-0.5)--(0.0,-0.5);
        \node at (-3.7,1.0) {${\chi_1}$};
        \node at (-3.7,-0.5) {$\overline{\chi}_1$};
        \node [right] at (-2.05,0.25) {$\chi_1/\chi_2$};
        \node at (0.21,1.0) {$Z'$};
        \node at (0.21,-0.5) {$Z'$};
        \end{tikzpicture}
        }
     \caption{Annihilation of $\chi_1$ to Majorana DM~$(N_1^0)$, and exotic gauge boson~$(Z')$ final states mediated via exotic scalars~$(H_j/A_1)$ and vector boson~$(Z')$. The diagram~\ref{fig:Feyn1g} are t-channel processes mediated via mass-states in the Dirac DM sector~$(\chi_1,\chi_2)$. Here, $j=1,2$.}
\label{fig:Feyn1}
\end{figure}
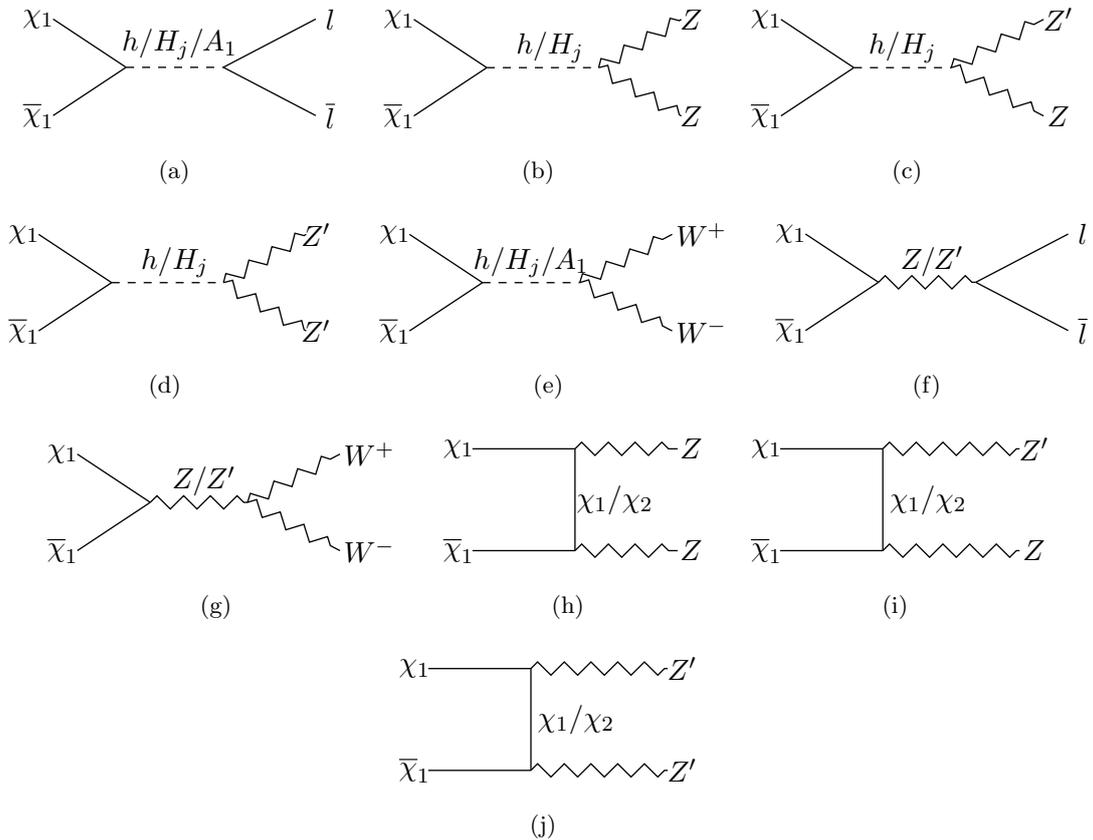
\begin{figure}[H]
		\centering
		\subfigure[]{
		\label{fig:Feyn2a}
		\begin{tikzpicture}[line width=0.5 pt, scale=0.64]
        \draw[solid] (5.0,1.0)--(6.5,0.0);
        \draw[solid] (5.0,-1.0)--(6.5,0.0);
        \draw[dashed] (6.5,0.0)--(8.5,0.0);
        \draw[solid] (8.5,0.0)--(10.4,1.0);
        \draw[solid] (8.5,0.0)--(10.4,-1.0);
        \node at (4.6,1.0) {${\chi_1}$};
        \node at (4.6,-1.0) {$\overline{\chi_2}$};
        \node [above] at (7.5,-0.1) {$H_j/A_1$};
        \node at (10.7,1.0) {$N^0_1$};
        \node at (10.7,-1.0) {$\overline{N^0_1}$};
        \end{tikzpicture}
        }
        \subfigure[]{
        \label{fig:Feyn2f}
        \begin{tikzpicture}[line width=0.5 pt, scale=0.64]
	    \draw[solid] (-3.0,1.0)--(-1.5,0.0);
        \draw[solid] (-3.0,-1.0)--(-1.5,0.0);
        \draw[dashed] (-1.5,0.0)--(0.8,0.0);
        \draw[snake] (0.8,0.0)--(2.5,1.0);
        \draw[snake] (0.8,0.0)--(2.5,-1.0);
        \node at (-3.4,1.0) {$\chi_1$};
        \node at (-3.4,-1.0) {$\overline{\chi_2}$};
        \node [above] at (-0.25,-0.1) {$H_j/A_1$};
        \node at (2.7,1.0) {$Z'$};
        \node at (2.7,-1.0) {$Z'$};
        \end{tikzpicture}
        }
        \subfigure[]{
        \label{fig:Feyn2l}
        \begin{tikzpicture}[line width=0.5 pt, scale=0.64]
        \draw[solid] (5.0,1.0)--(6.5,0.0);
        \draw[solid] (5.0,-1.0)--(6.5,0.0);
        \draw[snake] (6.5,0.0)--(8.5,0.0);
        \draw[solid] (8.5,0.0)--(10.4,1.0);
        \draw[solid] (8.5,0.0)--(10.4,-1.0);
        \node at (4.6,1.0) {$\chi_1$};
        \node at (4.6,-1.0) {$\overline{\chi_2}$};
        \node [above] at (7.6,-0.08) {$Z'$};
        \node at (10.7,1.0) {$N^0_1$};
        \node at (10.7,-1.0) {$\overline{N^0_1}$};
     \end{tikzpicture}
        }
        \subfigure[]{
        \label{fig:Feyn2g}
        \begin{tikzpicture}[line width=0.5 pt, scale=0.9]
	    \draw[solid] (-3.5,1.0)--(-2.0,1.0);
        \draw[solid] (-3.5,-0.5)--(-2.0,-0.5);
        \draw[solid](-2.0,1.0)--(-2.0,-0.5);
        \draw[snake] (-2.0,1.0)--(0.0,1.0);
        \draw[snake] (-2.0,-0.5)--(0.0,-0.5);
        \node at (-3.8,1.0) {$\chi_1$};
        \node at (-3.8,-0.5) {$\overline{\chi_2}$};
        \node [right] at (-2.05,0.25) {$\chi_1/\chi_2$};
        \node at (0.21,1.0) {$Z'$};
        \node at (0.21,-0.5) {$Z'$};
        \end{tikzpicture}
        }
     \caption{Co-annihilations of $\chi_1$ with the other neutral fermion state~$(\chi_2)$ to Majorana DM~$(N^1_0)$ and exotic gauge boson~$(Z')$ final states mediated via exotic scalars~$(H_j/A_1)$ and vector boson~$(Z')$. The diagram~\ref{fig:Feyn2g} is a t-channel processes to $Z'$ final states mediated via mass-states in the Dirac DM sector~$(\chi_1,\chi_2)$. Here, $j=1,2$.}
\label{fig:Feyn2}
\end{figure}
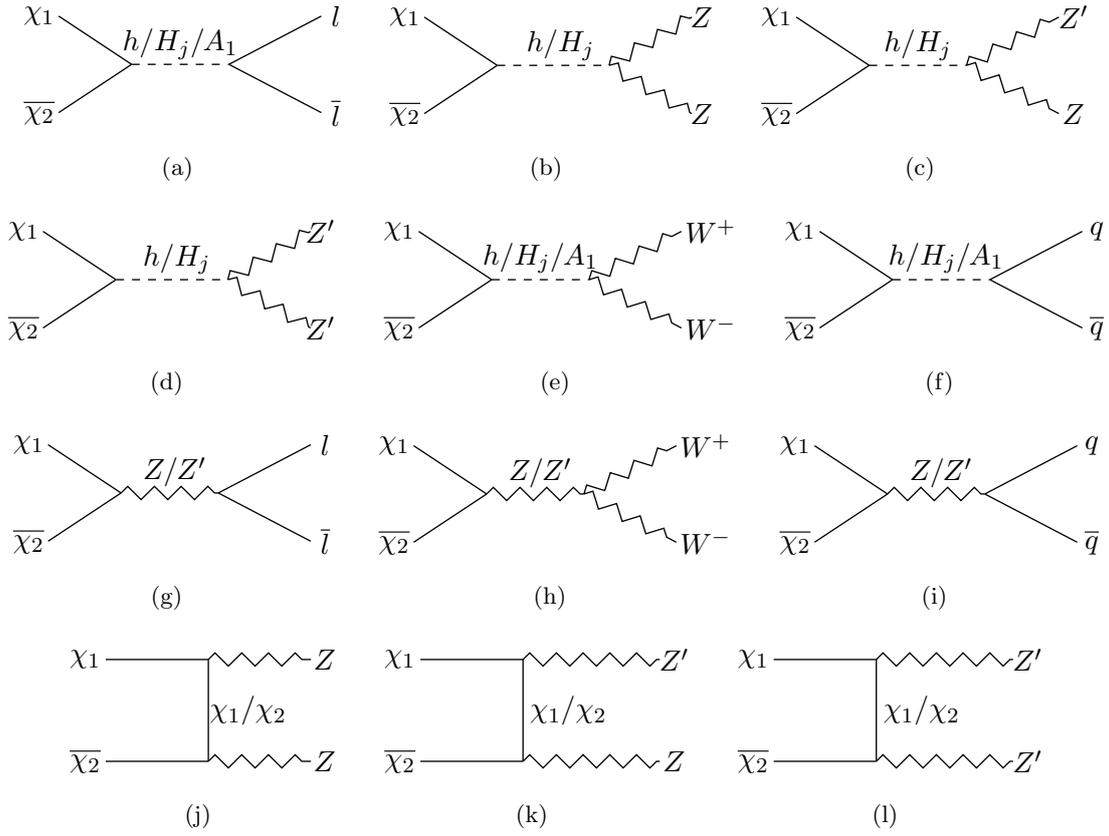
\newpage
\subsection{Majorana DM~$\mathbf{(N_1^0)}$}
\label{app:MDM}
The Feynman diagrams for the phenomenologically relevant annihilations and co-annihilations due to neutral exotic fermions of the Majorana DM candidate~$N_1^0$ have been shown in Figs.~\ref{fig:Feyn3} and~\ref{fig:Feyn4}, respectively. In comparison to the $\chi_1$ phenomenology, the presence of charged exotic fermions~$(\Psi_1^+,\Psi_2^+)$ in the Majorana DM sector opens up the possibility of new co-annihilation channels due to these charged exotic states for $N_1^0$. The diagrams relevant to these co-annihilations have been shown in Figure~\ref{fig:Feyn5}.
\begin{figure}[H]
		\centering
		\subfigure[]{
		\label{fig:Feyn3a}
		\begin{tikzpicture}[line width=0.5 pt, scale=0.64]
        \draw[solid] (5.0,1.0)--(6.5,0.0);
        \draw[solid] (5.0,-1.0)--(6.5,0.0);
        \draw[dashed] (6.5,0.0)--(8.5,0.0);
        \draw[solid] (8.5,0.0)--(10.4,1.0);
        \draw[solid] (8.5,0.0)--(10.4,-1.0);
        \node at (4.6,1.0) {${N_1^0}$};
        \node at (4.6,-1.0) {$\overline{N_1^0}$};
        \node [above] at (7.5,-0.05) {$h/H_j/A_1$};
        \node at (10.7,1.0) {$l$};
        \node at (10.7,-1.0) {$\overline{l}$};
        \end{tikzpicture}
        }
        \subfigure[]{
        \label{fig:Feyn3c}
        \begin{tikzpicture}[line width=0.5 pt, scale=0.64]
	    \draw[solid] (-3.0,1.0)--(-1.5,0.0);
        \draw[solid] (-3.0,-1.0)--(-1.5,0.0);
        \draw[dashed] (-1.5,0.0)--(0.8,0.0);
        \draw[snake] (0.8,0.0)--(2.5,1.0);
        \draw[snake] (0.8,0.0)--(2.5,-1.0);
        \node at (-3.4,1.0) {${N_1^0}$};
        \node at (-3.4,-1.0) {$\overline{N_1^0}$};
        \node [above] at (-0.2,-0.05) {$h/H_j$};
        \node at (2.7,1.0) {$Z$};
        \node at (2.7,-1.0) {$Z$};
        \end{tikzpicture}
        }
        \subfigure[]{
        \label{fig:Feyn3d}
        \begin{tikzpicture}[line width=0.5 pt, scale=0.64]
        \draw[solid] (5.0,1.0)--(6.5,0.0);
        \draw[solid] (5.0,-1.0)--(6.5,0.0);
        \draw[dashed] (6.5,0.0)--(8.5,0.0);
        \draw[snake] (8.5,0.0)--(10.4,1.0);
        \draw[snake] (8.5,0.0)--(10.4,-1.0);
        \node at (4.6,1.0) {${N_1^0}$};
        \node at (4.6,-1.0) {$\overline{N_1^0}$};
        \node [above] at (7.5,-0.05) {$h/H_j$};
        \node at (10.7,1.0) {$Z'$};
        \node at (10.7,-1.0) {$Z$};
        \end{tikzpicture}
        }
        \subfigure[]{
        \label{fig:Feyn3f}
        \begin{tikzpicture}[line width=0.5 pt, scale=0.64]
	    \draw[solid] (-3.0,1.0)--(-1.5,0.0);
        \draw[solid] (-3.0,-1.0)--(-1.5,0.0);
        \draw[dashed] (-1.5,0.0)--(0.8,0.0);
        \draw[snake] (0.8,0.0)--(2.5,1.0);
        \draw[snake] (0.8,0.0)--(2.5,-1.0);
        \node at (-3.4,1.0) {${N_1^0}$};
        \node at (-3.4,-1.0) {$\overline{N_1^0}$};
        \node [above] at (-0.3,-0.05) {$h/H_j$};
        \node at (2.7,1.0) {$Z'$};
        \node at (2.7,-1.0) {$Z'$};
        \end{tikzpicture}
        }
        \subfigure[]{
        \label{fig:Feyn3h}
        \begin{tikzpicture}[line width=0.5 pt, scale=0.64]
        \draw[solid] (5.0,1.0)--(6.5,0.0);
        \draw[solid] (5.0,-1.0)--(6.5,0.0);
        \draw[dashed] (6.5,0.0)--(8.5,0.0);
        \draw[snake] (8.5,0.0)--(10.4,1.0);
        \draw[snake] (8.5,0.0)--(10.4,-1.0);
        \node at (4.6,1.0) {${N_1^0}$};
        \node at (4.6,-1.0) {$\overline{N_1^0}$};
        \node [above] at (7.5,-0.05) {$h/H_j/A_1$};
        \node at (11.0,1.0) {$W^+$};
        \node at (11.0,-1.0) {$W^-$};
     \end{tikzpicture}
        }
        \subfigure[]{
		\label{fig:Feyn3i}
		\begin{tikzpicture}[line width=0.5 pt, scale=0.64]
        \draw[solid] (5.0,1.0)--(6.5,0.0);
        \draw[solid] (5.0,-1.0)--(6.5,0.0);
        \draw[snake] (6.5,0.0)--(8.5,0.0);
        \draw[solid] (8.5,0.0)--(10.4,1.0);
        \draw[solid] (8.5,0.0)--(10.4,-1.0);
        \node at (4.6,1.0) {${N_1^0}$};
        \node at (4.6,-1.0) {$\overline{N_1^0}$};
        \node [above] at (7.6,-0.08) {$Z/Z'$};
        \node at (10.7,1.0) {$l$};
        \node at (10.7,-1.0) {$\overline{l}$};
        \end{tikzpicture}
        }
        \subfigure[]{
        \label{fig:Feyn3j}
        \begin{tikzpicture}[line width=0.5 pt, scale=0.64]
        \draw[solid] (5.0,1.0)--(6.5,0.0);
        \draw[solid] (5.0,-1.0)--(6.5,0.0);
        \draw[snake] (6.5,0.0)--(8.5,0.0);
        \draw[snake] (8.5,0.0)--(10.4,1.0);
        \draw[snake] (8.5,0.0)--(10.4,-1.0);
        \node at (4.6,1.0) {${N_1^0}$};
        \node at (4.6,-1.0) {$\overline{N_1^0}$};
        \node [above] at (7.6,-0.08) {$Z/Z'$};
        \node at (11.0,1.0) {$W^+$};
        \node at (11.0,-1.0) {$W^-$};
     \end{tikzpicture}
        }
        \subfigure[]{
        \label{fig:Feyn3b}
        \begin{tikzpicture}[line width=0.5 pt, scale=0.9]
        \draw[solid] (1.7,1.0)--(3.2,1.0);
        \draw[solid] (1.7,-0.5)--(3.2,-0.5);
        \draw[solid](3.2,1.0)--(3.2,-0.5);
        \draw[snake] (3.2,1.0)--(4.7,1.0);
        \draw[snake] (3.2,-0.5)--(4.7,-0.5);
        \node at (1.4,1.0) {$N_1^0$};
        \node at (1.4,-0.5) {$\overline{N_1^0}$};
        \node [right] at (3.07,0.25) {$N_k^0$};
        \node at (4.9,1.0) {$Z$};
        \node at (4.9,-0.5) {$Z$};
        \end{tikzpicture}
        }
        \subfigure[]{
		\label{fig:Feyn3e}
		\begin{tikzpicture}[line width=0.5 pt, scale=0.9]
	    \draw[solid] (-3.5,1.0)--(-2.0,1.0);
        \draw[solid] (-3.5,-0.5)--(-2.0,-0.5);
        \draw[solid](-2.0,1.0)--(-2.0,-0.5);
        \draw[snake] (-2.0,1.0)--(0.0,1.0);
        \draw[snake] (-2.0,-0.5)--(0.0,-0.5);
        \node at (-3.8,1.0) {${N_1^0}$};
        \node at (-3.8,-0.5) {$\overline{N_1^0}$};
        \node [right] at (-2.05,0.25) {$N_k^0$};
        \node at (0.21,1.0) {$Z'$};
        \node at (0.21,-0.5) {$Z$};
        \end{tikzpicture}
        }
        \subfigure[]{
        \label{fig:Feyn3g}
        \begin{tikzpicture}[line width=0.5 pt, scale=0.9]
	    \draw[solid] (-3.5,1.0)--(-2.0,1.0);
        \draw[solid] (-3.5,-0.5)--(-2.0,-0.5);
        \draw[solid](-2.0,1.0)--(-2.0,-0.5);
        \draw[snake] (-2.0,1.0)--(0.0,1.0);
        \draw[snake] (-2.0,-0.5)--(0.0,-0.5);
        \node at (-3.8,1.0) {${N_1^0}$};
        \node at (-3.8,-0.5) {$\overline{N_1^0}$};
        \node [right] at (-2.05,0.25) {$N_k^0$};
        \node at (0.21,1.0) {$Z'$};
        \node at (0.21,-0.5) {$Z'$};
        \end{tikzpicture}
        }
        \subfigure[]{
        \label{fig:Feyn3k}
        \begin{tikzpicture}[line width=0.5 pt, scale=0.9]
	    \draw[solid] (-3.5,1.0)--(-2.0,1.0);
        \draw[solid] (-3.5,-0.5)--(-2.0,-0.5);
        \draw[solid](-2.0,1.0)--(-2.0,-0.5);
        \draw[snake] (-2.0,1.0)--(0.0,1.0);
        \draw[snake] (-2.0,-0.5)--(0.0,-0.5);
        \node at (-3.8,1.0) {${N_1^0}$};
        \node at (-3.8,-0.5) {$\overline{N_1^0}$};
        \node [right] at (-2.05,0.25) {$\Psi_1^+/\Psi_2^+$};
        \node at (0.38,1.0) {$W^+$};
        \node at (0.38,-0.5) {$W^-$};
        \end{tikzpicture}
        }
     \caption{Annihilation of $N_1^0$ to leptons~$(l)$, SM and exotic gauge boson~$(Z,Z',W^+,W^-)$ final states mediated via Standard model Higgs~$(h)$ and $Z$ boson, and also via exotic scalars~$(H_j/A_1)$ and vector boson~$(Z')$. The diagrams~\ref{fig:Feyn3b}-\ref{fig:Feyn3g} are t-channel processes mediated via mass-states in the Majorana DM sector~$(N_k^0)$ with $k=1,2,3,4$ and diagram~\ref{fig:Feyn3k} is a t-channel process mediated via charged exotic fermions $\Psi_1^+, \Psi_2^+$ to charged gauge bosons~$(W^+,W^-)$. Here, $j=1,2$.}
\label{fig:Feyn3}
\end{figure}
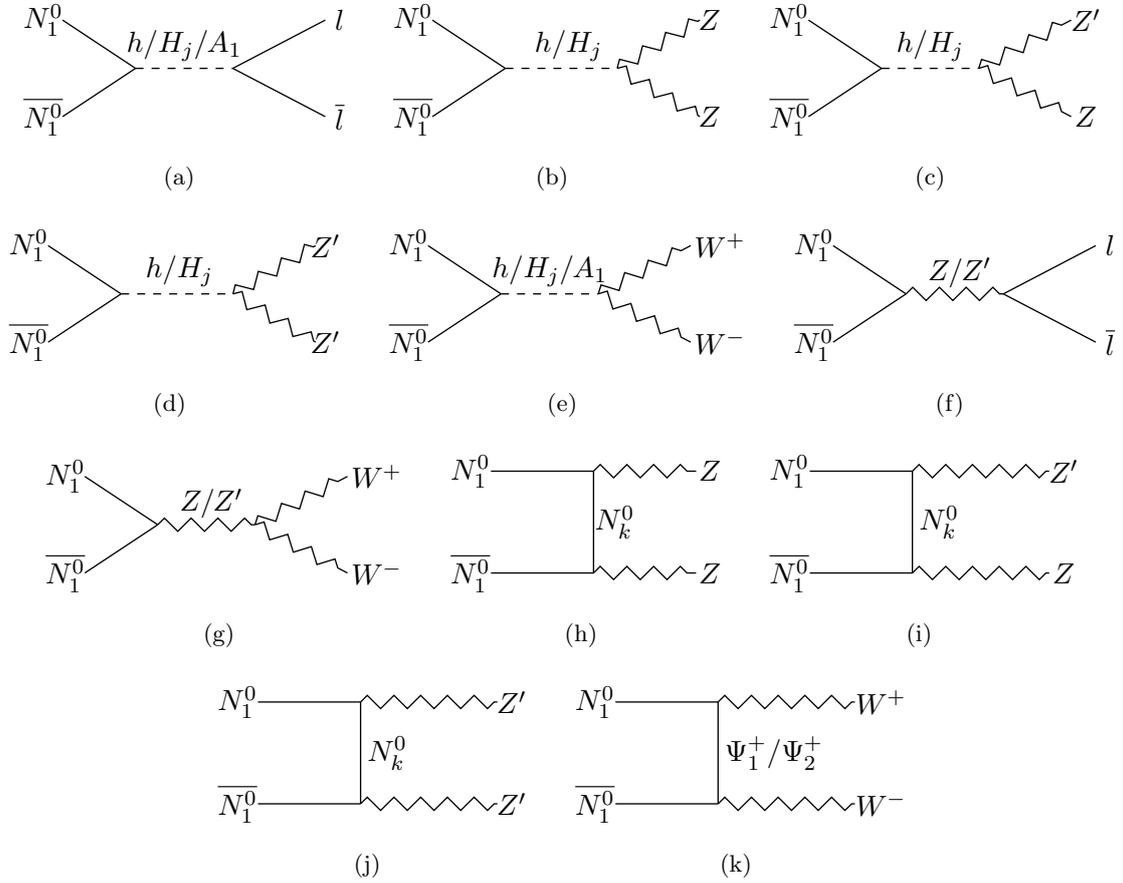

\begin{figure}[H]
		\centering
		\subfigure[]{
		\label{fig:Feyn4a}
		\begin{tikzpicture}[line width=0.5 pt, scale=0.64]
        \draw[solid] (5.0,1.0)--(6.5,0.0);
        \draw[solid] (5.0,-1.0)--(6.5,0.0);
        \draw[dashed] (6.5,0.0)--(8.5,0.0);
        \draw[solid] (8.5,0.0)--(10.4,1.0);
        \draw[solid] (8.5,0.0)--(10.4,-1.0);
        \node at (4.6,1.0) {${N_1^0}$};
        \node at (4.6,-1.0) {$\overline{N_k^0}$};
        \node [above] at (7.5,-0.1) {$h/H_j/A_1$};
        \node at (10.7,1.0) {$l$};
        \node at (10.7,-1.0) {$\overline{l}$};
        \end{tikzpicture}
        }
        \subfigure[]{
        \label{fig:Feyn4c}
        \begin{tikzpicture}[line width=0.5 pt, scale=0.64]
	    \draw[solid] (-3.0,1.0)--(-1.5,0.0);
        \draw[solid] (-3.0,-1.0)--(-1.5,0.0);
        \draw[dashed] (-1.5,0.0)--(0.8,0.0);
        \draw[snake] (0.8,0.0)--(2.5,1.0);
        \draw[snake] (0.8,0.0)--(2.5,-1.0);
        \node at (-3.4,1.0) {${N_1^0}$};
        \node at (-3.4,-1.0) {$\overline{N_k^0}$};
        \node [above] at (-0.2,-0.1) {$h/H_j$};
        \node at (2.7,1.0) {$Z$};
        \node at (2.7,-1.0) {$Z$};
        \end{tikzpicture}
        }
        \subfigure[]{
        \label{fig:Feyn4d}
        \begin{tikzpicture}[line width=0.5 pt, scale=0.64]
        \draw[solid] (5.0,1.0)--(6.5,0.0);
        \draw[solid] (5.0,-1.0)--(6.5,0.0);
        \draw[dashed] (6.5,0.0)--(8.5,0.0);
        \draw[snake] (8.5,0.0)--(10.4,1.0);
        \draw[snake] (8.5,0.0)--(10.4,-1.0);
        \node at (4.6,1.0) {${N_1^0}$};
        \node at (4.6,-1.0) {$\overline{N_k^0}$};
        \node [above] at (7.5,-0.1) {$h/H_j$};
        \node at (10.7,1.0) {$Z'$};
        \node at (10.7,-1.0) {$Z$};
        \end{tikzpicture}
        }
        \subfigure[]{
        \label{fig:Feyn4f}
        \begin{tikzpicture}[line width=0.5 pt, scale=0.64]
	    \draw[solid] (-3.0,1.0)--(-1.5,0.0);
        \draw[solid] (-3.0,-1.0)--(-1.5,0.0);
        \draw[dashed] (-1.5,0.0)--(0.8,0.0);
        \draw[snake] (0.8,0.0)--(2.5,1.0);
        \draw[snake] (0.8,0.0)--(2.5,-1.0);
        \node at (-3.4,1.0) {${N_1^0}$};
        \node at (-3.4,-1.0) {$\overline{N_k^0}$};
        \node [above] at (-0.25,-0.1) {$h/H_j$};
        \node at (2.7,1.0) {$Z'$};
        \node at (2.7,-1.0) {$Z'$};
        \end{tikzpicture}
        }
        \subfigure[]{
        \label{fig:Feyn4h}
        \begin{tikzpicture}[line width=0.5 pt, scale=0.64]
        \draw[solid] (5.0,1.0)--(6.5,0.0);
        \draw[solid] (5.0,-1.0)--(6.5,0.0);
        \draw[dashed] (6.5,0.0)--(8.5,0.0);
        \draw[snake] (8.5,0.0)--(10.4,1.0);
        \draw[snake] (8.5,0.0)--(10.4,-1.0);
        \node at (4.6,1.0) {${N_1^0}$};
        \node at (4.6,-1.0) {$\overline{N_k^0}$};
        \node [above] at (7.5,-0.08) {$h/H_j/A_1$};
        \node at (11.0,1.0) {$W^+$};
        \node at (11.0,-1.0) {$W^-$};
        \end{tikzpicture}
        }
        \subfigure[]{
        \label{fig:Feyn4l}
        \begin{tikzpicture}[line width=0.5 pt, scale=0.64]
        \draw[solid] (5.0,1.0)--(6.5,0.0);
        \draw[solid] (5.0,-1.0)--(6.5,0.0);
        \draw[dashed] (6.5,0.0)--(8.5,0.0);
        \draw[solid] (8.5,0.0)--(10.4,1.0);
        \draw[solid] (8.5,0.0)--(10.4,-1.0);
        \node at (4.6,1.0) {${N_1^0}$};
        \node at (4.6,-1.0) {$\overline{N_k^0}$};
        \node [above] at (7.6,-0.08) {$h/H_j/A_1$};
        \node at (10.7,1.0) {$q$};
        \node at (10.7,-1.0) {$\overline{q}$};
        \end{tikzpicture}
        }
        \subfigure[]{
		\label{fig:Feyn4i}
		\begin{tikzpicture}[line width=0.5 pt, scale=0.64]
        \draw[solid] (5.0,1.0)--(6.5,0.0);
        \draw[solid] (5.0,-1.0)--(6.5,0.0);
        \draw[snake] (6.5,0.0)--(8.5,0.0);
        \draw[solid] (8.5,0.0)--(10.4,1.0);
        \draw[solid] (8.5,0.0)--(10.4,-1.0);
        \node at (4.6,1.0) {${N_1^0}$};
        \node at (4.6,-1.0) {$\overline{N_k^0}$};
        \node [above] at (7.6,-0.08) {$Z/Z'$};
        \node at (10.7,1.0) {$l$};
        \node at (10.7,-1.0) {$\overline{l}$};
        \end{tikzpicture}
        }
        \subfigure[]{
        \label{fig:Feyn4j}
        \begin{tikzpicture}[line width=0.5 pt, scale=0.64]
        \draw[solid] (5.0,1.0)--(6.5,0.0);
        \draw[solid] (5.0,-1.0)--(6.5,0.0);
        \draw[snake] (6.5,0.0)--(8.5,0.0);
        \draw[snake] (8.5,0.0)--(10.4,1.0);
        \draw[snake] (8.5,0.0)--(10.4,-1.0);
        \node at (4.6,1.0) {${N_1^0}$};
        \node at (4.6,-1.0) {$\overline{N_k^0}$};
        \node [above] at (7.6,-0.08) {$Z/Z'$};
        \node at (11.0,1.0) {$W^+$};
        \node at (11.0,-1.0) {$W^-$};
     \end{tikzpicture}
        }
        \subfigure[]{
        \label{fig:Feyn4m}
        \begin{tikzpicture}[line width=0.5 pt, scale=0.64]
        \draw[solid] (5.0,1.0)--(6.5,0.0);
        \draw[solid] (5.0,-1.0)--(6.5,0.0);
        \draw[snake] (6.5,0.0)--(8.5,0.0);
        \draw[solid] (8.5,0.0)--(10.4,1.0);
        \draw[solid] (8.5,0.0)--(10.4,-1.0);
        \node at (4.6,1.0) {${N_1^0}$};
        \node at (4.6,-1.0) {$\overline{N_k^0}$};
        \node [above] at (7.6,-0.08) {$Z/Z'$};
        \node at (10.7,1.0) {$q$};
        \node at (10.7,-1.0) {$\overline{q}$};
        \end{tikzpicture}
        }
        \subfigure[]{
        \label{fig:Feyn4b}
        \begin{tikzpicture}[line width=0.5 pt, scale=0.9]
        \draw[solid] (1.7,1.0)--(3.2,1.0);
        \draw[solid] (1.7,-0.5)--(3.2,-0.5);
        \draw[solid](3.2,1.0)--(3.2,-0.5);
        \draw[snake] (3.2,1.0)--(4.7,1.0);
        \draw[snake] (3.2,-0.5)--(4.7,-0.5);
        \node at (1.4,1.0) {$N_1^0$};
        \node at (1.4,-0.5) {$\overline{N_k^0}$};
        \node [right] at (3.07,0.25) {$N_k^0$};
        \node at (4.9,1.0) {$Z$};
        \node at (4.9,-0.5) {$Z$};
        \end{tikzpicture}
        }
        \subfigure[]{
		\label{fig:Feyn4e}
		\begin{tikzpicture}[line width=0.5 pt, scale=0.9]
	    \draw[solid] (-3.5,1.0)--(-2.0,1.0);
        \draw[solid] (-3.5,-0.5)--(-2.0,-0.5);
        \draw[solid](-2.0,1.0)--(-2.0,-0.5);
        \draw[snake] (-2.0,1.0)--(0.0,1.0);
        \draw[snake] (-2.0,-0.5)--(0.0,-0.5);
        \node at (-3.8,1.0) {${N_1^0}$};
        \node at (-3.8,-0.5) {$\overline{N_k^0}$};
        \node [right] at (-2.05,0.25) {$N_k^0$};
        \node at (0.21,1.0) {$Z'$};
        \node at (0.21,-0.5) {$Z$};
        \end{tikzpicture}
        }
        \subfigure[]{
        \label{fig:Feyn4g}
        \begin{tikzpicture}[line width=0.5 pt, scale=0.9]
	    \draw[solid] (-3.5,1.0)--(-2.0,1.0);
        \draw[solid] (-3.5,-0.5)--(-2.0,-0.5);
        \draw[solid](-2.0,1.0)--(-2.0,-0.5);
        \draw[snake] (-2.0,1.0)--(0.0,1.0);
        \draw[snake] (-2.0,-0.5)--(0.0,-0.5);
        \node at (-3.8,1.0) {${N_1^0}$};
        \node at (-3.8,-0.5) {$\overline{N_k^0}$};
        \node [right] at (-2.05,0.25) {$N_k^0$};
        \node at (0.21,1.0) {$Z'$};
        \node at (0.21,-0.5) {$Z'$};
        \end{tikzpicture}
        }
        \subfigure[]{
        \label{fig:Feyn4k}
        \begin{tikzpicture}[line width=0.5 pt, scale=0.9]
	    \draw[solid] (-3.5,1.0)--(-2.0,1.0);
        \draw[solid] (-3.5,-0.5)--(-2.0,-0.5);
        \draw[solid](-2.0,1.0)--(-2.0,-0.5);
        \draw[snake] (-2.0,1.0)--(0.0,1.0);
        \draw[snake] (-2.0,-0.5)--(0.0,-0.5);
        \node at (-3.8,1.0) {${N_1^0}$};
        \node at (-3.8,-0.5) {$\overline{N_k^0}$};
        \node [right] at (-2.05,0.25) {$\Psi_1^+/\Psi_2^+$};
        \node at (0.38,1.0) {$W^+$};
        \node at (0.38,-0.5) {$W^-$};
        \end{tikzpicture}
        }
     \caption{Co-annihilations of $N_1^0$ with the other neutral fermion states~$(N_k^0)$ to leptons~$(l)$, quarks~$(q)$, SM and exotic gauge boson~$(Z,Z',W^+,W^-)$ final states mediated via Standard model Higgs~$(h)$ and $Z$ boson, and also via exotic scalars~$(H_j/A_1)$ and vector boson~$(Z')$. The diagrams~\ref{fig:Feyn4b}-\ref{fig:Feyn4g} are t-channel processes mediated via mass-states in the Majorana DM sector~$(N_k^0)$ with $k=1,2,3,4$ and diagram~\ref{fig:Feyn4k} is a t-channel process mediated via charged exotic fermions $\Psi_1^+, \Psi_2^+$ to charged gauge bosons~$(W^+,W^-)$. Here, $j=1,2$.}
\label{fig:Feyn4}
\end{figure}
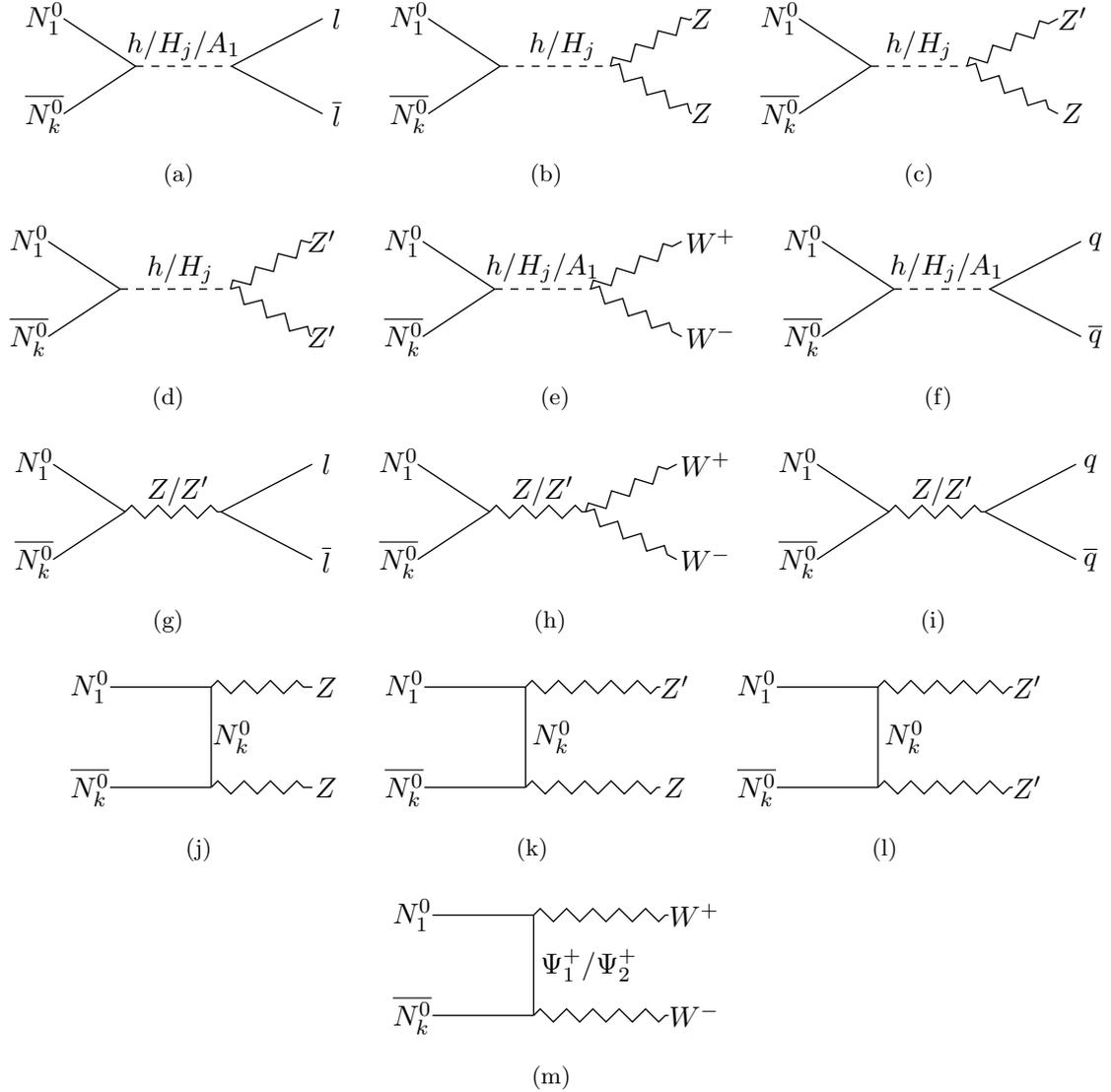
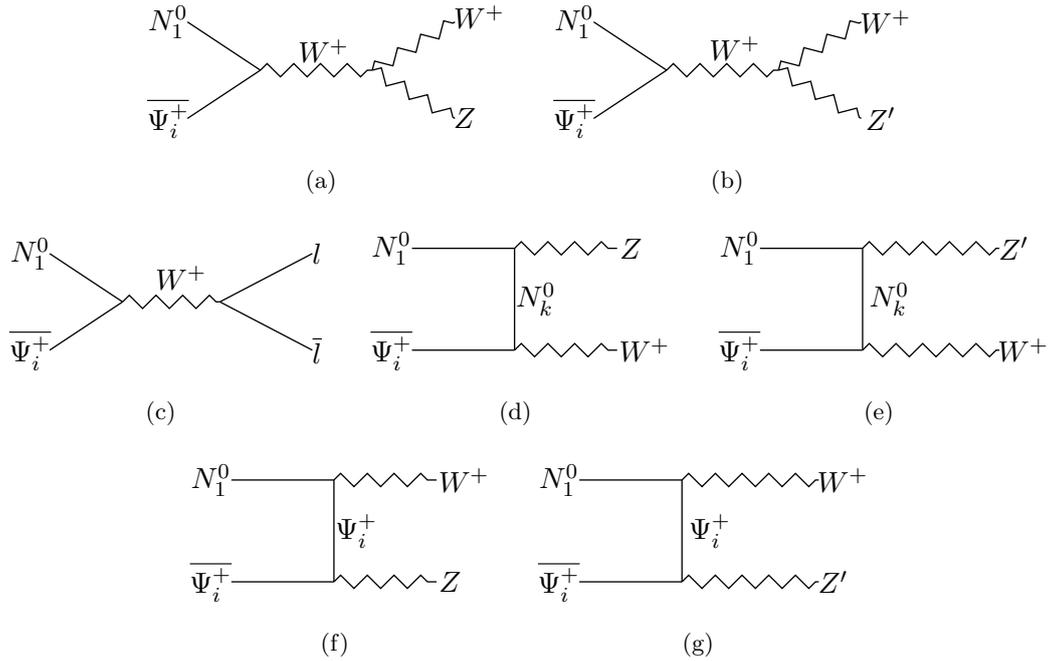
\begin{figure}[H]
        \subfigure[]{
        \label{fig:Feyn5c}
        \begin{tikzpicture}[line width=0.5 pt, scale=0.64]
	    \draw[solid] (-3.0,1.0)--(-1.5,0.0);
        \draw[solid] (-3.0,-1.0)--(-1.5,0.0);
        \draw[snake] (-1.5,0.0)--(0.8,0.0);
        \draw[snake] (0.8,0.0)--(2.5,1.0);
        \draw[snake] (0.8,0.0)--(2.5,-1.0);
        \node at (-3.4,1.0) {${N_1^0}$};
        \node at (-3.4,-1.0) {$\overline{\Psi_i^+}$};
        \node [above] at (-0.2,-0.05) {$W^+$};
        \node at (3.0,1.1) {$W^+$};
        \node at (2.7,-1.0) {$Z$};
        \end{tikzpicture}
        }
        \subfigure[]{
        \label{fig:Feyn5f}
        \begin{tikzpicture}[line width=0.5 pt, scale=0.64]
	    \draw[solid] (-3.0,1.0)--(-1.5,0.0);
        \draw[solid] (-3.0,-1.0)--(-1.5,0.0);
        \draw[snake] (-1.5,0.0)--(0.8,0.0);
        \draw[snake] (0.8,0.0)--(2.5,1.0);
        \draw[snake] (0.8,0.0)--(2.5,-1.0);
        \node at (-3.4,1.0) {${N_1^0}$};
        \node at (-3.4,-1.0) {$\overline{\Psi_i^+}$};
        \node [above] at (-0.1,-0.0) {$W^+$};
        \node at (3.0,1.0) {$W^+$};
        \node at (2.9,-1.0) {$Z'$};
        \end{tikzpicture}
        }
        \subfigure[]{
        \label{fig:Feyn5g}
        \begin{tikzpicture}[line width=0.5 pt, scale=0.64]
	    \draw[solid] (5.0,1.0)--(6.5,0.0);
        \draw[solid] (5.0,-1.0)--(6.5,0.0);
        \draw[snake] (6.5,0.0)--(8.5,0.0);
        \draw[solid] (8.5,0.0)--(10.4,1.0);
        \draw[solid] (8.5,0.0)--(10.4,-1.0);
        \node at (4.6,1.0) {${N_1^0}$};
        \node at (4.6,-1.0) {$\overline{\Psi_i^+}$};
        \node [above] at (7.7,0.05) {$W^+$};
        \node at (10.5,1.0) {$l$};
        \node at (10.5,-1.0) {$\overline{l}$};
        \end{tikzpicture}
        }
        \centering
		\subfigure[]{
		\label{fig:Feyn5a}
		\begin{tikzpicture}[line width=0.5 pt, scale=0.9]
        \draw[solid] (1.7,1.0)--(3.2,1.0);
        \draw[solid] (1.7,-0.5)--(3.2,-0.5);
        \draw[solid](3.2,1.0)--(3.2,-0.5);
        \draw[snake] (3.2,1.0)--(4.7,1.0);
        \draw[snake] (3.2,-0.5)--(4.7,-0.5);
        \node at (1.4,1.0) {$N_1^0$};
        \node at (1.4,-0.5) {$\overline{\Psi_i^+}$};
        \node [right] at (3.07,0.25) {$N_k^0$};
        \node at (4.9,1.0) {$Z$};
        \node at (5.1,-0.5) {$W^+$};
        \end{tikzpicture}
        }
        \subfigure[]{
        \label{fig:Feyn5d}
        \begin{tikzpicture}[line width=0.5 pt, scale=0.9]
        \draw[solid] (-3.5,1.0)--(-2.0,1.0);
        \draw[solid] (-3.5,-0.5)--(-2.0,-0.5);
        \draw[solid](-2.0,1.0)--(-2.0,-0.5);
        \draw[snake] (-2.0,1.0)--(0.0,1.0);
        \draw[snake] (-2.0,-0.5)--(0.0,-0.5);
        \node at (-3.8,1.0) {${N_1^0}$};
        \node at (-3.8,-0.5) {$\overline{\Psi_i^+}$};
        \node [right] at (-2.05,0.25) {$N_k^0$};
        \node at (0.21,1.0) {$Z'$};
        \node at (0.35,-0.5) {$W^+$};
        \end{tikzpicture}
        }
        \subfigure[]{
        \label{fig:Feyn5b}
        \begin{tikzpicture}[line width=0.5 pt, scale=0.9]
        \draw[solid] (1.7,1.0)--(3.2,1.0);
        \draw[solid] (1.7,-0.5)--(3.2,-0.5);
        \draw[solid](3.2,1.0)--(3.2,-0.5);
        \draw[snake] (3.2,1.0)--(4.7,1.0);
        \draw[snake] (3.2,-0.5)--(4.7,-0.5);
        \node at (1.4,1.0) {$N_1^0$};
        \node at (1.4,-0.5) {$\overline{\Psi_i^+}$};
        \node [right] at (3.07,0.25) {$\Psi_i^+$};
        \node at (5.1,1.0) {$W^+$};
        \node at (4.9,-0.5) {$Z$};
        \end{tikzpicture}
        }
        \subfigure[]{
		\label{fig:Feyn5e}
		\begin{tikzpicture}[line width=0.5 pt, scale=0.9]
	    \draw[solid] (-3.5,1.0)--(-2.0,1.0);
        \draw[solid] (-3.5,-0.5)--(-2.0,-0.5);
        \draw[solid](-2.0,1.0)--(-2.0,-0.5);
        \draw[snake] (-2.0,1.0)--(0.0,1.0);
        \draw[snake] (-2.0,-0.5)--(0.0,-0.5);
        \node at (-3.8,1.0) {${N_1^0}$};
        \node at (-3.8,-0.5) {$\overline{\Psi_i^+}$};
        \node [right] at (-2.05,0.25) {$\Psi_i^+$};
        \node at (0.35,1.0) {$W^+$};
        \node at (0.21,-0.5) {$Z'$};
        \end{tikzpicture}
        }
     \caption{Co-annihilations of $N_1^0$ with the charged exotic fermion states~$(\Psi_i^+)$ to leptons~$(l)$, SM and exotic gauge boson~$(Z,Z',W^+)$ final states mediated via $W^+$ boson. The diagrams~\ref{fig:Feyn5a} and~\ref{fig:Feyn5d} are t-channel processes mediated via mass-states in the Majorana DM sector~$(N_k^0)$ with $k=1,2,3,4$ and diagrams~\ref{fig:Feyn5b} and~\ref{fig:Feyn5e} are t-channel processes mediated via charged exotic fermions $\Psi_1^+, \Psi_2^+$ to final states containing combinations of charged and neutral gauge bosons~$(W^+,Z,Z')$. Here, $i=1,2$.}
\label{fig:Feyn5}
\end{figure}
\bibliographystyle{utcaps_mod}
\bibliography{U1L_dm}
\end{document}